\documentclass[apjl]{emulateapj}
\usepackage{comment}
\usepackage{ifthen}

\newcommand{\forloop}[5][1]%
{%
\setcounter{#2}{#3}%
\ifthenelse{#4}%
	{%
	#5%
	\addtocounter{#2}{#1}%
	\forloop[#1]{#2}{\value{#2}}{#4}{#5}%
	}%
	{%
	}%
}%

\newcommand{\ctbd}[1]{}

\newcommand{\lc}{light curve}
\newcommand{\lcs}{light curves}
\newcommand{\Lc}{Light curve}

\newcommand{\band}[1]{\ensuremath{#1}~band}

\newcommand{\kms}{\ensuremath{\rm km\,s^{-1}}}
\newcommand{\ms}{\ensuremath{\rm m\,s^{-1}}}

\newcommand{\gcmc}{\ensuremath{\rm g\,cm^{-3}}}
\newcommand{\ergscmsq}{\ensuremath{\rm erg\,s^{-1}\,cm^{-2}}}

\newcommand{\vsini}{\ensuremath{v \sin{i}}}
\newcommand{\feh}{\ensuremath{\rm [Fe/H]}}

\newcommand{\vmac}{\ensuremath{v_{\rm mac}}}
\newcommand{\vmic}{\ensuremath{v_{\rm mic}}}

\newcommand{\vic}{\ensuremath{V\!-\!I_C}}

\newcommand{\rsun}{\ensuremath{R_\sun}}
\newcommand{\msun}{\ensuremath{M_\sun}}
\newcommand{\lsun}{\ensuremath{L_\sun}}

\newcommand{\rstar}{\ensuremath{R_\star}}
\newcommand{\mstar}{\ensuremath{M_\star}}
\newcommand{\lstar}{\ensuremath{L_\star}}

\newcommand{\teffstar}{\ensuremath{T_{\rm eff\star}}}
\newcommand{\rhostar}{\ensuremath{\rho_\star}}
\newcommand{\loggstar}{\ensuremath{\log{g_{\star}}}}

\newcommand{\rpl}{\ensuremath{R_{p}}}
\newcommand{\mpl}{\ensuremath{M_{p}}}

\newcommand{\rhopl}{\ensuremath{\rho_{p}}}

\newcommand{\arstar}{\ensuremath{a/\rstar}}
\newcommand{\zrstar}{\ensuremath{\zeta/\rstar}}

\newcommand{\rjup}{\ensuremath{R_{\rm J}}}
\newcommand{\mjup}{\ensuremath{M_{\rm J}}}

\newcommand{\reffigl}[1]{Figure~\ref{fig:#1}}
\newcommand{\refsecl}[1]{\mbox{Section \ref{sec:#1}}}

\newcommand{\reftabl}[1]{Table~\ref{tab:#1}}

\newcommand{\reffigls}[2]{Figures~\ref{fig:#1}-\ref{fig:#2}}
\newcommand{\reftabls}[2]{Tables~\ref{tab:#1}-\ref{tab:#2}}

\newcommand{\flwof}{\mbox{FLWO 1.2\,m}}

\newcommand{\flwos}{\mbox{FLWO 1.5\,m}}

\newcommand{\loopand}{\ifnum\value{planetcounter}=2 and \else\fi}
\newcommand{\loopcomma}{\ifnum\value{planetcounter}<2 ,\else. \fi}
\newcommand{\loopcommanoperiod}{\ifnum\value{planetcounter}<2 ,\else \space\fi}
\newcommand{\loopcommanospace}{\ifnum\value{planetcounter}<2 ,\else \fi}

\newcommand{\hatcurhtrxxxA}{HTR125-001}                                
\newcommand{\hatcurfieldxxxA}{125}                                     
\newcommand{\hatcurCCraxxxA}{\ensuremath{02^{\mathrm h}04^{\mathrm m}10.27{\mathrm s}}}                              
\newcommand{\hatcurCCdecxxxA}{\ensuremath{+46{\arcdeg}41{\arcmin}16.2{\arcsec}}}                             
\newcommand{\hatcurCCmagxxxA}{11.289}                                  
\newcommand{\hatcurCCtwomassxxxA}{2MASS~02041028+4641162}              
\newcommand{\hatcurCCgscxxxA}{GSC~3281-00800}                          
\newcommand{\hatcurCCtassmvxxxA}{11.289}                               
\newcommand{\hatcurCCtwomassJmagxxxA}{\ensuremath{10.251\pm0.022}}     
\newcommand{\hatcurCCtwomassHmagxxxA}{\ensuremath{10.024\pm0.022}}     
\newcommand{\hatcurCCtwomassKmagxxxA}{\ensuremath{9.990\pm0.022}}      
\newcommand{\hatcurCCcitJmagxxxA}{\ensuremath{10.273\pm0.023}}         
\newcommand{\hatcurCCcitHmagxxxA}{\ensuremath{10.019\pm0.023}}         
\newcommand{\hatcurCCcitKmagxxxA}{\ensuremath{10.014\pm0.022}}         
\newcommand{\hatcurCCbbJmagxxxA}{\ensuremath{10.313\pm0.023}}          
\newcommand{\hatcurCCbbHmagxxxA}{\ensuremath{10.040\pm0.023}}          
\newcommand{\hatcurCCbbKmagxxxA}{\ensuremath{10.034\pm0.022}}          
\newcommand{\hatcurCCesoJmagxxxA}{\ensuremath{10.315\pm0.024}}         
\newcommand{\hatcurCCesoHmagxxxA}{\ensuremath{10.034\pm0.023}}         
\newcommand{\hatcurCCesoKmagxxxA}{\ensuremath{10.034\pm0.022}}         
\newcommand{\hatcurCCesoJHmagxxxA}{\ensuremath{0.281\pm0.005}}         
\newcommand{\hatcurCCesoJKmagxxxA}{\ensuremath{0.281\pm0.008}}         
\newcommand{\hatcurCCesoHKmagxxxA}{\ensuremath{0.000\pm0.006}}         
\newcommand{\hatcurLCdipxxxA}{\ensuremath{20.3}}                       
\newcommand{\hatcurLCrprstarxxxA}{\ensuremath{0.1508\pm0.0004}}        
\newcommand{\hatcurLCbsqxxxA}{\ensuremath{0.014_{-0.008}^{+0.014}}}    
\newcommand{\hatcurLCimpxxxA}{\ensuremath{0.117_{-0.047}^{+0.045}}}    
\newcommand{\hatcurLCzetaxxxA}{\ensuremath{17.80\pm0.03}}              
\newcommand{\hatcurLCdurxxxA}{\ensuremath{0.1295\pm0.0003}}            
\newcommand{\hatcurLCdurshortxxxA}{\ensuremath{0.1295}}                
\newcommand{\hatcurLCdurhrxxxA}{\ensuremath{3.109\pm0.007}}            
\newcommand{\hatcurLCdurhrshortxxxA}{\ensuremath{3.109}}               
\newcommand{\hatcurLCqxxxA}{\ensuremath{0.0602\pm0.0001}}              
\newcommand{\hatcurLCqshortxxxA}{\ensuremath{0.060}}                   
\newcommand{\hatcurLCingdurxxxA}{\ensuremath{0.0172\pm0.0002}}         
\newcommand{\hatcurLCPxxxA}{\ensuremath{2.150008\pm0.000001}}          
\newcommand{\hatcurLCPprecxxxA}{\ensuremath{2.1500085}}                
\newcommand{\hatcurLCPshortxxxA}{\ensuremath{2.1500}}                  
\newcommand{\hatcurLCTxxxA}{\ensuremath{2454420.44637\pm0.00009}}      
\newcommand{\hatcurLCTAxxxA}{\ensuremath{2453364.79222\pm0.00066}}     
\newcommand{\hatcurLCTBxxxA}{\ensuremath{2454439.79645\pm0.00009}}     
\newcommand{\hatcurLChatnetmAxxxA}{\ensuremath{10.5995\pm0.0002}}      
\newcommand{\hatcurLCiblendAxxxA}{\ensuremath{0.69\pm0.03}}            
\newcommand{\hatcurLChatnetmBxxxA}{\ensuremath{10.5988\pm0.0001}}      
\newcommand{\hatcurLCiblendBxxxA}{\ensuremath{0.87\pm0.02}}            
\newcommand{\hatcurSMEiteffxxxA}{\ensuremath{6001\pm88}}               
\newcommand{\hatcurSMEizfehxxxA}{\ensuremath{-0.16\pm0.08}}            
\newcommand{\hatcurSMEizfehshortxxxA}{\ensuremath{-0.16}}              
\newcommand{\hatcurSMEiloggxxxA}{\ensuremath{4.02\pm0.07}}             
\newcommand{\hatcurSMEivsinxxxA}{\ensuremath{21\pm0.5}}                
\newcommand{\hatcurSMEivmacxxxA}{\ensuremath{4.3}}                     
\newcommand{\hatcurSMEivmicxxxA}{\ensuremath{0.0}}                     
\newcommand{\hatcurSMEiiteffxxxA}{\ensuremath{6207\pm88}}              
\newcommand{\hatcurSMEiizfehxxxA}{\ensuremath{-0.04\pm0.08}}           
\newcommand{\hatcurSMEiizfehshortxxxA}{\ensuremath{-0.04}}             
\newcommand{\hatcurSMEiiloggxxxA}{\ensuremath{4.31\pm0.06}}            
\newcommand{\hatcurSMEiivsinxxxA}{\ensuremath{20.7\pm0.5}}             
\newcommand{\hatcurSMEiivmacxxxA}{\ensuremath{4.69}}                   
\newcommand{\hatcurSMEiivmicxxxA}{\ensuremath{0.85}}                   
\newcommand{\hatcurDSteffxxxA}{\ensuremath{6500\pm100}}                
\newcommand{\hatcurDSzfehxxxA}{\ensuremath{0.0\pm0.0}}                 
\newcommand{\hatcurDSloggxxxA}{\ensuremath{4.5\pm0.25}}                
\newcommand{\hatcurDSvsinixxxA}{\ensuremath{21.9\pm1.0}}               
\newcommand{\hatcurDSgammaxxxA}{\ensuremath{-23.21\pm0.26}}            
\newcommand{\hatcurDSnumspecxxxA}{\ensuremath{15}}                     
\newcommand{\hatcurDSspanxxxA}{\ensuremath{738}}                       
\newcommand{\hatcurDSrvrmsxxxA}{\ensuremath{0.97}}                     
\newcommand{\hatcurTRESteffxxxA}{\ensuremath{NULL\pmNULL}}             
\newcommand{\hatcurTRESzfehxxxA}{\ensuremath{NULL\pmNULL}}             
\newcommand{\hatcurTRESloggxxxA}{\ensuremath{NULL\pmNULL}}             
\newcommand{\hatcurTRESvsinixxxA}{\ensuremath{NULL\pmNULL}}            
\newcommand{\hatcurTRESgammaxxxA}{\ensuremath{NULL\pmNULL}}            
\newcommand{\hatcurTRESnumspecxxxA}{\ensuremath{0}}                    
\newcommand{\hatcurTRESspanxxxA}{\ensuremath{0}}                       
\newcommand{\hatcurTRESrvrmsxxxA}{\ensuremath{0.00}}                   
\newcommand{\hatcurFIESteffxxxA}{\ensuremath{NULL\pmNULL}}             
\newcommand{\hatcurFIESzfehxxxA}{\ensuremath{NULL\pmNULL}}             
\newcommand{\hatcurFIESloggxxxA}{\ensuremath{NULL\pmNULL}}             
\newcommand{\hatcurFIESvsinixxxA}{\ensuremath{NULL\pmNULL}}            
\newcommand{\hatcurFIESgammaxxxA}{\ensuremath{NULL\pmNULL}}            
\newcommand{\hatcurFIESnumspecxxxA}{\ensuremath{0}}                    
\newcommand{\hatcurFIESspanxxxA}{\ensuremath{0}}                       
\newcommand{\hatcurFIESrvrmsxxxA}{\ensuremath{0.00}}                   
\newcommand{\hatcurLBizxxxA}{\ensuremath{0.1527}}                      
\newcommand{\hatcurLBiizxxxA}{\ensuremath{0.3513}}                     
\newcommand{\hatcurLBiixxxA}{\ensuremath{0.2045}}                      
\newcommand{\hatcurLBiiixxxA}{\ensuremath{0.3593}}                     
\newcommand{\hatcurLBiIxxxA}{\ensuremath{0.1863}}                      
\newcommand{\hatcurLBiiIxxxA}{\ensuremath{0.3577}}                     
\newcommand{\hatcurLBigxxxA}{\ensuremath{0.4460}}                      
\newcommand{\hatcurLBiigxxxA}{\ensuremath{0.3107}}                     
\newcommand{\hatcurLBikepxxxA}{\ensuremath{0.1000}}                    
\newcommand{\hatcurLBiikepxxxA}{\ensuremath{0.1000}}                   
\newcommand{\hatcurISOmxxxA}{\ensuremath{1.16\pm0.04}}                 
\newcommand{\hatcurISOmshortxxxA}{\ensuremath{1.16}}                   
\newcommand{\hatcurISOmlongxxxA}{\ensuremath{1.160\pm0.041}}           
\newcommand{\hatcurISOrxxxA}{\ensuremath{1.22\pm0.02}}                 
\newcommand{\hatcurISOrshortxxxA}{\ensuremath{1.22}}                   
\newcommand{\hatcurISOrlongxxxA}{\ensuremath{1.219\pm0.016}}           
\newcommand{\hatcurISOrhoxxxA}{\ensuremath{0.91_{-0.02}^{+0.01}}}      
\newcommand{\hatcurISOloggxxxA}{\ensuremath{4.33\pm0.01}}              
\newcommand{\hatcurISOlumxxxA}{\ensuremath{1.97\pm0.15}}               
\newcommand{\hatcurISOlumshortxxxA}{\ensuremath{1.97}}                 
\newcommand{\hatcurISOmvxxxA}{\ensuremath{4.04\pm0.10}}                
\newcommand{\hatcurISOvixxxA}{\ensuremath{0.574\pm0.023}}              
\newcommand{\hatcurISOagexxxA}{\ensuremath{2.7\pm0.8}}                 
\newcommand{\hatcurISOsigmaxxxA}{\ensuremath{0.00080\pm0.00015}}       
\newcommand{\hatcurISOMJxxxA}{\ensuremath{3.09\pm0.05}}                
\newcommand{\hatcurISOMHxxxA}{\ensuremath{2.82\pm0.04}}                
\newcommand{\hatcurISOMKxxxA}{\ensuremath{2.77\pm0.04}}                
\newcommand{\hatcurISOJKxxxA}{\ensuremath{0.32\pm0.02}}                
\newcommand{\hatcurISOspecxxxA}{G0}                                    
\newcommand{\hatcurRVKxxxA}{\ensuremath{122.8\pm23.2}}                 
\newcommand{\hatcurRVkxxxA}{\ensuremath{0.000\pm0.000}}                
\newcommand{\hatcurRVhxxxA}{\ensuremath{0.000\pm0.000}}                
\newcommand{\hatcurRVgammaxxxA}{\ensuremath{54.9\pm16.8}}              
\newcommand{\hatcurRVjitterxxxA}{\ensuremath{78.7}}                    
\newcommand{\hatcurRVeccenxxxA}{\ensuremath{0.000\pm0.000}}            
\newcommand{\hatcurRVomegaxxxA}{\ensuremath{0\pm0}}                    
\newcommand{\hatcurRVfitrmsxxxA}{\ensuremath{80.3}}                    
\newcommand{\hatcurPPixxxA}{\ensuremath{88.9\pm0.4}}                   
\newcommand{\hatcurPPgxxxA}{\ensuremath{6.7\pm1.3}}                    
\newcommand{\hatcurPPloggxxxA}{\ensuremath{2.82_{-0.10}^{+0.07}}}      
\newcommand{\hatcurPParxxxA}{\ensuremath{6.05_{-0.04}^{+0.03}}}        
\newcommand{\hatcurPParelxxxA}{\ensuremath{0.0343\pm0.0004}}           
\newcommand{\hatcurPPrhoxxxA}{\ensuremath{0.19\pm0.04}}                
\newcommand{\hatcurPPmxxxA}{\ensuremath{0.86\pm0.16}}                  
\newcommand{\hatcurPPmshortxxxA}{\ensuremath{0.86}}                    
\newcommand{\hatcurPPmlongxxxA}{\ensuremath{0.860\pm0.164}}            
\newcommand{\hatcurPPmexxxA}{\ensuremath{273.3\pm52.1}}                
\newcommand{\hatcurPPmeshortxxxA}{\ensuremath{273.3}}                  
\newcommand{\hatcurPPmelongxxxA}{\ensuremath{273.31\pm52.12}}          
\newcommand{\hatcurPPrxxxA}{\ensuremath{1.79\pm0.03}}                  
\newcommand{\hatcurPPrshortxxxA}{\ensuremath{1.79}}                    
\newcommand{\hatcurPPrlongxxxA}{\ensuremath{1.789\pm0.025}}            
\newcommand{\hatcurPPrexxxA}{\ensuremath{20.1\pm0.3}}                  
\newcommand{\hatcurPPreshortxxxA}{\ensuremath{20.1}}                   
\newcommand{\hatcurPPrelongxxxA}{\ensuremath{20.06\pm0.28}}            
\newcommand{\hatcurPPmrcorrxxxA}{\ensuremath{0.10}}                    
\newcommand{\hatcurPPteffxxxA}{\ensuremath{1786\pm26}}                 
\newcommand{\hatcurPPthetaxxxA}{\ensuremath{0.028\pm0.005}}            
\newcommand{\hatcurPPfluxperixxxA}{\ensuremath{2.29\pm0.13}}           
\newcommand{\hatcurPPfluxperidimxxxA}{\ensuremath{9}}                  
\newcommand{\hatcurPPfluxapxxxA}{\ensuremath{2.29\pm0.13}}             
\newcommand{\hatcurPPfluxapdimxxxA}{\ensuremath{9}}                    
\newcommand{\hatcurPPfluxavgxxxA}{\ensuremath{2.29\pm0.13}}            
\newcommand{\hatcurPPfluxavgdimxxxA}{\ensuremath{9}}                   
\newcommand{\hatcurXsecphasexxxA}{\ensuremath{0.5000\pm0.0000}}        
\newcommand{\hatcurXsecondaryxxxA}{\ensuremath{2454421.521\pm0.000}}   
\newcommand{\hatcurXsecdurxxxA}{\ensuremath{0.1295\pm0.0003}}          
\newcommand{\hatcurXsecingdurxxxA}{\ensuremath{0.0172\pm0.0002}}       
\newcommand{\hatcurPPphiconjxxxA}{\ensuremath{0.2500\pm0.0000}}        
\newcommand{\hatcurPPperixxxA}{\ensuremath{2454419.91\pm0.00}}         
\newcommand{\hatcurPPaequivxxxA}{\ensuremath{0.0244\pm0.0007}}         
\newcommand{\hatcurPPtcircxxxA}{\ensuremath{4.3\pm0.8}}                
\newcommand{\hatcurPPtinfallxxxA}{\ensuremath{221.9_{-33.1}^{+64.3}}}  
\newcommand{\hatcurXdistxxxA}{\ensuremath{283\pm5}}                    
\newcommand{\hatcurCCpmraxxxA}{\ensuremath{-12.3\pm1.3}}               
\newcommand{\hatcurCCpmdecxxxA}{\ensuremath{4.6\pm2.4}}                
\newcommand{\hatcurCCpmxxxA}{\ensuremath{13.132\pm2.72947}}            
\newcommand{\hatcurhtrxxxAe}{HTR125-001}                               
\newcommand{\hatcurfieldxxxAe}{125}                                    
\newcommand{\hatcurCCraxxxAe}{\ensuremath{02^{\mathrm h}04^{\mathrm m}10.27{\mathrm s}}}                             
\newcommand{\hatcurCCdecxxxAe}{\ensuremath{+46{\arcdeg}41{\arcmin}16.2{\arcsec}}}                            
\newcommand{\hatcurCCmagxxxAe}{11.289}                                 
\newcommand{\hatcurCCtwomassxxxAe}{2MASS~02041028+4641162}             
\newcommand{\hatcurCCgscxxxAe}{GSC~3281-00800}                         
\newcommand{\hatcurCCtassmvxxxAe}{11.289}                              
\newcommand{\hatcurCCtwomassJmagxxxAe}{\ensuremath{10.251\pm0.022}}    
\newcommand{\hatcurCCtwomassHmagxxxAe}{\ensuremath{10.024\pm0.022}}    
\newcommand{\hatcurCCtwomassKmagxxxAe}{\ensuremath{9.990\pm0.022}}     
\newcommand{\hatcurCCcitJmagxxxAe}{\ensuremath{10.273\pm0.023}}        
\newcommand{\hatcurCCcitHmagxxxAe}{\ensuremath{10.019\pm0.023}}        
\newcommand{\hatcurCCcitKmagxxxAe}{\ensuremath{10.014\pm0.022}}        
\newcommand{\hatcurCCbbJmagxxxAe}{\ensuremath{10.313\pm0.023}}         
\newcommand{\hatcurCCbbHmagxxxAe}{\ensuremath{10.040\pm0.023}}         
\newcommand{\hatcurCCbbKmagxxxAe}{\ensuremath{10.034\pm0.022}}         
\newcommand{\hatcurCCesoJmagxxxAe}{\ensuremath{10.315\pm0.024}}        
\newcommand{\hatcurCCesoHmagxxxAe}{\ensuremath{10.034\pm0.023}}        
\newcommand{\hatcurCCesoKmagxxxAe}{\ensuremath{10.034\pm0.022}}        
\newcommand{\hatcurCCesoJHmagxxxAe}{\ensuremath{0.281\pm0.005}}        
\newcommand{\hatcurCCesoJKmagxxxAe}{\ensuremath{0.281\pm0.008}}        
\newcommand{\hatcurCCesoHKmagxxxAe}{\ensuremath{0.000\pm0.006}}        
\newcommand{\hatcurLCdipxxxAe}{\ensuremath{20.3}}                      
\newcommand{\hatcurLCrprstarxxxAe}{\ensuremath{0.1508\pm0.0004}}       
\newcommand{\hatcurLCbsqxxxAe}{\ensuremath{0.011_{-0.007}^{+0.012}}}   
\newcommand{\hatcurLCimpxxxAe}{\ensuremath{0.106_{-0.044}^{+0.043}}}   
\newcommand{\hatcurLCzetaxxxAe}{\ensuremath{17.85\pm0.04}}             
\newcommand{\hatcurLCdurxxxAe}{\ensuremath{0.1291\pm0.0003}}           
\newcommand{\hatcurLCdurshortxxxAe}{\ensuremath{0.1291}}               
\newcommand{\hatcurLCdurhrxxxAe}{\ensuremath{3.099\pm0.008}}           
\newcommand{\hatcurLCdurhrshortxxxAe}{\ensuremath{3.099}}              
\newcommand{\hatcurLCqxxxAe}{\ensuremath{0.0601\pm0.0002}}             
\newcommand{\hatcurLCqshortxxxAe}{\ensuremath{0.060}}                  
\newcommand{\hatcurLCingdurxxxAe}{\ensuremath{0.0171\pm0.0002}}        
\newcommand{\hatcurLCPxxxAe}{\ensuremath{2.150009\pm0.000001}}         
\newcommand{\hatcurLCPprecxxxAe}{\ensuremath{2.1500085}}               
\newcommand{\hatcurLCPshortxxxAe}{\ensuremath{2.1500}}                 
\newcommand{\hatcurLCTxxxAe}{\ensuremath{2454416.14639\pm0.00009}}     
\newcommand{\hatcurLCTAxxxAe}{\ensuremath{2453364.79222\pm0.00064}}    
\newcommand{\hatcurLCTBxxxAe}{\ensuremath{2454439.79649\pm0.00010}}    
\newcommand{\hatcurLChatnetmAxxxAe}{\ensuremath{10.5995\pm0.0002}}     
\newcommand{\hatcurLCiblendAxxxAe}{\ensuremath{0.69\pm0.03}}           
\newcommand{\hatcurLChatnetmBxxxAe}{\ensuremath{10.5988\pm0.0001}}     
\newcommand{\hatcurLCiblendBxxxAe}{\ensuremath{0.87\pm0.02}}           
\newcommand{\hatcurSMEiteffxxxAe}{\ensuremath{6001\pm88}}              
\newcommand{\hatcurSMEizfehxxxAe}{\ensuremath{-0.16\pm0.08}}           
\newcommand{\hatcurSMEizfehshortxxxAe}{\ensuremath{-0.16}}             
\newcommand{\hatcurSMEiloggxxxAe}{\ensuremath{4.02\pm0.07}}            
\newcommand{\hatcurSMEivsinxxxAe}{\ensuremath{21\pm0.5}}               
\newcommand{\hatcurSMEivmacxxxAe}{\ensuremath{4.3}}                    
\newcommand{\hatcurSMEivmicxxxAe}{\ensuremath{0.0}}                    
\newcommand{\hatcurSMEiiteffxxxAe}{\ensuremath{6136\pm88}}             
\newcommand{\hatcurSMEiizfehxxxAe}{\ensuremath{-0.08\pm0.08}}          
\newcommand{\hatcurSMEiizfehshortxxxAe}{\ensuremath{-0.08}}            
\newcommand{\hatcurSMEiiloggxxxAe}{\ensuremath{4.21\pm0.04}}           
\newcommand{\hatcurSMEiivsinxxxAe}{\ensuremath{20.8\pm0.5}}            
\newcommand{\hatcurSMEiivmacxxxAe}{\ensuremath{4.69}}                  
\newcommand{\hatcurSMEiivmicxxxAe}{\ensuremath{0.85}}                  
\newcommand{\hatcurDSteffxxxAe}{\ensuremath{6500\pm100}}               
\newcommand{\hatcurDSzfehxxxAe}{\ensuremath{0.0\pm0.0}}                
\newcommand{\hatcurDSloggxxxAe}{\ensuremath{4.5\pm0.25}}               
\newcommand{\hatcurDSvsinixxxAe}{\ensuremath{21.9\pm1.0}}              
\newcommand{\hatcurDSgammaxxxAe}{\ensuremath{-23.21\pm0.26}}           
\newcommand{\hatcurDSnumspecxxxAe}{\ensuremath{15}}                    
\newcommand{\hatcurDSspanxxxAe}{\ensuremath{738}}                      
\newcommand{\hatcurDSrvrmsxxxAe}{\ensuremath{0.97}}                    
\newcommand{\hatcurTRESteffxxxAe}{\ensuremath{NULL\pmNULL}}            
\newcommand{\hatcurTRESzfehxxxAe}{\ensuremath{NULL\pmNULL}}            
\newcommand{\hatcurTRESloggxxxAe}{\ensuremath{NULL\pmNULL}}            
\newcommand{\hatcurTRESvsinixxxAe}{\ensuremath{NULL\pmNULL}}           
\newcommand{\hatcurTRESgammaxxxAe}{\ensuremath{NULL\pmNULL}}           
\newcommand{\hatcurTRESnumspecxxxAe}{\ensuremath{0}}                   
\newcommand{\hatcurTRESspanxxxAe}{\ensuremath{0}}                      
\newcommand{\hatcurTRESrvrmsxxxAe}{\ensuremath{0.00}}                  
\newcommand{\hatcurFIESteffxxxAe}{\ensuremath{NULL\pmNULL}}            
\newcommand{\hatcurFIESzfehxxxAe}{\ensuremath{NULL\pmNULL}}            
\newcommand{\hatcurFIESloggxxxAe}{\ensuremath{NULL\pmNULL}}            
\newcommand{\hatcurFIESvsinixxxAe}{\ensuremath{NULL\pmNULL}}           
\newcommand{\hatcurFIESgammaxxxAe}{\ensuremath{NULL\pmNULL}}           
\newcommand{\hatcurFIESnumspecxxxAe}{\ensuremath{0}}                   
\newcommand{\hatcurFIESspanxxxAe}{\ensuremath{0}}                      
\newcommand{\hatcurFIESrvrmsxxxAe}{\ensuremath{0.00}}                  
\newcommand{\hatcurLBizxxxAe}{\ensuremath{0.1580}}                     
\newcommand{\hatcurLBiizxxxAe}{\ensuremath{0.3476}}                    
\newcommand{\hatcurLBiixxxAe}{\ensuremath{0.2098}}                     
\newcommand{\hatcurLBiiixxxAe}{\ensuremath{0.3562}}                    
\newcommand{\hatcurLBiIxxxAe}{\ensuremath{0.1915}}                     
\newcommand{\hatcurLBiiIxxxAe}{\ensuremath{0.3543}}                    
\newcommand{\hatcurLBigxxxAe}{\ensuremath{0.4564}}                     
\newcommand{\hatcurLBiigxxxAe}{\ensuremath{0.3027}}                    
\newcommand{\hatcurLBikepxxxAe}{\ensuremath{0.1000}}                   
\newcommand{\hatcurLBiikepxxxAe}{\ensuremath{0.1000}}                  
\newcommand{\hatcurISOmxxxAe}{\ensuremath{1.18_{-0.07}^{+0.05}}}       
\newcommand{\hatcurISOmshortxxxAe}{\ensuremath{1.18}}                  
\newcommand{\hatcurISOmlongxxxAe}{\ensuremath{1.179_{-0.070}^{+0.049}}} 
\newcommand{\hatcurISOrxxxAe}{\ensuremath{1.40\pm0.08}}                
\newcommand{\hatcurISOrshortxxxAe}{\ensuremath{1.40}}                  
\newcommand{\hatcurISOrlongxxxAe}{\ensuremath{1.401\pm0.082}}          
\newcommand{\hatcurISOrhoxxxAe}{\ensuremath{0.60\pm0.09}}              
\newcommand{\hatcurISOloggxxxAe}{\ensuremath{4.21\pm0.04}}             
\newcommand{\hatcurISOlumxxxAe}{\ensuremath{2.49_{-0.30}^{+0.40}}}     
\newcommand{\hatcurISOlumshortxxxAe}{\ensuremath{2.49}}                
\newcommand{\hatcurISOmvxxxAe}{\ensuremath{3.80\pm0.16}}               
\newcommand{\hatcurISOvixxxAe}{\ensuremath{0.592\pm0.024}}             
\newcommand{\hatcurISOagexxxAe}{\ensuremath{3.8_{-0.5}^{+1.5}}}        
\newcommand{\hatcurISOsigmaxxxAe}{\ensuremath{0.00060\pm0.00013}}      
\newcommand{\hatcurISOMJxxxAe}{\ensuremath{2.82\pm0.14}}               
\newcommand{\hatcurISOMHxxxAe}{\ensuremath{2.53\pm0.13}}               
\newcommand{\hatcurISOMKxxxAe}{\ensuremath{2.48\pm0.13}}               
\newcommand{\hatcurISOJKxxxAe}{\ensuremath{0.34\pm0.02}}               
\newcommand{\hatcurISOspecxxxAe}{G0}                                   
\newcommand{\hatcurRVKxxxAe}{\ensuremath{135.3\pm24.2}}                
\newcommand{\hatcurRVkxxxAe}{\ensuremath{0.114\pm0.094}}               
\newcommand{\hatcurRVhxxxAe}{\ensuremath{0.132\pm0.041}}               
\newcommand{\hatcurRVgammaxxxAe}{\ensuremath{63.6\pm16.3}}             
\newcommand{\hatcurRVjitterxxxAe}{\ensuremath{73.3}}                   
\newcommand{\hatcurRVeccenxxxAe}{\ensuremath{0.177\pm0.079}}           
\newcommand{\hatcurRVomegaxxxAe}{\ensuremath{48\pm28}}                 
\newcommand{\hatcurRVfitrmsxxxAe}{\ensuremath{75.0}}                   
\newcommand{\hatcurPPixxxAe}{\ensuremath{88.7\pm0.6}}                  
\newcommand{\hatcurPPgxxxAe}{\ensuremath{5.5\pm1.0}}                   
\newcommand{\hatcurPPloggxxxAe}{\ensuremath{2.74\pm0.08}}              
\newcommand{\hatcurPParxxxAe}{\ensuremath{5.27\pm0.26}}                
\newcommand{\hatcurPParelxxxAe}{\ensuremath{0.0344_{-0.0007}^{+0.0005}}} 
\newcommand{\hatcurPPrhoxxxAe}{\ensuremath{0.13\pm0.03}}               
\newcommand{\hatcurPPmxxxAe}{\ensuremath{0.93\pm0.17}}                 
\newcommand{\hatcurPPmshortxxxAe}{\ensuremath{0.93}}                   
\newcommand{\hatcurPPmlongxxxAe}{\ensuremath{0.934\pm0.169}}           
\newcommand{\hatcurPPmexxxAe}{\ensuremath{296.7\pm53.6}}               
\newcommand{\hatcurPPmeshortxxxAe}{\ensuremath{296.7}}                 
\newcommand{\hatcurPPmelongxxxAe}{\ensuremath{296.73\pm53.58}}         
\newcommand{\hatcurPPrxxxAe}{\ensuremath{2.06\pm0.12}}                 
\newcommand{\hatcurPPrshortxxxAe}{\ensuremath{2.06}}                   
\newcommand{\hatcurPPrlongxxxAe}{\ensuremath{2.056\pm0.121}}           
\newcommand{\hatcurPPrexxxAe}{\ensuremath{23.0\pm1.4}}                 
\newcommand{\hatcurPPreshortxxxAe}{\ensuremath{23.0}}                  
\newcommand{\hatcurPPrelongxxxAe}{\ensuremath{23.05\pm1.35}}           
\newcommand{\hatcurPPmrcorrxxxAe}{\ensuremath{0.27}}                   
\newcommand{\hatcurPPteffxxxAe}{\ensuremath{1899\pm62}}                
\newcommand{\hatcurPPthetaxxxAe}{\ensuremath{0.027\pm0.005}}           
\newcommand{\hatcurPPfluxperixxxAe}{\ensuremath{4.27_{-0.91}^{+2.04}}} 
\newcommand{\hatcurPPfluxperidimxxxAe}{\ensuremath{9}}                 
\newcommand{\hatcurPPfluxapxxxAe}{\ensuremath{2.07\pm0.16}}            
\newcommand{\hatcurPPfluxapdimxxxAe}{\ensuremath{9}}                   
\newcommand{\hatcurPPfluxavgxxxAe}{\ensuremath{2.93_{-0.32}^{+0.46}}}  
\newcommand{\hatcurPPfluxavgdimxxxAe}{\ensuremath{9}}                  
\newcommand{\hatcurXsecphasexxxAe}{\ensuremath{0.5732\pm0.0602}}       
\newcommand{\hatcurXsecondaryxxxAe}{\ensuremath{2454417.379\pm0.130}}  
\newcommand{\hatcurXsecdurxxxAe}{\ensuremath{0.1677\pm0.0136}}         
\newcommand{\hatcurXsecingdurxxxAe}{\ensuremath{0.0224\pm0.0019}}      
\newcommand{\hatcurPPphiconjxxxAe}{\ensuremath{0.0753_{-0.0789}^{+0.0193}}} 
\newcommand{\hatcurPPperixxxAe}{\ensuremath{2454415.98\pm0.12}}        
\newcommand{\hatcurPPaequivxxxAe}{\ensuremath{0.0218\pm0.0013}}        
\newcommand{\hatcurPPtcircxxxAe}{\ensuremath{3.0\pm0.9}}               
\newcommand{\hatcurPPtinfallxxxAe}{\ensuremath{103.8_{-26.2}^{+48.4}}} 
\newcommand{\hatcurXdistxxxAe}{\ensuremath{324\pm19}}                  
\newcommand{\hatcurCCpmraxxxAe}{\ensuremath{-12.3\pm1.3}}              
\newcommand{\hatcurCCpmdecxxxAe}{\ensuremath{4.6\pm2.4}}               
\newcommand{\hatcurCCpmxxxAe}{\ensuremath{13.132\pm2.72947}}           
\newcommand{\hatcurhtrxxxAer}{HTR125-001}                              
\newcommand{\hatcurfieldxxxAer}{125}                                   
\newcommand{\hatcurCCraxxxAer}{\ensuremath{02^{\mathrm h}04^{\mathrm m}10.27{\mathrm s}}}                            
\newcommand{\hatcurCCdecxxxAer}{\ensuremath{+46{\arcdeg}41{\arcmin}16.2{\arcsec}}}                           
\newcommand{\hatcurCCmagxxxAer}{11.289}                                
\newcommand{\hatcurCCtwomassxxxAer}{2MASS~02041028+4641162}            
\newcommand{\hatcurCCgscxxxAer}{GSC~3281-00800}                        
\newcommand{\hatcurCCtassmvxxxAer}{11.289}                             
\newcommand{\hatcurCCtwomassJmagxxxAer}{\ensuremath{10.251\pm0.022}}   
\newcommand{\hatcurCCtwomassHmagxxxAer}{\ensuremath{10.024\pm0.022}}   
\newcommand{\hatcurCCtwomassKmagxxxAer}{\ensuremath{9.990\pm0.022}}    
\newcommand{\hatcurCCcitJmagxxxAer}{\ensuremath{10.273\pm0.023}}       
\newcommand{\hatcurCCcitHmagxxxAer}{\ensuremath{10.019\pm0.023}}       
\newcommand{\hatcurCCcitKmagxxxAer}{\ensuremath{10.014\pm0.022}}       
\newcommand{\hatcurCCbbJmagxxxAer}{\ensuremath{10.313\pm0.023}}        
\newcommand{\hatcurCCbbHmagxxxAer}{\ensuremath{10.040\pm0.023}}        
\newcommand{\hatcurCCbbKmagxxxAer}{\ensuremath{10.034\pm0.022}}        
\newcommand{\hatcurCCesoJmagxxxAer}{\ensuremath{10.315\pm0.024}}       
\newcommand{\hatcurCCesoHmagxxxAer}{\ensuremath{10.034\pm0.023}}       
\newcommand{\hatcurCCesoKmagxxxAer}{\ensuremath{10.034\pm0.022}}       
\newcommand{\hatcurCCesoJHmagxxxAer}{\ensuremath{0.281\pm0.005}}       
\newcommand{\hatcurCCesoJKmagxxxAer}{\ensuremath{0.281\pm0.008}}       
\newcommand{\hatcurCCesoHKmagxxxAer}{\ensuremath{0.000\pm0.006}}       
\newcommand{\hatcurLCdipxxxAer}{\ensuremath{20.3}}                     
\newcommand{\hatcurLCrprstarxxxAer}{\ensuremath{0.1508\pm0.0004}}      
\newcommand{\hatcurLCbsqxxxAer}{\ensuremath{0.012_{-0.007}^{+0.012}}}  
\newcommand{\hatcurLCimpxxxAer}{\ensuremath{0.108_{-0.044}^{+0.043}}}  
\newcommand{\hatcurLCzetaxxxAer}{\ensuremath{17.84\pm0.03}}            
\newcommand{\hatcurLCdurxxxAer}{\ensuremath{0.1292\pm0.0003}}          
\newcommand{\hatcurLCdurshortxxxAer}{\ensuremath{0.1292}}              
\newcommand{\hatcurLCdurhrxxxAer}{\ensuremath{3.101\pm0.007}}          
\newcommand{\hatcurLCdurhrshortxxxAer}{\ensuremath{3.101}}             
\newcommand{\hatcurLCqxxxAer}{\ensuremath{0.0601\pm0.0001}}            
\newcommand{\hatcurLCqshortxxxAer}{\ensuremath{0.060}}                 
\newcommand{\hatcurLCingdurxxxAer}{\ensuremath{0.0171\pm0.0002}}       
\newcommand{\hatcurLCPxxxAer}{\ensuremath{2.150009\pm0.000001}}        
\newcommand{\hatcurLCPprecxxxAer}{\ensuremath{2.1500085}}              
\newcommand{\hatcurLCPshortxxxAer}{\ensuremath{2.1500}}                
\newcommand{\hatcurLCTxxxAer}{\ensuremath{2454416.14639\pm0.00009}}    
\newcommand{\hatcurLCTAxxxAer}{\ensuremath{2453364.79222\pm0.00065}}   
\newcommand{\hatcurLCTBxxxAer}{\ensuremath{2454439.79648\pm0.00009}}   
\newcommand{\hatcurLChatnetmAxxxAer}{\ensuremath{10.5995\pm0.0002}}    
\newcommand{\hatcurLCiblendAxxxAer}{\ensuremath{0.69\pm0.03}}          
\newcommand{\hatcurLChatnetmBxxxAer}{\ensuremath{10.5988\pm0.0001}}    
\newcommand{\hatcurLCiblendBxxxAer}{\ensuremath{0.87\pm0.02}}          
\newcommand{\hatcurSMEiteffxxxAer}{\ensuremath{6001\pm88}}             
\newcommand{\hatcurSMEizfehxxxAer}{\ensuremath{-0.16\pm0.08}}          
\newcommand{\hatcurSMEizfehshortxxxAer}{\ensuremath{-0.16}}            
\newcommand{\hatcurSMEiloggxxxAer}{\ensuremath{4.02\pm0.07}}           
\newcommand{\hatcurSMEivsinxxxAer}{\ensuremath{21\pm0.5}}              
\newcommand{\hatcurSMEivmacxxxAer}{\ensuremath{4.3}}                   
\newcommand{\hatcurSMEivmicxxxAer}{\ensuremath{0.0}}                   
\newcommand{\hatcurSMEiiteffxxxAer}{\ensuremath{6136\pm88}}            
\newcommand{\hatcurSMEiizfehxxxAer}{\ensuremath{-0.08\pm0.08}}         
\newcommand{\hatcurSMEiizfehshortxxxAer}{\ensuremath{-0.08}}           
\newcommand{\hatcurSMEiiloggxxxAer}{\ensuremath{4.21\pm0.04}}          
\newcommand{\hatcurSMEiivsinxxxAer}{\ensuremath{20.8\pm0.5}}           
\newcommand{\hatcurSMEiivmacxxxAer}{\ensuremath{4.69}}                 
\newcommand{\hatcurSMEiivmicxxxAer}{\ensuremath{0.85}}                 
\newcommand{\hatcurDSteffxxxAer}{\ensuremath{6500\pm100}}              
\newcommand{\hatcurDSzfehxxxAer}{\ensuremath{0.0\pm0.0}}               
\newcommand{\hatcurDSloggxxxAer}{\ensuremath{4.5\pm0.25}}              
\newcommand{\hatcurDSvsinixxxAer}{\ensuremath{21.9\pm1.0}}             
\newcommand{\hatcurDSgammaxxxAer}{\ensuremath{-23.21\pm0.26}}          
\newcommand{\hatcurDSnumspecxxxAer}{\ensuremath{15}}                   
\newcommand{\hatcurDSspanxxxAer}{\ensuremath{738}}                     
\newcommand{\hatcurDSrvrmsxxxAer}{\ensuremath{0.97}}                   
\newcommand{\hatcurTRESteffxxxAer}{\ensuremath{NULL\pmNULL}}           
\newcommand{\hatcurTRESzfehxxxAer}{\ensuremath{NULL\pmNULL}}           
\newcommand{\hatcurTRESloggxxxAer}{\ensuremath{NULL\pmNULL}}           
\newcommand{\hatcurTRESvsinixxxAer}{\ensuremath{NULL\pmNULL}}          
\newcommand{\hatcurTRESgammaxxxAer}{\ensuremath{NULL\pmNULL}}          
\newcommand{\hatcurTRESnumspecxxxAer}{\ensuremath{0}}                  
\newcommand{\hatcurTRESspanxxxAer}{\ensuremath{0}}                     
\newcommand{\hatcurTRESrvrmsxxxAer}{\ensuremath{0.00}}                 
\newcommand{\hatcurFIESteffxxxAer}{\ensuremath{NULL\pmNULL}}           
\newcommand{\hatcurFIESzfehxxxAer}{\ensuremath{NULL\pmNULL}}           
\newcommand{\hatcurFIESloggxxxAer}{\ensuremath{NULL\pmNULL}}           
\newcommand{\hatcurFIESvsinixxxAer}{\ensuremath{NULL\pmNULL}}          
\newcommand{\hatcurFIESgammaxxxAer}{\ensuremath{NULL\pmNULL}}          
\newcommand{\hatcurFIESnumspecxxxAer}{\ensuremath{0}}                  
\newcommand{\hatcurFIESspanxxxAer}{\ensuremath{0}}                     
\newcommand{\hatcurFIESrvrmsxxxAer}{\ensuremath{0.00}}                 
\newcommand{\hatcurLBizxxxAer}{\ensuremath{0.1580}}                    
\newcommand{\hatcurLBiizxxxAer}{\ensuremath{0.3476}}                   
\newcommand{\hatcurLBiixxxAer}{\ensuremath{0.2098}}                    
\newcommand{\hatcurLBiiixxxAer}{\ensuremath{0.3562}}                   
\newcommand{\hatcurLBiIxxxAer}{\ensuremath{0.1915}}                    
\newcommand{\hatcurLBiiIxxxAer}{\ensuremath{0.3543}}                   
\newcommand{\hatcurLBigxxxAer}{\ensuremath{0.4564}}                    
\newcommand{\hatcurLBiigxxxAer}{\ensuremath{0.3027}}                   
\newcommand{\hatcurLBikepxxxAer}{\ensuremath{0.1000}}                  
\newcommand{\hatcurLBiikepxxxAer}{\ensuremath{0.1000}}                 
\newcommand{\hatcurISOmxxxAer}{\ensuremath{1.18_{-0.07}^{+0.04}}}      
\newcommand{\hatcurISOmshortxxxAer}{\ensuremath{1.18}}                 
\newcommand{\hatcurISOmlongxxxAer}{\ensuremath{1.176_{-0.070}^{+0.043}}} 
\newcommand{\hatcurISOrxxxAer}{\ensuremath{1.39\pm0.07}}               
\newcommand{\hatcurISOrshortxxxAer}{\ensuremath{1.39}}                 
\newcommand{\hatcurISOrlongxxxAer}{\ensuremath{1.387\pm0.067}}         
\newcommand{\hatcurISOrhoxxxAer}{\ensuremath{0.61_{-0.07}^{+0.09}}}    
\newcommand{\hatcurISOloggxxxAer}{\ensuremath{4.22\pm0.04}}            
\newcommand{\hatcurISOlumxxxAer}{\ensuremath{2.43\pm0.30}}             
\newcommand{\hatcurISOlumshortxxxAer}{\ensuremath{2.43}}               
\newcommand{\hatcurISOmvxxxAer}{\ensuremath{3.82\pm0.14}}              
\newcommand{\hatcurISOvixxxAer}{\ensuremath{0.592\pm0.024}}            
\newcommand{\hatcurISOagexxxAer}{\ensuremath{3.8_{-0.5}^{+1.5}}}       
\newcommand{\hatcurISOsigmaxxxAer}{\ensuremath{0.00060\pm0.00013}}     
\newcommand{\hatcurISOMJxxxAer}{\ensuremath{2.84\pm0.12}}              
\newcommand{\hatcurISOMHxxxAer}{\ensuremath{2.55\pm0.11}}              
\newcommand{\hatcurISOMKxxxAer}{\ensuremath{2.50\pm0.11}}              
\newcommand{\hatcurISOJKxxxAer}{\ensuremath{0.34\pm0.02}}              
\newcommand{\hatcurISOspecxxxAer}{G0}                                  
\newcommand{\hatcurRVKxxxAer}{\ensuremath{136.1\pm23.8}}               
\newcommand{\hatcurRVkxxxAer}{\ensuremath{0.099\pm0.080}}              
\newcommand{\hatcurRVhxxxAer}{\ensuremath{0.124\pm0.037}}              
\newcommand{\hatcurRVgammaxxxAer}{\ensuremath{61.9\pm15.7}}            
\newcommand{\hatcurRVjitterxxxAer}{\ensuremath{73.3}}                  
\newcommand{\hatcurRVeccenxxxAer}{\ensuremath{0.163\pm0.061}}          
\newcommand{\hatcurRVomegaxxxAer}{\ensuremath{52\pm29}}                
\newcommand{\hatcurRVfitrmsxxxAer}{\ensuremath{75.0}}                  
\newcommand{\hatcurPPixxxAer}{\ensuremath{88.7\pm0.6}}                 
\newcommand{\hatcurPPgxxxAer}{\ensuremath{5.6\pm0.9}}                  
\newcommand{\hatcurPPloggxxxAer}{\ensuremath{2.75\pm0.07}}             
\newcommand{\hatcurPParxxxAer}{\ensuremath{5.32\pm0.22}}               
\newcommand{\hatcurPParelxxxAer}{\ensuremath{0.0344_{-0.0007}^{+0.0004}}} 
\newcommand{\hatcurPPrhoxxxAer}{\ensuremath{0.14_{-0.02}^{+0.03}}}     
\newcommand{\hatcurPPmxxxAer}{\ensuremath{0.94\pm0.17}}                
\newcommand{\hatcurPPmshortxxxAer}{\ensuremath{0.94}}                  
\newcommand{\hatcurPPmlongxxxAer}{\ensuremath{0.941\pm0.166}}          
\newcommand{\hatcurPPmexxxAer}{\ensuremath{298.9\pm52.8}}              
\newcommand{\hatcurPPmeshortxxxAer}{\ensuremath{298.9}}                
\newcommand{\hatcurPPmelongxxxAer}{\ensuremath{298.91\pm52.79}}        
\newcommand{\hatcurPPrxxxAer}{\ensuremath{2.04\pm0.10}}                
\newcommand{\hatcurPPrshortxxxAer}{\ensuremath{2.04}}                  
\newcommand{\hatcurPPrlongxxxAer}{\ensuremath{2.037\pm0.099}}          
\newcommand{\hatcurPPrexxxAer}{\ensuremath{22.8\pm1.1}}                
\newcommand{\hatcurPPreshortxxxAer}{\ensuremath{22.8}}                 
\newcommand{\hatcurPPrelongxxxAer}{\ensuremath{22.83\pm1.11}}          
\newcommand{\hatcurPPmrcorrxxxAer}{\ensuremath{0.27}}                  
\newcommand{\hatcurPPteffxxxAer}{\ensuremath{1888\pm51}}               
\newcommand{\hatcurPPthetaxxxAer}{\ensuremath{0.027\pm0.004}}          
\newcommand{\hatcurPPfluxperixxxAer}{\ensuremath{4.04_{-0.78}^{+1.17}}} 
\newcommand{\hatcurPPfluxperidimxxxAer}{\ensuremath{9}}                
\newcommand{\hatcurPPfluxapxxxAer}{\ensuremath{2.08\pm0.15}}           
\newcommand{\hatcurPPfluxapdimxxxAer}{\ensuremath{9}}                  
\newcommand{\hatcurPPfluxavgxxxAer}{\ensuremath{2.86\pm0.31}}          
\newcommand{\hatcurPPfluxavgdimxxxAer}{\ensuremath{9}}                 
\newcommand{\hatcurXsecphasexxxAer}{\ensuremath{0.5632\pm0.0509}}      
\newcommand{\hatcurXsecondaryxxxAer}{\ensuremath{2454417.357\pm0.109}} 
\newcommand{\hatcurXsecdurxxxAer}{\ensuremath{0.1653\pm0.0120}}        
\newcommand{\hatcurXsecingdurxxxAer}{\ensuremath{0.0221\pm0.0017}}     
\newcommand{\hatcurPPphiconjxxxAer}{\ensuremath{0.0738_{-0.0839}^{+0.0214}}} 
\newcommand{\hatcurPPperixxxAer}{\ensuremath{2454415.99\pm0.13}}       
\newcommand{\hatcurPPaequivxxxAer}{\ensuremath{0.0220\pm0.0011}}       
\newcommand{\hatcurPPtcircxxxAer}{\ensuremath{3.1_{-0.6}^{+0.9}}}      
\newcommand{\hatcurPPtinfallxxxAer}{\ensuremath{108.3_{-25.4}^{+47.9}}} 
\newcommand{\hatcurXdistxxxAer}{\ensuremath{320\pm16}}                 
\newcommand{\hatcurCCpmraxxxAer}{\ensuremath{-12.3\pm1.3}}             
\newcommand{\hatcurCCpmdecxxxAer}{\ensuremath{4.6\pm2.4}}              
\newcommand{\hatcurCCpmxxxAer}{\ensuremath{13.132\pm2.72947}}          
\newcommand{\hatcurhtrxxxBbs}{HTR176-001}                              
\newcommand{\hatcurfieldxxxBbs}{176}                                   
\newcommand{\hatcurCCraxxxBbs}{\ensuremath{07^{\mathrm h}32^{\mathrm m}44.20{\mathrm s}}}                            
\newcommand{\hatcurCCdecxxxBbs}{\ensuremath{+33{\arcdeg}50{\arcmin}06.2{\arcsec}}}                           
\newcommand{\hatcurCCmagxxxBbs}{11.188}                                
\newcommand{\hatcurCCtwomassxxxBbs}{2MASS~07324421+3350061}            
\newcommand{\hatcurCCgscxxxBbs}{GSC~2461-00988}                        
\newcommand{\hatcurCCtassmvxxxBbs}{11.188}                             
\newcommand{\hatcurCCtwomassJmagxxxBbs}{\ensuremath{10.263\pm0.021}}   
\newcommand{\hatcurCCtwomassHmagxxxBbs}{\ensuremath{10.061\pm0.024}}   
\newcommand{\hatcurCCtwomassKmagxxxBbs}{\ensuremath{10.004\pm0.018}}   
\newcommand{\hatcurCCcitJmagxxxBbs}{\ensuremath{10.286\pm0.021}}       
\newcommand{\hatcurCCcitHmagxxxBbs}{\ensuremath{10.056\pm0.024}}       
\newcommand{\hatcurCCcitKmagxxxBbs}{\ensuremath{10.028\pm0.018}}       
\newcommand{\hatcurCCbbJmagxxxBbs}{\ensuremath{10.326\pm0.023}}        
\newcommand{\hatcurCCbbHmagxxxBbs}{\ensuremath{10.077\pm0.025}}        
\newcommand{\hatcurCCbbKmagxxxBbs}{\ensuremath{10.048\pm0.018}}        
\newcommand{\hatcurCCesoJmagxxxBbs}{\ensuremath{10.327\pm0.024}}       
\newcommand{\hatcurCCesoHmagxxxBbs}{\ensuremath{10.072\pm0.028}}       
\newcommand{\hatcurCCesoKmagxxxBbs}{\ensuremath{10.047\pm0.019}}       
\newcommand{\hatcurCCesoJHmagxxxBbs}{\ensuremath{0.254\pm0.034}}       
\newcommand{\hatcurCCesoJKmagxxxBbs}{\ensuremath{0.280\pm0.030}}       
\newcommand{\hatcurCCesoHKmagxxxBbs}{\ensuremath{0.025\pm0.033}}       
\newcommand{\hatcurLCdipxxxBbs}{\ensuremath{8.5}}                      
\newcommand{\hatcurLCrprstarxxxBbs}{\ensuremath{0.1058\pm0.0011}}      
\newcommand{\hatcurLCbsqxxxBbs}{\ensuremath{0.106_{-0.001}^{+0.001}}}  
\newcommand{\hatcurLCimpxxxBbs}{\ensuremath{0.325_{-0.002}^{+0.002}}}  
\newcommand{\hatcurLCzetaxxxBbs}{\ensuremath{12.16\pm0.03}}            
\newcommand{\hatcurLCdurxxxBbs}{\ensuremath{0.1839\pm0.0005}}          
\newcommand{\hatcurLCdurshortxxxBbs}{\ensuremath{0.1839}}              
\newcommand{\hatcurLCdurhrxxxBbs}{\ensuremath{4.414\pm0.013}}          
\newcommand{\hatcurLCdurhrshortxxxBbs}{\ensuremath{4.414}}             
\newcommand{\hatcurLCqxxxBbs}{\ensuremath{0.0529\pm0.0002}}            
\newcommand{\hatcurLCqshortxxxBbs}{\ensuremath{0.053}}                 
\newcommand{\hatcurLCingdurxxxBbs}{\ensuremath{0.0195\pm0.0002}}       
\newcommand{\hatcurLCPxxxBbs}{\ensuremath{3.474474\pm0.000001}}        
\newcommand{\hatcurLCPprecxxxBbs}{\ensuremath{3.4744742}}              
\newcommand{\hatcurLCPshortxxxBbs}{\ensuremath{3.4745}}                
\newcommand{\hatcurLCTxxxBbs}{\ensuremath{2455110.92595\pm0.00022}}    
\newcommand{\hatcurLCTAxxxBbs}{\ensuremath{2453064.46062\pm0.00069}}   
\newcommand{\hatcurLCTBxxxBbs}{\ensuremath{2455614.72471\pm0.00034}}   
\newcommand{\hatcurLChatnetmAxxxBbs}{\ensuremath{10.6184\pm0.0001}}    
\newcommand{\hatcurLCiblendAxxxBbs}{\ensuremath{0.68\pm0.04}}          
\newcommand{\hatcurLChatnetmBxxxBbs}{\ensuremath{10.6181\pm0.0002}}    
\newcommand{\hatcurLCiblendBxxxBbs}{\ensuremath{0.79\pm0.06}}          
\newcommand{\hatcurSMEiteffxxxBbs}{\ensuremath{6234\pm114}}            
\newcommand{\hatcurSMEizfehxxxBbs}{\ensuremath{-0.04\pm0.08}}          
\newcommand{\hatcurSMEizfehshortxxxBbs}{\ensuremath{-0.04}}            
\newcommand{\hatcurSMEiloggxxxBbs}{\ensuremath{3.86\pm0.10}}           
\newcommand{\hatcurSMEivsinxxxBbs}{\ensuremath{13.9\pm0.5}}            
\newcommand{\hatcurSMEivmacxxxBbs}{\ensuremath{4.73}}                  
\newcommand{\hatcurSMEivmicxxxBbs}{\ensuremath{0.85}}                  
\newcommand{\hatcurSMEiiteffxxxBbs}{\ensuremath{6446\pm88}}            
\newcommand{\hatcurSMEiizfehxxxBbs}{\ensuremath{0.07\pm0.08}}          
\newcommand{\hatcurSMEiizfehshortxxxBbs}{\ensuremath{0.07}}            
\newcommand{\hatcurSMEiiloggxxxBbs}{\ensuremath{4.14\pm0.06}}          
\newcommand{\hatcurSMEiivsinxxxBbs}{\ensuremath{13.7\pm0.5}}           
\newcommand{\hatcurSMEiivmacxxxBbs}{\ensuremath{5.06}}                 
\newcommand{\hatcurSMEiivmicxxxBbs}{\ensuremath{0.85}}                 
\newcommand{\hatcurDSteffxxxBbs}{\ensuremath{6500\pm100}}              
\newcommand{\hatcurDSzfehxxxBbs}{\ensuremath{0.0\pm0.0}}               
\newcommand{\hatcurDSloggxxxBbs}{\ensuremath{4.0\pm0.25}}              
\newcommand{\hatcurDSvsinixxxBbs}{\ensuremath{15.6\pm1.0}}             
\newcommand{\hatcurDSgammaxxxBbs}{\ensuremath{23.03\pm0.28}}           
\newcommand{\hatcurDSnumspecxxxBbs}{\ensuremath{9}}                    
\newcommand{\hatcurDSspanxxxBbs}{\ensuremath{184}}                     
\newcommand{\hatcurDSrvrmsxxxBbs}{\ensuremath{0.84}}                   
\newcommand{\hatcurTRESteffxxxBbs}{\ensuremath{NULL\pmNULL}}           
\newcommand{\hatcurTRESzfehxxxBbs}{\ensuremath{NULL\pmNULL}}           
\newcommand{\hatcurTRESloggxxxBbs}{\ensuremath{NULL\pmNULL}}           
\newcommand{\hatcurTRESvsinixxxBbs}{\ensuremath{NULL\pmNULL}}          
\newcommand{\hatcurTRESgammaxxxBbs}{\ensuremath{NULL\pmNULL}}          
\newcommand{\hatcurTRESnumspecxxxBbs}{\ensuremath{0}}                  
\newcommand{\hatcurTRESspanxxxBbs}{\ensuremath{0}}                     
\newcommand{\hatcurTRESrvrmsxxxBbs}{\ensuremath{0.00}}                 
\newcommand{\hatcurFIESteffxxxBbs}{\ensuremath{NULL\pmNULL}}           
\newcommand{\hatcurFIESzfehxxxBbs}{\ensuremath{NULL\pmNULL}}           
\newcommand{\hatcurFIESloggxxxBbs}{\ensuremath{NULL\pmNULL}}           
\newcommand{\hatcurFIESvsinixxxBbs}{\ensuremath{NULL\pmNULL}}          
\newcommand{\hatcurFIESgammaxxxBbs}{\ensuremath{NULL\pmNULL}}          
\newcommand{\hatcurFIESnumspecxxxBbs}{\ensuremath{0}}                  
\newcommand{\hatcurFIESspanxxxBbs}{\ensuremath{0}}                     
\newcommand{\hatcurFIESrvrmsxxxBbs}{\ensuremath{0.00}}                 
\newcommand{\hatcurLBizxxxBbs}{\ensuremath{0.1260}}                    
\newcommand{\hatcurLBiizxxxBbs}{\ensuremath{0.3671}}                   
\newcommand{\hatcurLBiixxxBbs}{\ensuremath{0.1762}}                    
\newcommand{\hatcurLBiiixxxBbs}{\ensuremath{0.3768}}                   
\newcommand{\hatcurLBiIxxxBbs}{\ensuremath{0.1578}}                    
\newcommand{\hatcurLBiiIxxxBbs}{\ensuremath{0.3749}}                   
\newcommand{\hatcurLBigxxxBbs}{\ensuremath{0.4149}}                    
\newcommand{\hatcurLBiigxxxBbs}{\ensuremath{0.3327}}                   
\newcommand{\hatcurLBikepxxxBbs}{\ensuremath{}}                
\newcommand{\hatcurLBiikepxxxBbs}{\ensuremath{}}               
\newcommand{\hatcurISOmxxxBbs}{\ensuremath{1.38\pm0.04}}               
\newcommand{\hatcurISOmshortxxxBbs}{\ensuremath{1.38}}                 
\newcommand{\hatcurISOmlongxxxBbs}{\ensuremath{1.375\pm0.040}}         
\newcommand{\hatcurISOrxxxBbs}{\ensuremath{1.64\pm0.03}}               
\newcommand{\hatcurISOrshortxxxBbs}{\ensuremath{1.64}}                 
\newcommand{\hatcurISOrlongxxxBbs}{\ensuremath{1.637\pm0.034}}         
\newcommand{\hatcurISOrhoxxxBbs}{\ensuremath{0.44_{-0.02}^{+0.02}}}    
\newcommand{\hatcurISOloggxxxBbs}{\ensuremath{4.15\pm0.01}}            
\newcommand{\hatcurISOlumxxxBbs}{\ensuremath{4.15\pm0.33}}             
\newcommand{\hatcurISOlumshortxxxBbs}{\ensuremath{4.15}}               
\newcommand{\hatcurISOmvxxxBbs}{\ensuremath{3.19\pm0.10}}              
\newcommand{\hatcurISOvixxxBbs}{\ensuremath{0.511\pm0.022}}            
\newcommand{\hatcurISOagexxxBbs}{\ensuremath{2.3\pm0.3}}               
\newcommand{\hatcurISOsigmaxxxBbs}{\ensuremath{0.00040\pm0.00006}}     
\newcommand{\hatcurISOMJxxxBbs}{\ensuremath{2.37\pm0.06}}              
\newcommand{\hatcurISOMHxxxBbs}{\ensuremath{2.15\pm0.05}}              
\newcommand{\hatcurISOMKxxxBbs}{\ensuremath{2.11\pm0.05}}              
\newcommand{\hatcurISOJKxxxBbs}{\ensuremath{0.27\pm0.02}}              
\newcommand{\hatcurISOspecxxxBbs}{G5}                                  
\newcommand{\hatcurRVKxxxBbs}{\ensuremath{82.7\pm10.8}}                
\newcommand{\hatcurRVkxxxBbs}{\ensuremath{0.000\pm0.000}}              
\newcommand{\hatcurRVhxxxBbs}{\ensuremath{0.000\pm0.000}}              
\newcommand{\hatcurRVBScorrxxxBbs}{\ensuremath{-0.814\pm0.164}}        
\newcommand{\hatcurRVgammaxxxBbs}{\ensuremath{-2.7\pm8.0}}             
\newcommand{\hatcurRVjitterxxxBbs}{\ensuremath{34.4}}                  
\newcommand{\hatcurRVeccenxxxBbs}{\ensuremath{0.000\pm0.000}}          
\newcommand{\hatcurRVomegaxxxBbs}{\ensuremath{0\pm0}}                  
\newcommand{\hatcurRVfitrmsxxxBbs}{\ensuremath{35.3}}                  
\newcommand{\hatcurPPixxxBbs}{\ensuremath{87.2_{-0.1}^{+0.0}}}         
\newcommand{\hatcurPPgxxxBbs}{\ensuremath{6.6\pm0.9}}                  
\newcommand{\hatcurPPloggxxxBbs}{\ensuremath{2.82\pm0.06}}             
\newcommand{\hatcurPParxxxBbs}{\ensuremath{6.56_{-0.12}^{+0.09}}}      
\newcommand{\hatcurPParelxxxBbs}{\ensuremath{0.0499\pm0.0005}}         
\newcommand{\hatcurPPrhoxxxBbs}{\ensuremath{0.20\pm0.03}}              
\newcommand{\hatcurPPmxxxBbs}{\ensuremath{0.76\pm0.10}}                
\newcommand{\hatcurPPmshortxxxBbs}{\ensuremath{0.76}}                  
\newcommand{\hatcurPPmlongxxxBbs}{\ensuremath{0.762\pm0.101}}          
\newcommand{\hatcurPPmexxxBbs}{\ensuremath{242.3\pm32.1}}              
\newcommand{\hatcurPPmeshortxxxBbs}{\ensuremath{242.3}}                
\newcommand{\hatcurPPmelongxxxBbs}{\ensuremath{242.32\pm32.13}}        
\newcommand{\hatcurPPrxxxBbs}{\ensuremath{1.69\pm0.04}}                
\newcommand{\hatcurPPrshortxxxBbs}{\ensuremath{1.69}}                  
\newcommand{\hatcurPPrlongxxxBbs}{\ensuremath{1.686\pm0.045}}          
\newcommand{\hatcurPPrexxxBbs}{\ensuremath{18.9\pm0.5}}                
\newcommand{\hatcurPPreshortxxxBbs}{\ensuremath{18.9}}                 
\newcommand{\hatcurPPrelongxxxBbs}{\ensuremath{18.90\pm0.50}}          
\newcommand{\hatcurPPmrcorrxxxBbs}{\ensuremath{0.10}}                  
\newcommand{\hatcurPPteffxxxBbs}{\ensuremath{1782\pm28}}               
\newcommand{\hatcurPPthetaxxxBbs}{\ensuremath{0.033\pm0.004}}          
\newcommand{\hatcurPPfluxperixxxBbs}{\ensuremath{2.27\pm0.14}}         
\newcommand{\hatcurPPfluxperidimxxxBbs}{\ensuremath{9}}                
\newcommand{\hatcurPPfluxapxxxBbs}{\ensuremath{2.27\pm0.14}}           
\newcommand{\hatcurPPfluxapdimxxxBbs}{\ensuremath{9}}                  
\newcommand{\hatcurPPfluxavgxxxBbs}{\ensuremath{2.27\pm0.14}}          
\newcommand{\hatcurPPfluxavgdimxxxBbs}{\ensuremath{9}}                 
\newcommand{\hatcurXsecphasexxxBbs}{\ensuremath{0.5000\pm0.0000}}      
\newcommand{\hatcurXsecondaryxxxBbs}{\ensuremath{2455112.663\pm0.000}} 
\newcommand{\hatcurXsecdurxxxBbs}{\ensuremath{0.1839\pm0.0005}}        
\newcommand{\hatcurXsecingdurxxxBbs}{\ensuremath{0.0195\pm0.0002}}     
\newcommand{\hatcurPPphiconjxxxBbs}{\ensuremath{0.2500\pm0.0000}}      
\newcommand{\hatcurPPperixxxBbs}{\ensuremath{2455110.06\pm0.00}}       
\newcommand{\hatcurPPaequivxxxBbs}{\ensuremath{0.0245\pm0.0008}}       
\newcommand{\hatcurPPtcircxxxBbs}{\ensuremath{43.8\pm7.0}}             
\newcommand{\hatcurPPtinfallxxxBbs}{\ensuremath{720.4_{-96.8}^{+134.5}}} 
\newcommand{\hatcurXdistxxxBbs}{\ensuremath{387\pm9}}                  
\newcommand{\hatcurCCpmraxxxBbs}{\ensuremath{-2.1\pm1.3}}              
\newcommand{\hatcurCCpmdecxxxBbs}{\ensuremath{-4.9\pm1.0}}             
\newcommand{\hatcurCCpmxxxBbs}{\ensuremath{5.33104\pm1.64012}}         
\newcommand{\hatcurhtrxxxBebs}{HTR176-001}                             
\newcommand{\hatcurfieldxxxBebs}{176}                                  
\newcommand{\hatcurCCraxxxBebs}{\ensuremath{07^{\mathrm h}32^{\mathrm m}44.20{\mathrm s}}}                           
\newcommand{\hatcurCCdecxxxBebs}{\ensuremath{+33{\arcdeg}50{\arcmin}06.2{\arcsec}}}                          
\newcommand{\hatcurCCmagxxxBebs}{11.188}                               
\newcommand{\hatcurCCtwomassxxxBebs}{2MASS~07324421+3350061}           
\newcommand{\hatcurCCgscxxxBebs}{GSC~2461-00988}                       
\newcommand{\hatcurCCtassmvxxxBebs}{11.188}                            
\newcommand{\hatcurCCtwomassJmagxxxBebs}{\ensuremath{10.263\pm0.021}}  
\newcommand{\hatcurCCtwomassHmagxxxBebs}{\ensuremath{10.061\pm0.024}}  
\newcommand{\hatcurCCtwomassKmagxxxBebs}{\ensuremath{10.004\pm0.018}}  
\newcommand{\hatcurCCcitJmagxxxBebs}{\ensuremath{10.286\pm0.021}}      
\newcommand{\hatcurCCcitHmagxxxBebs}{\ensuremath{10.056\pm0.024}}      
\newcommand{\hatcurCCcitKmagxxxBebs}{\ensuremath{10.028\pm0.018}}      
\newcommand{\hatcurCCbbJmagxxxBebs}{\ensuremath{10.326\pm0.023}}       
\newcommand{\hatcurCCbbHmagxxxBebs}{\ensuremath{10.077\pm0.025}}       
\newcommand{\hatcurCCbbKmagxxxBebs}{\ensuremath{10.048\pm0.018}}       
\newcommand{\hatcurCCesoJmagxxxBebs}{\ensuremath{10.327\pm0.024}}      
\newcommand{\hatcurCCesoHmagxxxBebs}{\ensuremath{10.072\pm0.028}}      
\newcommand{\hatcurCCesoKmagxxxBebs}{\ensuremath{10.047\pm0.019}}      
\newcommand{\hatcurCCesoJHmagxxxBebs}{\ensuremath{0.254\pm0.034}}      
\newcommand{\hatcurCCesoJKmagxxxBebs}{\ensuremath{0.280\pm0.030}}      
\newcommand{\hatcurCCesoHKmagxxxBebs}{\ensuremath{0.025\pm0.033}}      
\newcommand{\hatcurLCdipxxxBebs}{\ensuremath{8.5}}                     
\newcommand{\hatcurLCrprstarxxxBebs}{\ensuremath{0.1057\pm0.0011}}     
\newcommand{\hatcurLCbsqxxxBebs}{\ensuremath{0.106_{-0.001}^{+0.001}}} 
\newcommand{\hatcurLCimpxxxBebs}{\ensuremath{0.325_{-0.002}^{+0.002}}} 
\newcommand{\hatcurLCzetaxxxBebs}{\ensuremath{12.17\pm0.05}}           
\newcommand{\hatcurLCdurxxxBebs}{\ensuremath{0.1836\pm0.0007}}         
\newcommand{\hatcurLCdurshortxxxBebs}{\ensuremath{0.1836}}             
\newcommand{\hatcurLCdurhrxxxBebs}{\ensuremath{4.406\pm0.018}}         
\newcommand{\hatcurLCdurhrshortxxxBebs}{\ensuremath{4.406}}            
\newcommand{\hatcurLCqxxxBebs}{\ensuremath{0.0528\pm0.0002}}           
\newcommand{\hatcurLCqshortxxxBebs}{\ensuremath{0.053}}                
\newcommand{\hatcurLCingdurxxxBebs}{\ensuremath{0.0194\pm0.0002}}      
\newcommand{\hatcurLCPxxxBebs}{\ensuremath{3.474474\pm0.000001}}       
\newcommand{\hatcurLCPprecxxxBebs}{\ensuremath{3.4744742}}             
\newcommand{\hatcurLCPshortxxxBebs}{\ensuremath{3.4745}}               
\newcommand{\hatcurLCTxxxBebs}{\ensuremath{2455100.50255\pm0.00023}}   
\newcommand{\hatcurLCTAxxxBebs}{\ensuremath{2453064.46066\pm0.00068}}  
\newcommand{\hatcurLCTBxxxBebs}{\ensuremath{2455614.72474\pm0.00034}}  
\newcommand{\hatcurLChatnetmAxxxBebs}{\ensuremath{10.6184\pm0.0001}}   
\newcommand{\hatcurLCiblendAxxxBebs}{\ensuremath{0.68\pm0.04}}         
\newcommand{\hatcurLChatnetmBxxxBebs}{\ensuremath{10.6181\pm0.0002}}   
\newcommand{\hatcurLCiblendBxxxBebs}{\ensuremath{0.79\pm0.06}}         
\newcommand{\hatcurSMEiteffxxxBebs}{\ensuremath{6234\pm114}}           
\newcommand{\hatcurSMEizfehxxxBebs}{\ensuremath{-0.04\pm0.08}}         
\newcommand{\hatcurSMEizfehshortxxxBebs}{\ensuremath{-0.04}}           
\newcommand{\hatcurSMEiloggxxxBebs}{\ensuremath{3.86\pm0.10}}          
\newcommand{\hatcurSMEivsinxxxBebs}{\ensuremath{13.9\pm0.5}}           
\newcommand{\hatcurSMEivmacxxxBebs}{\ensuremath{4.73}}                 
\newcommand{\hatcurSMEivmicxxxBebs}{\ensuremath{0.85}}                 
\newcommand{\hatcurSMEiiteffxxxBebs}{\ensuremath{6401\pm88}}           
\newcommand{\hatcurSMEiizfehxxxBebs}{\ensuremath{0.05\pm0.08}}         
\newcommand{\hatcurSMEiizfehshortxxxBebs}{\ensuremath{0.05}}           
\newcommand{\hatcurSMEiiloggxxxBebs}{\ensuremath{4.08\pm0.11}}         
\newcommand{\hatcurSMEiivsinxxxBebs}{\ensuremath{13.8\pm0.5}}          
\newcommand{\hatcurSMEiivmacxxxBebs}{\ensuremath{4.99}}                
\newcommand{\hatcurSMEiivmicxxxBebs}{\ensuremath{0.85}}                
\newcommand{\hatcurDSteffxxxBebs}{\ensuremath{6500\pm100}}             
\newcommand{\hatcurDSzfehxxxBebs}{\ensuremath{0.0\pm0.0}}              
\newcommand{\hatcurDSloggxxxBebs}{\ensuremath{4.0\pm0.25}}             
\newcommand{\hatcurDSvsinixxxBebs}{\ensuremath{15.6\pm1.0}}            
\newcommand{\hatcurDSgammaxxxBebs}{\ensuremath{23.03\pm0.28}}          
\newcommand{\hatcurDSnumspecxxxBebs}{\ensuremath{9}}                   
\newcommand{\hatcurDSspanxxxBebs}{\ensuremath{184}}                    
\newcommand{\hatcurDSrvrmsxxxBebs}{\ensuremath{0.84}}                  
\newcommand{\hatcurTRESteffxxxBebs}{\ensuremath{NULL\pmNULL}}          
\newcommand{\hatcurTRESzfehxxxBebs}{\ensuremath{NULL\pmNULL}}          
\newcommand{\hatcurTRESloggxxxBebs}{\ensuremath{NULL\pmNULL}}          
\newcommand{\hatcurTRESvsinixxxBebs}{\ensuremath{NULL\pmNULL}}         
\newcommand{\hatcurTRESgammaxxxBebs}{\ensuremath{NULL\pmNULL}}         
\newcommand{\hatcurTRESnumspecxxxBebs}{\ensuremath{0}}                 
\newcommand{\hatcurTRESspanxxxBebs}{\ensuremath{0}}                    
\newcommand{\hatcurTRESrvrmsxxxBebs}{\ensuremath{0.00}}                
\newcommand{\hatcurFIESteffxxxBebs}{\ensuremath{NULL\pmNULL}}          
\newcommand{\hatcurFIESzfehxxxBebs}{\ensuremath{NULL\pmNULL}}          
\newcommand{\hatcurFIESloggxxxBebs}{\ensuremath{NULL\pmNULL}}          
\newcommand{\hatcurFIESvsinixxxBebs}{\ensuremath{NULL\pmNULL}}         
\newcommand{\hatcurFIESgammaxxxBebs}{\ensuremath{NULL\pmNULL}}         
\newcommand{\hatcurFIESnumspecxxxBebs}{\ensuremath{0}}                 
\newcommand{\hatcurFIESspanxxxBebs}{\ensuremath{0}}                    
\newcommand{\hatcurFIESrvrmsxxxBebs}{\ensuremath{0.00}}                
\newcommand{\hatcurLBizxxxBebs}{\ensuremath{0.1294}}                   
\newcommand{\hatcurLBiizxxxBebs}{\ensuremath{0.3656}}                  
\newcommand{\hatcurLBiixxxBebs}{\ensuremath{0.1799}}                   
\newcommand{\hatcurLBiiixxxBebs}{\ensuremath{0.3748}}                  
\newcommand{\hatcurLBiIxxxBebs}{\ensuremath{0.1613}}                   
\newcommand{\hatcurLBiiIxxxBebs}{\ensuremath{0.3732}}                  
\newcommand{\hatcurLBigxxxBebs}{\ensuremath{0.4216}}                   
\newcommand{\hatcurLBiigxxxBebs}{\ensuremath{0.3278}}                  
\newcommand{\hatcurLBikepxxxBebs}{\ensuremath{}}               
\newcommand{\hatcurLBiikepxxxBebs}{\ensuremath{}}              
\newcommand{\hatcurISOmxxxBebs}{\ensuremath{1.40\pm0.10}}              
\newcommand{\hatcurISOmshortxxxBebs}{\ensuremath{1.40}}                
\newcommand{\hatcurISOmlongxxxBebs}{\ensuremath{1.403\pm0.096}}        
\newcommand{\hatcurISOrxxxBebs}{\ensuremath{1.78\pm0.28}}              
\newcommand{\hatcurISOrshortxxxBebs}{\ensuremath{1.78}}                
\newcommand{\hatcurISOrlongxxxBebs}{\ensuremath{1.777\pm0.280}}        
\newcommand{\hatcurISOrhoxxxBebs}{\ensuremath{0.35_{-0.11}^{+0.22}}}   
\newcommand{\hatcurISOloggxxxBebs}{\ensuremath{4.09\pm0.11}}           
\newcommand{\hatcurISOlumxxxBebs}{\ensuremath{4.73_{-1.25}^{+1.87}}}   
\newcommand{\hatcurISOlumshortxxxBebs}{\ensuremath{4.73}}              
\newcommand{\hatcurISOmvxxxBebs}{\ensuremath{3.06\pm0.35}}             
\newcommand{\hatcurISOvixxxBebs}{\ensuremath{0.522\pm0.022}}           
\newcommand{\hatcurISOagexxxBebs}{\ensuremath{2.4\pm0.4}}              
\newcommand{\hatcurISOsigmaxxxBebs}{\ensuremath{0.00030\pm0.00012}}    
\newcommand{\hatcurISOMJxxxBebs}{\ensuremath{2.21\pm0.34}}             
\newcommand{\hatcurISOMHxxxBebs}{\ensuremath{1.98\pm0.34}}             
\newcommand{\hatcurISOMKxxxBebs}{\ensuremath{1.93\pm0.34}}             
\newcommand{\hatcurISOJKxxxBebs}{\ensuremath{0.28\pm0.02}}             
\newcommand{\hatcurISOspecxxxBebs}{G5}                                 
\newcommand{\hatcurRVKxxxBebs}{\ensuremath{82.8\pm12.0}}               
\newcommand{\hatcurRVkxxxBebs}{\ensuremath{0.040\pm0.078}}             
\newcommand{\hatcurRVhxxxBebs}{\ensuremath{0.073\pm0.138}}             
\newcommand{\hatcurRVBScorrxxxBebs}{\ensuremath{-0.794\pm0.179}}       
\newcommand{\hatcurRVgammaxxxBebs}{\ensuremath{-3.5\pm8.7}}            
\newcommand{\hatcurRVjitterxxxBebs}{\ensuremath{36.0}}                 
\newcommand{\hatcurRVeccenxxxBebs}{\ensuremath{0.148\pm0.081}}         
\newcommand{\hatcurRVomegaxxxBebs}{\ensuremath{96\pm119}}              
\newcommand{\hatcurRVfitrmsxxxBebs}{\ensuremath{36.8}}                 
\newcommand{\hatcurPPixxxBebs}{\ensuremath{86.7_{-1.2}^{+0.8}}}        
\newcommand{\hatcurPPgxxxBebs}{\ensuremath{5.6_{-1.3}^{+2.3}}}         
\newcommand{\hatcurPPloggxxxBebs}{\ensuremath{2.75\pm0.13}}            
\newcommand{\hatcurPParxxxBebs}{\ensuremath{6.08_{-0.72}^{+0.98}}}     
\newcommand{\hatcurPParelxxxBebs}{\ensuremath{0.0503\pm0.0011}}        
\newcommand{\hatcurPPrhoxxxBebs}{\ensuremath{0.15_{-0.05}^{+0.11}}}    
\newcommand{\hatcurPPmxxxBebs}{\ensuremath{0.76\pm0.12}}               
\newcommand{\hatcurPPmshortxxxBebs}{\ensuremath{0.76}}                 
\newcommand{\hatcurPPmlongxxxBebs}{\ensuremath{0.763\pm0.117}}         
\newcommand{\hatcurPPmexxxBebs}{\ensuremath{242.6\pm37.2}}             
\newcommand{\hatcurPPmeshortxxxBebs}{\ensuremath{242.6}}               
\newcommand{\hatcurPPmelongxxxBebs}{\ensuremath{242.64\pm37.24}}       
\newcommand{\hatcurPPrxxxBebs}{\ensuremath{1.83\pm0.29}}               
\newcommand{\hatcurPPrshortxxxBebs}{\ensuremath{1.83}}                 
\newcommand{\hatcurPPrlongxxxBebs}{\ensuremath{1.827\pm0.290}}         
\newcommand{\hatcurPPrexxxBebs}{\ensuremath{20.5\pm3.2}}               
\newcommand{\hatcurPPreshortxxxBebs}{\ensuremath{20.5}}                
\newcommand{\hatcurPPrelongxxxBebs}{\ensuremath{20.48\pm3.25}}         
\newcommand{\hatcurPPmrcorrxxxBebs}{\ensuremath{0.34}}                 
\newcommand{\hatcurPPteffxxxBebs}{\ensuremath{1838\pm133}}             
\newcommand{\hatcurPPthetaxxxBebs}{\ensuremath{0.030_{-0.005}^{+0.007}}} 
\newcommand{\hatcurPPfluxperixxxBebs}{\ensuremath{3.28_{-0.77}^{+2.77}}} 
\newcommand{\hatcurPPfluxperidimxxxBebs}{\ensuremath{9}}               
\newcommand{\hatcurPPfluxapxxxBebs}{\ensuremath{2.06_{-0.54}^{+0.23}}} 
\newcommand{\hatcurPPfluxapdimxxxBebs}{\ensuremath{9}}                 
\newcommand{\hatcurPPfluxavgxxxBebs}{\ensuremath{2.58_{-0.61}^{+0.93}}} 
\newcommand{\hatcurPPfluxavgdimxxxBebs}{\ensuremath{9}}                
\newcommand{\hatcurXsecphasexxxBebs}{\ensuremath{0.5260\pm0.0504}}     
\newcommand{\hatcurXsecondaryxxxBebs}{\ensuremath{2455102.330\pm0.175}} 
\newcommand{\hatcurXsecdurxxxBebs}{\ensuremath{0.2090\pm0.0480}}       
\newcommand{\hatcurXsecingdurxxxBebs}{\ensuremath{0.0230\pm0.0085}}    
\newcommand{\hatcurPPphiconjxxxBebs}{\ensuremath{0.0463\pm0.2119}}     
\newcommand{\hatcurPPperixxxBebs}{\ensuremath{2455100.34\pm0.74}}      
\newcommand{\hatcurPPaequivxxxBebs}{\ensuremath{0.0231_{-0.0028}^{+0.0038}}} 
\newcommand{\hatcurPPtcircxxxBebs}{\ensuremath{36.5_{-16.4}^{+53.3}}}  
\newcommand{\hatcurPPtinfallxxxBebs}{\ensuremath{511.9_{-222.1}^{+719.5}}} 
\newcommand{\hatcurXdistxxxBebs}{\ensuremath{419\pm66}}                
\newcommand{\hatcurCCpmraxxxBebs}{\ensuremath{-2.1\pm1.3}}             
\newcommand{\hatcurCCpmdecxxxBebs}{\ensuremath{-4.9\pm1.0}}            
\newcommand{\hatcurCCpmxxxBebs}{\ensuremath{5.33104\pm1.64012}}        
\newcommand{\hatcurCCbbHmag}[1]{\ifnum#1=32 %
\hatcurCCbbHmagxxxA
\else
\ifnum#1=34 %
\hatcurCCbbHmagxxxAe
\else
\ifnum#1=36 %
\hatcurCCbbHmagxxxAer
\else
\ifnum#1=33 %
\hatcurCCbbHmagxxxBbs
\else
\ifnum#1=35 %
\hatcurCCbbHmagxxxBebs
\else
??????\fi
\fi
\fi
\fi
\fi
}
\newcommand{\hatcurCCbbJmag}[1]{\ifnum#1=32 %
\hatcurCCbbJmagxxxA
\else
\ifnum#1=34 %
\hatcurCCbbJmagxxxAe
\else
\ifnum#1=36 %
\hatcurCCbbJmagxxxAer
\else
\ifnum#1=33 %
\hatcurCCbbJmagxxxBbs
\else
\ifnum#1=35 %
\hatcurCCbbJmagxxxBebs
\else
??????\fi
\fi
\fi
\fi
\fi
}
\newcommand{\hatcurCCbbKmag}[1]{\ifnum#1=32 %
\hatcurCCbbKmagxxxA
\else
\ifnum#1=34 %
\hatcurCCbbKmagxxxAe
\else
\ifnum#1=36 %
\hatcurCCbbKmagxxxAer
\else
\ifnum#1=33 %
\hatcurCCbbKmagxxxBbs
\else
\ifnum#1=35 %
\hatcurCCbbKmagxxxBebs
\else
??????\fi
\fi
\fi
\fi
\fi
}
\newcommand{\hatcurCCcitHmag}[1]{\ifnum#1=32 %
\hatcurCCcitHmagxxxA
\else
\ifnum#1=34 %
\hatcurCCcitHmagxxxAe
\else
\ifnum#1=36 %
\hatcurCCcitHmagxxxAer
\else
\ifnum#1=33 %
\hatcurCCcitHmagxxxBbs
\else
\ifnum#1=35 %
\hatcurCCcitHmagxxxBebs
\else
??????\fi
\fi
\fi
\fi
\fi
}
\newcommand{\hatcurCCcitJmag}[1]{\ifnum#1=32 %
\hatcurCCcitJmagxxxA
\else
\ifnum#1=34 %
\hatcurCCcitJmagxxxAe
\else
\ifnum#1=36 %
\hatcurCCcitJmagxxxAer
\else
\ifnum#1=33 %
\hatcurCCcitJmagxxxBbs
\else
\ifnum#1=35 %
\hatcurCCcitJmagxxxBebs
\else
??????\fi
\fi
\fi
\fi
\fi
}
\newcommand{\hatcurCCcitKmag}[1]{\ifnum#1=32 %
\hatcurCCcitKmagxxxA
\else
\ifnum#1=34 %
\hatcurCCcitKmagxxxAe
\else
\ifnum#1=36 %
\hatcurCCcitKmagxxxAer
\else
\ifnum#1=33 %
\hatcurCCcitKmagxxxBbs
\else
\ifnum#1=35 %
\hatcurCCcitKmagxxxBebs
\else
??????\fi
\fi
\fi
\fi
\fi
}
\newcommand{\hatcurCCdec}[1]{\ifnum#1=32 %
\hatcurCCdecxxxA
\else
\ifnum#1=34 %
\hatcurCCdecxxxAe
\else
\ifnum#1=36 %
\hatcurCCdecxxxAer
\else
\ifnum#1=33 %
\hatcurCCdecxxxBbs
\else
\ifnum#1=35 %
\hatcurCCdecxxxBebs
\else
??????\fi
\fi
\fi
\fi
\fi
}
\newcommand{\hatcurCCesoHKmag}[1]{\ifnum#1=32 %
\hatcurCCesoHKmagxxxA
\else
\ifnum#1=34 %
\hatcurCCesoHKmagxxxAe
\else
\ifnum#1=36 %
\hatcurCCesoHKmagxxxAer
\else
\ifnum#1=33 %
\hatcurCCesoHKmagxxxBbs
\else
\ifnum#1=35 %
\hatcurCCesoHKmagxxxBebs
\else
??????\fi
\fi
\fi
\fi
\fi
}
\newcommand{\hatcurCCesoHmag}[1]{\ifnum#1=32 %
\hatcurCCesoHmagxxxA
\else
\ifnum#1=34 %
\hatcurCCesoHmagxxxAe
\else
\ifnum#1=36 %
\hatcurCCesoHmagxxxAer
\else
\ifnum#1=33 %
\hatcurCCesoHmagxxxBbs
\else
\ifnum#1=35 %
\hatcurCCesoHmagxxxBebs
\else
??????\fi
\fi
\fi
\fi
\fi
}
\newcommand{\hatcurCCesoJHmag}[1]{\ifnum#1=32 %
\hatcurCCesoJHmagxxxA
\else
\ifnum#1=34 %
\hatcurCCesoJHmagxxxAe
\else
\ifnum#1=36 %
\hatcurCCesoJHmagxxxAer
\else
\ifnum#1=33 %
\hatcurCCesoJHmagxxxBbs
\else
\ifnum#1=35 %
\hatcurCCesoJHmagxxxBebs
\else
??????\fi
\fi
\fi
\fi
\fi
}
\newcommand{\hatcurCCesoJKmag}[1]{\ifnum#1=32 %
\hatcurCCesoJKmagxxxA
\else
\ifnum#1=34 %
\hatcurCCesoJKmagxxxAe
\else
\ifnum#1=36 %
\hatcurCCesoJKmagxxxAer
\else
\ifnum#1=33 %
\hatcurCCesoJKmagxxxBbs
\else
\ifnum#1=35 %
\hatcurCCesoJKmagxxxBebs
\else
??????\fi
\fi
\fi
\fi
\fi
}
\newcommand{\hatcurCCesoJmag}[1]{\ifnum#1=32 %
\hatcurCCesoJmagxxxA
\else
\ifnum#1=34 %
\hatcurCCesoJmagxxxAe
\else
\ifnum#1=36 %
\hatcurCCesoJmagxxxAer
\else
\ifnum#1=33 %
\hatcurCCesoJmagxxxBbs
\else
\ifnum#1=35 %
\hatcurCCesoJmagxxxBebs
\else
??????\fi
\fi
\fi
\fi
\fi
}
\newcommand{\hatcurCCesoKmag}[1]{\ifnum#1=32 %
\hatcurCCesoKmagxxxA
\else
\ifnum#1=34 %
\hatcurCCesoKmagxxxAe
\else
\ifnum#1=36 %
\hatcurCCesoKmagxxxAer
\else
\ifnum#1=33 %
\hatcurCCesoKmagxxxBbs
\else
\ifnum#1=35 %
\hatcurCCesoKmagxxxBebs
\else
??????\fi
\fi
\fi
\fi
\fi
}
\newcommand{\hatcurCCgsc}[1]{\ifnum#1=32 %
\hatcurCCgscxxxA
\else
\ifnum#1=34 %
\hatcurCCgscxxxAe
\else
\ifnum#1=36 %
\hatcurCCgscxxxAer
\else
\ifnum#1=33 %
\hatcurCCgscxxxBbs
\else
\ifnum#1=35 %
\hatcurCCgscxxxBebs
\else
??????\fi
\fi
\fi
\fi
\fi
}
\newcommand{\hatcurCCmag}[1]{\ifnum#1=32 %
\hatcurCCmagxxxA
\else
\ifnum#1=34 %
\hatcurCCmagxxxAe
\else
\ifnum#1=36 %
\hatcurCCmagxxxAer
\else
\ifnum#1=33 %
\hatcurCCmagxxxBbs
\else
\ifnum#1=35 %
\hatcurCCmagxxxBebs
\else
??????\fi
\fi
\fi
\fi
\fi
}
\newcommand{\hatcurCCpm}[1]{\ifnum#1=32 %
\hatcurCCpmxxxA
\else
\ifnum#1=34 %
\hatcurCCpmxxxAe
\else
\ifnum#1=36 %
\hatcurCCpmxxxAer
\else
\ifnum#1=33 %
\hatcurCCpmxxxBbs
\else
\ifnum#1=35 %
\hatcurCCpmxxxBebs
\else
??????\fi
\fi
\fi
\fi
\fi
}
\newcommand{\hatcurCCpmdec}[1]{\ifnum#1=32 %
\hatcurCCpmdecxxxA
\else
\ifnum#1=34 %
\hatcurCCpmdecxxxAe
\else
\ifnum#1=36 %
\hatcurCCpmdecxxxAer
\else
\ifnum#1=33 %
\hatcurCCpmdecxxxBbs
\else
\ifnum#1=35 %
\hatcurCCpmdecxxxBebs
\else
??????\fi
\fi
\fi
\fi
\fi
}
\newcommand{\hatcurCCpmra}[1]{\ifnum#1=32 %
\hatcurCCpmraxxxA
\else
\ifnum#1=34 %
\hatcurCCpmraxxxAe
\else
\ifnum#1=36 %
\hatcurCCpmraxxxAer
\else
\ifnum#1=33 %
\hatcurCCpmraxxxBbs
\else
\ifnum#1=35 %
\hatcurCCpmraxxxBebs
\else
??????\fi
\fi
\fi
\fi
\fi
}
\newcommand{\hatcurCCra}[1]{\ifnum#1=32 %
\hatcurCCraxxxA
\else
\ifnum#1=34 %
\hatcurCCraxxxAe
\else
\ifnum#1=36 %
\hatcurCCraxxxAer
\else
\ifnum#1=33 %
\hatcurCCraxxxBbs
\else
\ifnum#1=35 %
\hatcurCCraxxxBebs
\else
??????\fi
\fi
\fi
\fi
\fi
}
\newcommand{\hatcurCCtassmv}[1]{\ifnum#1=32 %
\hatcurCCtassmvxxxA
\else
\ifnum#1=34 %
\hatcurCCtassmvxxxAe
\else
\ifnum#1=36 %
\hatcurCCtassmvxxxAer
\else
\ifnum#1=33 %
\hatcurCCtassmvxxxBbs
\else
\ifnum#1=35 %
\hatcurCCtassmvxxxBebs
\else
??????\fi
\fi
\fi
\fi
\fi
}
\newcommand{\hatcurCCtwomass}[1]{\ifnum#1=32 %
\hatcurCCtwomassxxxA
\else
\ifnum#1=34 %
\hatcurCCtwomassxxxAe
\else
\ifnum#1=36 %
\hatcurCCtwomassxxxAer
\else
\ifnum#1=33 %
\hatcurCCtwomassxxxBbs
\else
\ifnum#1=35 %
\hatcurCCtwomassxxxBebs
\else
??????\fi
\fi
\fi
\fi
\fi
}
\newcommand{\hatcurCCtwomassHmag}[1]{\ifnum#1=32 %
\hatcurCCtwomassHmagxxxA
\else
\ifnum#1=34 %
\hatcurCCtwomassHmagxxxAe
\else
\ifnum#1=36 %
\hatcurCCtwomassHmagxxxAer
\else
\ifnum#1=33 %
\hatcurCCtwomassHmagxxxBbs
\else
\ifnum#1=35 %
\hatcurCCtwomassHmagxxxBebs
\else
??????\fi
\fi
\fi
\fi
\fi
}
\newcommand{\hatcurCCtwomassJmag}[1]{\ifnum#1=32 %
\hatcurCCtwomassJmagxxxA
\else
\ifnum#1=34 %
\hatcurCCtwomassJmagxxxAe
\else
\ifnum#1=36 %
\hatcurCCtwomassJmagxxxAer
\else
\ifnum#1=33 %
\hatcurCCtwomassJmagxxxBbs
\else
\ifnum#1=35 %
\hatcurCCtwomassJmagxxxBebs
\else
??????\fi
\fi
\fi
\fi
\fi
}
\newcommand{\hatcurCCtwomassKmag}[1]{\ifnum#1=32 %
\hatcurCCtwomassKmagxxxA
\else
\ifnum#1=34 %
\hatcurCCtwomassKmagxxxAe
\else
\ifnum#1=36 %
\hatcurCCtwomassKmagxxxAer
\else
\ifnum#1=33 %
\hatcurCCtwomassKmagxxxBbs
\else
\ifnum#1=35 %
\hatcurCCtwomassKmagxxxBebs
\else
??????\fi
\fi
\fi
\fi
\fi
}
\newcommand{\hatcurDSgamma}[1]{\ifnum#1=32 %
\hatcurDSgammaxxxA
\else
\ifnum#1=34 %
\hatcurDSgammaxxxAe
\else
\ifnum#1=36 %
\hatcurDSgammaxxxAer
\else
\ifnum#1=33 %
\hatcurDSgammaxxxBbs
\else
\ifnum#1=35 %
\hatcurDSgammaxxxBebs
\else
??????\fi
\fi
\fi
\fi
\fi
}
\newcommand{\hatcurDSlogg}[1]{\ifnum#1=32 %
\hatcurDSloggxxxA
\else
\ifnum#1=34 %
\hatcurDSloggxxxAe
\else
\ifnum#1=36 %
\hatcurDSloggxxxAer
\else
\ifnum#1=33 %
\hatcurDSloggxxxBbs
\else
\ifnum#1=35 %
\hatcurDSloggxxxBebs
\else
??????\fi
\fi
\fi
\fi
\fi
}
\newcommand{\hatcurDSnumspec}[1]{\ifnum#1=32 %
\hatcurDSnumspecxxxA
\else
\ifnum#1=34 %
\hatcurDSnumspecxxxAe
\else
\ifnum#1=36 %
\hatcurDSnumspecxxxAer
\else
\ifnum#1=33 %
\hatcurDSnumspecxxxBbs
\else
\ifnum#1=35 %
\hatcurDSnumspecxxxBebs
\else
??????\fi
\fi
\fi
\fi
\fi
}
\newcommand{\hatcurDSrvrms}[1]{\ifnum#1=32 %
\hatcurDSrvrmsxxxA
\else
\ifnum#1=34 %
\hatcurDSrvrmsxxxAe
\else
\ifnum#1=36 %
\hatcurDSrvrmsxxxAer
\else
\ifnum#1=33 %
\hatcurDSrvrmsxxxBbs
\else
\ifnum#1=35 %
\hatcurDSrvrmsxxxBebs
\else
??????\fi
\fi
\fi
\fi
\fi
}
\newcommand{\hatcurDSspan}[1]{\ifnum#1=32 %
\hatcurDSspanxxxA
\else
\ifnum#1=34 %
\hatcurDSspanxxxAe
\else
\ifnum#1=36 %
\hatcurDSspanxxxAer
\else
\ifnum#1=33 %
\hatcurDSspanxxxBbs
\else
\ifnum#1=35 %
\hatcurDSspanxxxBebs
\else
??????\fi
\fi
\fi
\fi
\fi
}
\newcommand{\hatcurDSteff}[1]{\ifnum#1=32 %
\hatcurDSteffxxxA
\else
\ifnum#1=34 %
\hatcurDSteffxxxAe
\else
\ifnum#1=36 %
\hatcurDSteffxxxAer
\else
\ifnum#1=33 %
\hatcurDSteffxxxBbs
\else
\ifnum#1=35 %
\hatcurDSteffxxxBebs
\else
??????\fi
\fi
\fi
\fi
\fi
}
\newcommand{\hatcurDSvsini}[1]{\ifnum#1=32 %
\hatcurDSvsinixxxA
\else
\ifnum#1=34 %
\hatcurDSvsinixxxAe
\else
\ifnum#1=36 %
\hatcurDSvsinixxxAer
\else
\ifnum#1=33 %
\hatcurDSvsinixxxBbs
\else
\ifnum#1=35 %
\hatcurDSvsinixxxBebs
\else
??????\fi
\fi
\fi
\fi
\fi
}
\newcommand{\hatcurDSzfeh}[1]{\ifnum#1=32 %
\hatcurDSzfehxxxA
\else
\ifnum#1=34 %
\hatcurDSzfehxxxAe
\else
\ifnum#1=36 %
\hatcurDSzfehxxxAer
\else
\ifnum#1=33 %
\hatcurDSzfehxxxBbs
\else
\ifnum#1=35 %
\hatcurDSzfehxxxBebs
\else
??????\fi
\fi
\fi
\fi
\fi
}
\newcommand{\hatcurfield}[1]{\ifnum#1=32 %
\hatcurfieldxxxA
\else
\ifnum#1=34 %
\hatcurfieldxxxAe
\else
\ifnum#1=36 %
\hatcurfieldxxxAer
\else
\ifnum#1=33 %
\hatcurfieldxxxBbs
\else
\ifnum#1=35 %
\hatcurfieldxxxBebs
\else
??????\fi
\fi
\fi
\fi
\fi
}
\newcommand{\hatcurFIESgamma}[1]{\ifnum#1=32 %
\hatcurFIESgammaxxxA
\else
\ifnum#1=34 %
\hatcurFIESgammaxxxAe
\else
\ifnum#1=36 %
\hatcurFIESgammaxxxAer
\else
\ifnum#1=33 %
\hatcurFIESgammaxxxBbs
\else
\ifnum#1=35 %
\hatcurFIESgammaxxxBebs
\else
??????\fi
\fi
\fi
\fi
\fi
}
\newcommand{\hatcurFIESlogg}[1]{\ifnum#1=32 %
\hatcurFIESloggxxxA
\else
\ifnum#1=34 %
\hatcurFIESloggxxxAe
\else
\ifnum#1=36 %
\hatcurFIESloggxxxAer
\else
\ifnum#1=33 %
\hatcurFIESloggxxxBbs
\else
\ifnum#1=35 %
\hatcurFIESloggxxxBebs
\else
??????\fi
\fi
\fi
\fi
\fi
}
\newcommand{\hatcurFIESnumspec}[1]{\ifnum#1=32 %
\hatcurFIESnumspecxxxA
\else
\ifnum#1=34 %
\hatcurFIESnumspecxxxAe
\else
\ifnum#1=36 %
\hatcurFIESnumspecxxxAer
\else
\ifnum#1=33 %
\hatcurFIESnumspecxxxBbs
\else
\ifnum#1=35 %
\hatcurFIESnumspecxxxBebs
\else
??????\fi
\fi
\fi
\fi
\fi
}
\newcommand{\hatcurFIESrvrms}[1]{\ifnum#1=32 %
\hatcurFIESrvrmsxxxA
\else
\ifnum#1=34 %
\hatcurFIESrvrmsxxxAe
\else
\ifnum#1=36 %
\hatcurFIESrvrmsxxxAer
\else
\ifnum#1=33 %
\hatcurFIESrvrmsxxxBbs
\else
\ifnum#1=35 %
\hatcurFIESrvrmsxxxBebs
\else
??????\fi
\fi
\fi
\fi
\fi
}
\newcommand{\hatcurFIESspan}[1]{\ifnum#1=32 %
\hatcurFIESspanxxxA
\else
\ifnum#1=34 %
\hatcurFIESspanxxxAe
\else
\ifnum#1=36 %
\hatcurFIESspanxxxAer
\else
\ifnum#1=33 %
\hatcurFIESspanxxxBbs
\else
\ifnum#1=35 %
\hatcurFIESspanxxxBebs
\else
??????\fi
\fi
\fi
\fi
\fi
}
\newcommand{\hatcurFIESteff}[1]{\ifnum#1=32 %
\hatcurFIESteffxxxA
\else
\ifnum#1=34 %
\hatcurFIESteffxxxAe
\else
\ifnum#1=36 %
\hatcurFIESteffxxxAer
\else
\ifnum#1=33 %
\hatcurFIESteffxxxBbs
\else
\ifnum#1=35 %
\hatcurFIESteffxxxBebs
\else
??????\fi
\fi
\fi
\fi
\fi
}
\newcommand{\hatcurFIESvsini}[1]{\ifnum#1=32 %
\hatcurFIESvsinixxxA
\else
\ifnum#1=34 %
\hatcurFIESvsinixxxAe
\else
\ifnum#1=36 %
\hatcurFIESvsinixxxAer
\else
\ifnum#1=33 %
\hatcurFIESvsinixxxBbs
\else
\ifnum#1=35 %
\hatcurFIESvsinixxxBebs
\else
??????\fi
\fi
\fi
\fi
\fi
}
\newcommand{\hatcurFIESzfeh}[1]{\ifnum#1=32 %
\hatcurFIESzfehxxxA
\else
\ifnum#1=34 %
\hatcurFIESzfehxxxAe
\else
\ifnum#1=36 %
\hatcurFIESzfehxxxAer
\else
\ifnum#1=33 %
\hatcurFIESzfehxxxBbs
\else
\ifnum#1=35 %
\hatcurFIESzfehxxxBebs
\else
??????\fi
\fi
\fi
\fi
\fi
}
\newcommand{\hatcurhtr}[1]{\ifnum#1=32 %
\hatcurhtrxxxA
\else
\ifnum#1=34 %
\hatcurhtrxxxAe
\else
\ifnum#1=36 %
\hatcurhtrxxxAer
\else
\ifnum#1=33 %
\hatcurhtrxxxBbs
\else
\ifnum#1=35 %
\hatcurhtrxxxBebs
\else
??????\fi
\fi
\fi
\fi
\fi
}
\newcommand{\hatcurISOage}[1]{\ifnum#1=32 %
\hatcurISOagexxxA
\else
\ifnum#1=34 %
\hatcurISOagexxxAe
\else
\ifnum#1=36 %
\hatcurISOagexxxAer
\else
\ifnum#1=33 %
\hatcurISOagexxxBbs
\else
\ifnum#1=35 %
\hatcurISOagexxxBebs
\else
??????\fi
\fi
\fi
\fi
\fi
}
\newcommand{\hatcurISOJK}[1]{\ifnum#1=32 %
\hatcurISOJKxxxA
\else
\ifnum#1=34 %
\hatcurISOJKxxxAe
\else
\ifnum#1=36 %
\hatcurISOJKxxxAer
\else
\ifnum#1=33 %
\hatcurISOJKxxxBbs
\else
\ifnum#1=35 %
\hatcurISOJKxxxBebs
\else
??????\fi
\fi
\fi
\fi
\fi
}
\newcommand{\hatcurISOlogg}[1]{\ifnum#1=32 %
\hatcurISOloggxxxA
\else
\ifnum#1=34 %
\hatcurISOloggxxxAe
\else
\ifnum#1=36 %
\hatcurISOloggxxxAer
\else
\ifnum#1=33 %
\hatcurISOloggxxxBbs
\else
\ifnum#1=35 %
\hatcurISOloggxxxBebs
\else
??????\fi
\fi
\fi
\fi
\fi
}
\newcommand{\hatcurISOlum}[1]{\ifnum#1=32 %
\hatcurISOlumxxxA
\else
\ifnum#1=34 %
\hatcurISOlumxxxAe
\else
\ifnum#1=36 %
\hatcurISOlumxxxAer
\else
\ifnum#1=33 %
\hatcurISOlumxxxBbs
\else
\ifnum#1=35 %
\hatcurISOlumxxxBebs
\else
??????\fi
\fi
\fi
\fi
\fi
}
\newcommand{\hatcurISOlumshort}[1]{\ifnum#1=32 %
\hatcurISOlumshortxxxA
\else
\ifnum#1=34 %
\hatcurISOlumshortxxxAe
\else
\ifnum#1=36 %
\hatcurISOlumshortxxxAer
\else
\ifnum#1=33 %
\hatcurISOlumshortxxxBbs
\else
\ifnum#1=35 %
\hatcurISOlumshortxxxBebs
\else
??????\fi
\fi
\fi
\fi
\fi
}
\newcommand{\hatcurISOm}[1]{\ifnum#1=32 %
\hatcurISOmxxxA
\else
\ifnum#1=34 %
\hatcurISOmxxxAe
\else
\ifnum#1=36 %
\hatcurISOmxxxAer
\else
\ifnum#1=33 %
\hatcurISOmxxxBbs
\else
\ifnum#1=35 %
\hatcurISOmxxxBebs
\else
??????\fi
\fi
\fi
\fi
\fi
}
\newcommand{\hatcurISOMH}[1]{\ifnum#1=32 %
\hatcurISOMHxxxA
\else
\ifnum#1=34 %
\hatcurISOMHxxxAe
\else
\ifnum#1=36 %
\hatcurISOMHxxxAer
\else
\ifnum#1=33 %
\hatcurISOMHxxxBbs
\else
\ifnum#1=35 %
\hatcurISOMHxxxBebs
\else
??????\fi
\fi
\fi
\fi
\fi
}
\newcommand{\hatcurISOMJ}[1]{\ifnum#1=32 %
\hatcurISOMJxxxA
\else
\ifnum#1=34 %
\hatcurISOMJxxxAe
\else
\ifnum#1=36 %
\hatcurISOMJxxxAer
\else
\ifnum#1=33 %
\hatcurISOMJxxxBbs
\else
\ifnum#1=35 %
\hatcurISOMJxxxBebs
\else
??????\fi
\fi
\fi
\fi
\fi
}
\newcommand{\hatcurISOMK}[1]{\ifnum#1=32 %
\hatcurISOMKxxxA
\else
\ifnum#1=34 %
\hatcurISOMKxxxAe
\else
\ifnum#1=36 %
\hatcurISOMKxxxAer
\else
\ifnum#1=33 %
\hatcurISOMKxxxBbs
\else
\ifnum#1=35 %
\hatcurISOMKxxxBebs
\else
??????\fi
\fi
\fi
\fi
\fi
}
\newcommand{\hatcurISOmlong}[1]{\ifnum#1=32 %
\hatcurISOmlongxxxA
\else
\ifnum#1=34 %
\hatcurISOmlongxxxAe
\else
\ifnum#1=36 %
\hatcurISOmlongxxxAer
\else
\ifnum#1=33 %
\hatcurISOmlongxxxBbs
\else
\ifnum#1=35 %
\hatcurISOmlongxxxBebs
\else
??????\fi
\fi
\fi
\fi
\fi
}
\newcommand{\hatcurISOmshort}[1]{\ifnum#1=32 %
\hatcurISOmshortxxxA
\else
\ifnum#1=34 %
\hatcurISOmshortxxxAe
\else
\ifnum#1=36 %
\hatcurISOmshortxxxAer
\else
\ifnum#1=33 %
\hatcurISOmshortxxxBbs
\else
\ifnum#1=35 %
\hatcurISOmshortxxxBebs
\else
??????\fi
\fi
\fi
\fi
\fi
}
\newcommand{\hatcurISOmv}[1]{\ifnum#1=32 %
\hatcurISOmvxxxA
\else
\ifnum#1=34 %
\hatcurISOmvxxxAe
\else
\ifnum#1=36 %
\hatcurISOmvxxxAer
\else
\ifnum#1=33 %
\hatcurISOmvxxxBbs
\else
\ifnum#1=35 %
\hatcurISOmvxxxBebs
\else
??????\fi
\fi
\fi
\fi
\fi
}
\newcommand{\hatcurISOr}[1]{\ifnum#1=32 %
\hatcurISOrxxxA
\else
\ifnum#1=34 %
\hatcurISOrxxxAe
\else
\ifnum#1=36 %
\hatcurISOrxxxAer
\else
\ifnum#1=33 %
\hatcurISOrxxxBbs
\else
\ifnum#1=35 %
\hatcurISOrxxxBebs
\else
??????\fi
\fi
\fi
\fi
\fi
}
\newcommand{\hatcurISOrho}[1]{\ifnum#1=32 %
\hatcurISOrhoxxxA
\else
\ifnum#1=34 %
\hatcurISOrhoxxxAe
\else
\ifnum#1=36 %
\hatcurISOrhoxxxAer
\else
\ifnum#1=33 %
\hatcurISOrhoxxxBbs
\else
\ifnum#1=35 %
\hatcurISOrhoxxxBebs
\else
??????\fi
\fi
\fi
\fi
\fi
}
\newcommand{\hatcurISOrlong}[1]{\ifnum#1=32 %
\hatcurISOrlongxxxA
\else
\ifnum#1=34 %
\hatcurISOrlongxxxAe
\else
\ifnum#1=36 %
\hatcurISOrlongxxxAer
\else
\ifnum#1=33 %
\hatcurISOrlongxxxBbs
\else
\ifnum#1=35 %
\hatcurISOrlongxxxBebs
\else
??????\fi
\fi
\fi
\fi
\fi
}
\newcommand{\hatcurISOrshort}[1]{\ifnum#1=32 %
\hatcurISOrshortxxxA
\else
\ifnum#1=34 %
\hatcurISOrshortxxxAe
\else
\ifnum#1=36 %
\hatcurISOrshortxxxAer
\else
\ifnum#1=33 %
\hatcurISOrshortxxxBbs
\else
\ifnum#1=35 %
\hatcurISOrshortxxxBebs
\else
??????\fi
\fi
\fi
\fi
\fi
}
\newcommand{\hatcurISOsigma}[1]{\ifnum#1=32 %
\hatcurISOsigmaxxxA
\else
\ifnum#1=34 %
\hatcurISOsigmaxxxAe
\else
\ifnum#1=36 %
\hatcurISOsigmaxxxAer
\else
\ifnum#1=33 %
\hatcurISOsigmaxxxBbs
\else
\ifnum#1=35 %
\hatcurISOsigmaxxxBebs
\else
??????\fi
\fi
\fi
\fi
\fi
}
\newcommand{\hatcurISOspec}[1]{\ifnum#1=32 %
\hatcurISOspecxxxA
\else
\ifnum#1=34 %
\hatcurISOspecxxxAe
\else
\ifnum#1=36 %
\hatcurISOspecxxxAer
\else
\ifnum#1=33 %
\hatcurISOspecxxxBbs
\else
\ifnum#1=35 %
\hatcurISOspecxxxBebs
\else
??????\fi
\fi
\fi
\fi
\fi
}
\newcommand{\hatcurISOvi}[1]{\ifnum#1=32 %
\hatcurISOvixxxA
\else
\ifnum#1=34 %
\hatcurISOvixxxAe
\else
\ifnum#1=36 %
\hatcurISOvixxxAer
\else
\ifnum#1=33 %
\hatcurISOvixxxBbs
\else
\ifnum#1=35 %
\hatcurISOvixxxBebs
\else
??????\fi
\fi
\fi
\fi
\fi
}
\newcommand{\hatcurLBig}[1]{\ifnum#1=32 %
\hatcurLBigxxxA
\else
\ifnum#1=34 %
\hatcurLBigxxxAe
\else
\ifnum#1=36 %
\hatcurLBigxxxAer
\else
\ifnum#1=33 %
\hatcurLBigxxxBbs
\else
\ifnum#1=35 %
\hatcurLBigxxxBebs
\else
??????\fi
\fi
\fi
\fi
\fi
}
\newcommand{\hatcurLBii}[1]{\ifnum#1=32 %
\hatcurLBiixxxA
\else
\ifnum#1=34 %
\hatcurLBiixxxAe
\else
\ifnum#1=36 %
\hatcurLBiixxxAer
\else
\ifnum#1=33 %
\hatcurLBiixxxBbs
\else
\ifnum#1=35 %
\hatcurLBiixxxBebs
\else
??????\fi
\fi
\fi
\fi
\fi
}
\newcommand{\hatcurLBiI}[1]{\ifnum#1=32 %
\hatcurLBiIxxxA
\else
\ifnum#1=34 %
\hatcurLBiIxxxAe
\else
\ifnum#1=36 %
\hatcurLBiIxxxAer
\else
\ifnum#1=33 %
\hatcurLBiIxxxBbs
\else
\ifnum#1=35 %
\hatcurLBiIxxxBebs
\else
??????\fi
\fi
\fi
\fi
\fi
}
\newcommand{\hatcurLBiig}[1]{\ifnum#1=32 %
\hatcurLBiigxxxA
\else
\ifnum#1=34 %
\hatcurLBiigxxxAe
\else
\ifnum#1=36 %
\hatcurLBiigxxxAer
\else
\ifnum#1=33 %
\hatcurLBiigxxxBbs
\else
\ifnum#1=35 %
\hatcurLBiigxxxBebs
\else
??????\fi
\fi
\fi
\fi
\fi
}
\newcommand{\hatcurLBiii}[1]{\ifnum#1=32 %
\hatcurLBiiixxxA
\else
\ifnum#1=34 %
\hatcurLBiiixxxAe
\else
\ifnum#1=36 %
\hatcurLBiiixxxAer
\else
\ifnum#1=33 %
\hatcurLBiiixxxBbs
\else
\ifnum#1=35 %
\hatcurLBiiixxxBebs
\else
??????\fi
\fi
\fi
\fi
\fi
}
\newcommand{\hatcurLBiiI}[1]{\ifnum#1=32 %
\hatcurLBiiIxxxA
\else
\ifnum#1=34 %
\hatcurLBiiIxxxAe
\else
\ifnum#1=36 %
\hatcurLBiiIxxxAer
\else
\ifnum#1=33 %
\hatcurLBiiIxxxBbs
\else
\ifnum#1=35 %
\hatcurLBiiIxxxBebs
\else
??????\fi
\fi
\fi
\fi
\fi
}
\newcommand{\hatcurLBiikep}[1]{\ifnum#1=32 %
\hatcurLBiikepxxxA
\else
\ifnum#1=34 %
\hatcurLBiikepxxxAe
\else
\ifnum#1=36 %
\hatcurLBiikepxxxAer
\else
\ifnum#1=33 %
\hatcurLBiikepxxxBbs
\else
\ifnum#1=35 %
\hatcurLBiikepxxxBebs
\else
??????\fi
\fi
\fi
\fi
\fi
}
\newcommand{\hatcurLBiiz}[1]{\ifnum#1=32 %
\hatcurLBiizxxxA
\else
\ifnum#1=34 %
\hatcurLBiizxxxAe
\else
\ifnum#1=36 %
\hatcurLBiizxxxAer
\else
\ifnum#1=33 %
\hatcurLBiizxxxBbs
\else
\ifnum#1=35 %
\hatcurLBiizxxxBebs
\else
??????\fi
\fi
\fi
\fi
\fi
}
\newcommand{\hatcurLBikep}[1]{\ifnum#1=32 %
\hatcurLBikepxxxA
\else
\ifnum#1=34 %
\hatcurLBikepxxxAe
\else
\ifnum#1=36 %
\hatcurLBikepxxxAer
\else
\ifnum#1=33 %
\hatcurLBikepxxxBbs
\else
\ifnum#1=35 %
\hatcurLBikepxxxBebs
\else
??????\fi
\fi
\fi
\fi
\fi
}
\newcommand{\hatcurLBiz}[1]{\ifnum#1=32 %
\hatcurLBizxxxA
\else
\ifnum#1=34 %
\hatcurLBizxxxAe
\else
\ifnum#1=36 %
\hatcurLBizxxxAer
\else
\ifnum#1=33 %
\hatcurLBizxxxBbs
\else
\ifnum#1=35 %
\hatcurLBizxxxBebs
\else
??????\fi
\fi
\fi
\fi
\fi
}
\newcommand{\hatcurLCbsq}[1]{\ifnum#1=32 %
\hatcurLCbsqxxxA
\else
\ifnum#1=34 %
\hatcurLCbsqxxxAe
\else
\ifnum#1=36 %
\hatcurLCbsqxxxAer
\else
\ifnum#1=33 %
\hatcurLCbsqxxxBbs
\else
\ifnum#1=35 %
\hatcurLCbsqxxxBebs
\else
??????\fi
\fi
\fi
\fi
\fi
}
\newcommand{\hatcurLCdip}[1]{\ifnum#1=32 %
\hatcurLCdipxxxA
\else
\ifnum#1=34 %
\hatcurLCdipxxxAe
\else
\ifnum#1=36 %
\hatcurLCdipxxxAer
\else
\ifnum#1=33 %
\hatcurLCdipxxxBbs
\else
\ifnum#1=35 %
\hatcurLCdipxxxBebs
\else
??????\fi
\fi
\fi
\fi
\fi
}
\newcommand{\hatcurLCdur}[1]{\ifnum#1=32 %
\hatcurLCdurxxxA
\else
\ifnum#1=34 %
\hatcurLCdurxxxAe
\else
\ifnum#1=36 %
\hatcurLCdurxxxAer
\else
\ifnum#1=33 %
\hatcurLCdurxxxBbs
\else
\ifnum#1=35 %
\hatcurLCdurxxxBebs
\else
??????\fi
\fi
\fi
\fi
\fi
}
\newcommand{\hatcurLCdurhr}[1]{\ifnum#1=32 %
\hatcurLCdurhrxxxA
\else
\ifnum#1=34 %
\hatcurLCdurhrxxxAe
\else
\ifnum#1=36 %
\hatcurLCdurhrxxxAer
\else
\ifnum#1=33 %
\hatcurLCdurhrxxxBbs
\else
\ifnum#1=35 %
\hatcurLCdurhrxxxBebs
\else
??????\fi
\fi
\fi
\fi
\fi
}
\newcommand{\hatcurLCdurhrshort}[1]{\ifnum#1=32 %
\hatcurLCdurhrshortxxxA
\else
\ifnum#1=34 %
\hatcurLCdurhrshortxxxAe
\else
\ifnum#1=36 %
\hatcurLCdurhrshortxxxAer
\else
\ifnum#1=33 %
\hatcurLCdurhrshortxxxBbs
\else
\ifnum#1=35 %
\hatcurLCdurhrshortxxxBebs
\else
??????\fi
\fi
\fi
\fi
\fi
}
\newcommand{\hatcurLCdurshort}[1]{\ifnum#1=32 %
\hatcurLCdurshortxxxA
\else
\ifnum#1=34 %
\hatcurLCdurshortxxxAe
\else
\ifnum#1=36 %
\hatcurLCdurshortxxxAer
\else
\ifnum#1=33 %
\hatcurLCdurshortxxxBbs
\else
\ifnum#1=35 %
\hatcurLCdurshortxxxBebs
\else
??????\fi
\fi
\fi
\fi
\fi
}
\newcommand{\hatcurLChatnetmA}[1]{\ifnum#1=32 %
\hatcurLChatnetmAxxxA
\else
\ifnum#1=34 %
\hatcurLChatnetmAxxxAe
\else
\ifnum#1=36 %
\hatcurLChatnetmAxxxAer
\else
\ifnum#1=33 %
\hatcurLChatnetmAxxxBbs
\else
\ifnum#1=35 %
\hatcurLChatnetmAxxxBebs
\else
??????\fi
\fi
\fi
\fi
\fi
}
\newcommand{\hatcurLChatnetmB}[1]{\ifnum#1=32 %
\hatcurLChatnetmBxxxA
\else
\ifnum#1=34 %
\hatcurLChatnetmBxxxAe
\else
\ifnum#1=36 %
\hatcurLChatnetmBxxxAer
\else
\ifnum#1=33 %
\hatcurLChatnetmBxxxBbs
\else
\ifnum#1=35 %
\hatcurLChatnetmBxxxBebs
\else
??????\fi
\fi
\fi
\fi
\fi
}
\newcommand{\hatcurLCiblendA}[1]{\ifnum#1=32 %
\hatcurLCiblendAxxxA
\else
\ifnum#1=34 %
\hatcurLCiblendAxxxAe
\else
\ifnum#1=36 %
\hatcurLCiblendAxxxAer
\else
\ifnum#1=33 %
\hatcurLCiblendAxxxBbs
\else
\ifnum#1=35 %
\hatcurLCiblendAxxxBebs
\else
??????\fi
\fi
\fi
\fi
\fi
}
\newcommand{\hatcurLCiblendB}[1]{\ifnum#1=32 %
\hatcurLCiblendBxxxA
\else
\ifnum#1=34 %
\hatcurLCiblendBxxxAe
\else
\ifnum#1=36 %
\hatcurLCiblendBxxxAer
\else
\ifnum#1=33 %
\hatcurLCiblendBxxxBbs
\else
\ifnum#1=35 %
\hatcurLCiblendBxxxBebs
\else
??????\fi
\fi
\fi
\fi
\fi
}
\newcommand{\hatcurLCimp}[1]{\ifnum#1=32 %
\hatcurLCimpxxxA
\else
\ifnum#1=34 %
\hatcurLCimpxxxAe
\else
\ifnum#1=36 %
\hatcurLCimpxxxAer
\else
\ifnum#1=33 %
\hatcurLCimpxxxBbs
\else
\ifnum#1=35 %
\hatcurLCimpxxxBebs
\else
??????\fi
\fi
\fi
\fi
\fi
}
\newcommand{\hatcurLCingdur}[1]{\ifnum#1=32 %
\hatcurLCingdurxxxA
\else
\ifnum#1=34 %
\hatcurLCingdurxxxAe
\else
\ifnum#1=36 %
\hatcurLCingdurxxxAer
\else
\ifnum#1=33 %
\hatcurLCingdurxxxBbs
\else
\ifnum#1=35 %
\hatcurLCingdurxxxBebs
\else
??????\fi
\fi
\fi
\fi
\fi
}
\newcommand{\hatcurLCP}[1]{\ifnum#1=32 %
\hatcurLCPxxxA
\else
\ifnum#1=34 %
\hatcurLCPxxxAe
\else
\ifnum#1=36 %
\hatcurLCPxxxAer
\else
\ifnum#1=33 %
\hatcurLCPxxxBbs
\else
\ifnum#1=35 %
\hatcurLCPxxxBebs
\else
??????\fi
\fi
\fi
\fi
\fi
}
\newcommand{\hatcurLCPprec}[1]{\ifnum#1=32 %
\hatcurLCPprecxxxA
\else
\ifnum#1=34 %
\hatcurLCPprecxxxAe
\else
\ifnum#1=36 %
\hatcurLCPprecxxxAer
\else
\ifnum#1=33 %
\hatcurLCPprecxxxBbs
\else
\ifnum#1=35 %
\hatcurLCPprecxxxBebs
\else
??????\fi
\fi
\fi
\fi
\fi
}
\newcommand{\hatcurLCPshort}[1]{\ifnum#1=32 %
\hatcurLCPshortxxxA
\else
\ifnum#1=34 %
\hatcurLCPshortxxxAe
\else
\ifnum#1=36 %
\hatcurLCPshortxxxAer
\else
\ifnum#1=33 %
\hatcurLCPshortxxxBbs
\else
\ifnum#1=35 %
\hatcurLCPshortxxxBebs
\else
??????\fi
\fi
\fi
\fi
\fi
}
\newcommand{\hatcurLCq}[1]{\ifnum#1=32 %
\hatcurLCqxxxA
\else
\ifnum#1=34 %
\hatcurLCqxxxAe
\else
\ifnum#1=36 %
\hatcurLCqxxxAer
\else
\ifnum#1=33 %
\hatcurLCqxxxBbs
\else
\ifnum#1=35 %
\hatcurLCqxxxBebs
\else
??????\fi
\fi
\fi
\fi
\fi
}
\newcommand{\hatcurLCqshort}[1]{\ifnum#1=32 %
\hatcurLCqshortxxxA
\else
\ifnum#1=34 %
\hatcurLCqshortxxxAe
\else
\ifnum#1=36 %
\hatcurLCqshortxxxAer
\else
\ifnum#1=33 %
\hatcurLCqshortxxxBbs
\else
\ifnum#1=35 %
\hatcurLCqshortxxxBebs
\else
??????\fi
\fi
\fi
\fi
\fi
}
\newcommand{\hatcurLCrprstar}[1]{\ifnum#1=32 %
\hatcurLCrprstarxxxA
\else
\ifnum#1=34 %
\hatcurLCrprstarxxxAe
\else
\ifnum#1=36 %
\hatcurLCrprstarxxxAer
\else
\ifnum#1=33 %
\hatcurLCrprstarxxxBbs
\else
\ifnum#1=35 %
\hatcurLCrprstarxxxBebs
\else
??????\fi
\fi
\fi
\fi
\fi
}
\newcommand{\hatcurLCT}[1]{\ifnum#1=32 %
\hatcurLCTxxxA
\else
\ifnum#1=34 %
\hatcurLCTxxxAe
\else
\ifnum#1=36 %
\hatcurLCTxxxAer
\else
\ifnum#1=33 %
\hatcurLCTxxxBbs
\else
\ifnum#1=35 %
\hatcurLCTxxxBebs
\else
??????\fi
\fi
\fi
\fi
\fi
}
\newcommand{\hatcurLCTA}[1]{\ifnum#1=32 %
\hatcurLCTAxxxA
\else
\ifnum#1=34 %
\hatcurLCTAxxxAe
\else
\ifnum#1=36 %
\hatcurLCTAxxxAer
\else
\ifnum#1=33 %
\hatcurLCTAxxxBbs
\else
\ifnum#1=35 %
\hatcurLCTAxxxBebs
\else
??????\fi
\fi
\fi
\fi
\fi
}
\newcommand{\hatcurLCTB}[1]{\ifnum#1=32 %
\hatcurLCTBxxxA
\else
\ifnum#1=34 %
\hatcurLCTBxxxAe
\else
\ifnum#1=36 %
\hatcurLCTBxxxAer
\else
\ifnum#1=33 %
\hatcurLCTBxxxBbs
\else
\ifnum#1=35 %
\hatcurLCTBxxxBebs
\else
??????\fi
\fi
\fi
\fi
\fi
}
\newcommand{\hatcurLCzeta}[1]{\ifnum#1=32 %
\hatcurLCzetaxxxA
\else
\ifnum#1=34 %
\hatcurLCzetaxxxAe
\else
\ifnum#1=36 %
\hatcurLCzetaxxxAer
\else
\ifnum#1=33 %
\hatcurLCzetaxxxBbs
\else
\ifnum#1=35 %
\hatcurLCzetaxxxBebs
\else
??????\fi
\fi
\fi
\fi
\fi
}
\newcommand{\hatcurPPaequiv}[1]{\ifnum#1=32 %
\hatcurPPaequivxxxA
\else
\ifnum#1=34 %
\hatcurPPaequivxxxAe
\else
\ifnum#1=36 %
\hatcurPPaequivxxxAer
\else
\ifnum#1=33 %
\hatcurPPaequivxxxBbs
\else
\ifnum#1=35 %
\hatcurPPaequivxxxBebs
\else
??????\fi
\fi
\fi
\fi
\fi
}
\newcommand{\hatcurPPar}[1]{\ifnum#1=32 %
\hatcurPParxxxA
\else
\ifnum#1=34 %
\hatcurPParxxxAe
\else
\ifnum#1=36 %
\hatcurPParxxxAer
\else
\ifnum#1=33 %
\hatcurPParxxxBbs
\else
\ifnum#1=35 %
\hatcurPParxxxBebs
\else
??????\fi
\fi
\fi
\fi
\fi
}
\newcommand{\hatcurPParel}[1]{\ifnum#1=32 %
\hatcurPParelxxxA
\else
\ifnum#1=34 %
\hatcurPParelxxxAe
\else
\ifnum#1=36 %
\hatcurPParelxxxAer
\else
\ifnum#1=33 %
\hatcurPParelxxxBbs
\else
\ifnum#1=35 %
\hatcurPParelxxxBebs
\else
??????\fi
\fi
\fi
\fi
\fi
}
\newcommand{\hatcurPPfluxap}[1]{\ifnum#1=32 %
\hatcurPPfluxapxxxA
\else
\ifnum#1=34 %
\hatcurPPfluxapxxxAe
\else
\ifnum#1=36 %
\hatcurPPfluxapxxxAer
\else
\ifnum#1=33 %
\hatcurPPfluxapxxxBbs
\else
\ifnum#1=35 %
\hatcurPPfluxapxxxBebs
\else
??????\fi
\fi
\fi
\fi
\fi
}
\newcommand{\hatcurPPfluxapdim}[1]{\ifnum#1=32 %
\hatcurPPfluxapdimxxxA
\else
\ifnum#1=34 %
\hatcurPPfluxapdimxxxAe
\else
\ifnum#1=36 %
\hatcurPPfluxapdimxxxAer
\else
\ifnum#1=33 %
\hatcurPPfluxapdimxxxBbs
\else
\ifnum#1=35 %
\hatcurPPfluxapdimxxxBebs
\else
??????\fi
\fi
\fi
\fi
\fi
}
\newcommand{\hatcurPPfluxavg}[1]{\ifnum#1=32 %
\hatcurPPfluxavgxxxA
\else
\ifnum#1=34 %
\hatcurPPfluxavgxxxAe
\else
\ifnum#1=36 %
\hatcurPPfluxavgxxxAer
\else
\ifnum#1=33 %
\hatcurPPfluxavgxxxBbs
\else
\ifnum#1=35 %
\hatcurPPfluxavgxxxBebs
\else
??????\fi
\fi
\fi
\fi
\fi
}
\newcommand{\hatcurPPfluxavgdim}[1]{\ifnum#1=32 %
\hatcurPPfluxavgdimxxxA
\else
\ifnum#1=34 %
\hatcurPPfluxavgdimxxxAe
\else
\ifnum#1=36 %
\hatcurPPfluxavgdimxxxAer
\else
\ifnum#1=33 %
\hatcurPPfluxavgdimxxxBbs
\else
\ifnum#1=35 %
\hatcurPPfluxavgdimxxxBebs
\else
??????\fi
\fi
\fi
\fi
\fi
}
\newcommand{\hatcurPPfluxperi}[1]{\ifnum#1=32 %
\hatcurPPfluxperixxxA
\else
\ifnum#1=34 %
\hatcurPPfluxperixxxAe
\else
\ifnum#1=36 %
\hatcurPPfluxperixxxAer
\else
\ifnum#1=33 %
\hatcurPPfluxperixxxBbs
\else
\ifnum#1=35 %
\hatcurPPfluxperixxxBebs
\else
??????\fi
\fi
\fi
\fi
\fi
}
\newcommand{\hatcurPPfluxperidim}[1]{\ifnum#1=32 %
\hatcurPPfluxperidimxxxA
\else
\ifnum#1=34 %
\hatcurPPfluxperidimxxxAe
\else
\ifnum#1=36 %
\hatcurPPfluxperidimxxxAer
\else
\ifnum#1=33 %
\hatcurPPfluxperidimxxxBbs
\else
\ifnum#1=35 %
\hatcurPPfluxperidimxxxBebs
\else
??????\fi
\fi
\fi
\fi
\fi
}
\newcommand{\hatcurPPg}[1]{\ifnum#1=32 %
\hatcurPPgxxxA
\else
\ifnum#1=34 %
\hatcurPPgxxxAe
\else
\ifnum#1=36 %
\hatcurPPgxxxAer
\else
\ifnum#1=33 %
\hatcurPPgxxxBbs
\else
\ifnum#1=35 %
\hatcurPPgxxxBebs
\else
??????\fi
\fi
\fi
\fi
\fi
}
\newcommand{\hatcurPPi}[1]{\ifnum#1=32 %
\hatcurPPixxxA
\else
\ifnum#1=34 %
\hatcurPPixxxAe
\else
\ifnum#1=36 %
\hatcurPPixxxAer
\else
\ifnum#1=33 %
\hatcurPPixxxBbs
\else
\ifnum#1=35 %
\hatcurPPixxxBebs
\else
??????\fi
\fi
\fi
\fi
\fi
}
\newcommand{\hatcurPPlogg}[1]{\ifnum#1=32 %
\hatcurPPloggxxxA
\else
\ifnum#1=34 %
\hatcurPPloggxxxAe
\else
\ifnum#1=36 %
\hatcurPPloggxxxAer
\else
\ifnum#1=33 %
\hatcurPPloggxxxBbs
\else
\ifnum#1=35 %
\hatcurPPloggxxxBebs
\else
??????\fi
\fi
\fi
\fi
\fi
}
\newcommand{\hatcurPPm}[1]{\ifnum#1=32 %
\hatcurPPmxxxA
\else
\ifnum#1=34 %
\hatcurPPmxxxAe
\else
\ifnum#1=36 %
\hatcurPPmxxxAer
\else
\ifnum#1=33 %
\hatcurPPmxxxBbs
\else
\ifnum#1=35 %
\hatcurPPmxxxBebs
\else
??????\fi
\fi
\fi
\fi
\fi
}
\newcommand{\hatcurPPme}[1]{\ifnum#1=32 %
\hatcurPPmexxxA
\else
\ifnum#1=34 %
\hatcurPPmexxxAe
\else
\ifnum#1=36 %
\hatcurPPmexxxAer
\else
\ifnum#1=33 %
\hatcurPPmexxxBbs
\else
\ifnum#1=35 %
\hatcurPPmexxxBebs
\else
??????\fi
\fi
\fi
\fi
\fi
}
\newcommand{\hatcurPPmelong}[1]{\ifnum#1=32 %
\hatcurPPmelongxxxA
\else
\ifnum#1=34 %
\hatcurPPmelongxxxAe
\else
\ifnum#1=36 %
\hatcurPPmelongxxxAer
\else
\ifnum#1=33 %
\hatcurPPmelongxxxBbs
\else
\ifnum#1=35 %
\hatcurPPmelongxxxBebs
\else
??????\fi
\fi
\fi
\fi
\fi
}
\newcommand{\hatcurPPmeshort}[1]{\ifnum#1=32 %
\hatcurPPmeshortxxxA
\else
\ifnum#1=34 %
\hatcurPPmeshortxxxAe
\else
\ifnum#1=36 %
\hatcurPPmeshortxxxAer
\else
\ifnum#1=33 %
\hatcurPPmeshortxxxBbs
\else
\ifnum#1=35 %
\hatcurPPmeshortxxxBebs
\else
??????\fi
\fi
\fi
\fi
\fi
}
\newcommand{\hatcurPPmlong}[1]{\ifnum#1=32 %
\hatcurPPmlongxxxA
\else
\ifnum#1=34 %
\hatcurPPmlongxxxAe
\else
\ifnum#1=36 %
\hatcurPPmlongxxxAer
\else
\ifnum#1=33 %
\hatcurPPmlongxxxBbs
\else
\ifnum#1=35 %
\hatcurPPmlongxxxBebs
\else
??????\fi
\fi
\fi
\fi
\fi
}
\newcommand{\hatcurPPmrcorr}[1]{\ifnum#1=32 %
\hatcurPPmrcorrxxxA
\else
\ifnum#1=34 %
\hatcurPPmrcorrxxxAe
\else
\ifnum#1=36 %
\hatcurPPmrcorrxxxAer
\else
\ifnum#1=33 %
\hatcurPPmrcorrxxxBbs
\else
\ifnum#1=35 %
\hatcurPPmrcorrxxxBebs
\else
??????\fi
\fi
\fi
\fi
\fi
}
\newcommand{\hatcurPPmshort}[1]{\ifnum#1=32 %
\hatcurPPmshortxxxA
\else
\ifnum#1=34 %
\hatcurPPmshortxxxAe
\else
\ifnum#1=36 %
\hatcurPPmshortxxxAer
\else
\ifnum#1=33 %
\hatcurPPmshortxxxBbs
\else
\ifnum#1=35 %
\hatcurPPmshortxxxBebs
\else
??????\fi
\fi
\fi
\fi
\fi
}
\newcommand{\hatcurPPperi}[1]{\ifnum#1=32 %
\hatcurPPperixxxA
\else
\ifnum#1=34 %
\hatcurPPperixxxAe
\else
\ifnum#1=36 %
\hatcurPPperixxxAer
\else
\ifnum#1=33 %
\hatcurPPperixxxBbs
\else
\ifnum#1=35 %
\hatcurPPperixxxBebs
\else
??????\fi
\fi
\fi
\fi
\fi
}
\newcommand{\hatcurPPphiconj}[1]{\ifnum#1=32 %
\hatcurPPphiconjxxxA
\else
\ifnum#1=34 %
\hatcurPPphiconjxxxAe
\else
\ifnum#1=36 %
\hatcurPPphiconjxxxAer
\else
\ifnum#1=33 %
\hatcurPPphiconjxxxBbs
\else
\ifnum#1=35 %
\hatcurPPphiconjxxxBebs
\else
??????\fi
\fi
\fi
\fi
\fi
}
\newcommand{\hatcurPPr}[1]{\ifnum#1=32 %
\hatcurPPrxxxA
\else
\ifnum#1=34 %
\hatcurPPrxxxAe
\else
\ifnum#1=36 %
\hatcurPPrxxxAer
\else
\ifnum#1=33 %
\hatcurPPrxxxBbs
\else
\ifnum#1=35 %
\hatcurPPrxxxBebs
\else
??????\fi
\fi
\fi
\fi
\fi
}
\newcommand{\hatcurPPre}[1]{\ifnum#1=32 %
\hatcurPPrexxxA
\else
\ifnum#1=34 %
\hatcurPPrexxxAe
\else
\ifnum#1=36 %
\hatcurPPrexxxAer
\else
\ifnum#1=33 %
\hatcurPPrexxxBbs
\else
\ifnum#1=35 %
\hatcurPPrexxxBebs
\else
??????\fi
\fi
\fi
\fi
\fi
}
\newcommand{\hatcurPPrelong}[1]{\ifnum#1=32 %
\hatcurPPrelongxxxA
\else
\ifnum#1=34 %
\hatcurPPrelongxxxAe
\else
\ifnum#1=36 %
\hatcurPPrelongxxxAer
\else
\ifnum#1=33 %
\hatcurPPrelongxxxBbs
\else
\ifnum#1=35 %
\hatcurPPrelongxxxBebs
\else
??????\fi
\fi
\fi
\fi
\fi
}
\newcommand{\hatcurPPreshort}[1]{\ifnum#1=32 %
\hatcurPPreshortxxxA
\else
\ifnum#1=34 %
\hatcurPPreshortxxxAe
\else
\ifnum#1=36 %
\hatcurPPreshortxxxAer
\else
\ifnum#1=33 %
\hatcurPPreshortxxxBbs
\else
\ifnum#1=35 %
\hatcurPPreshortxxxBebs
\else
??????\fi
\fi
\fi
\fi
\fi
}
\newcommand{\hatcurPPrho}[1]{\ifnum#1=32 %
\hatcurPPrhoxxxA
\else
\ifnum#1=34 %
\hatcurPPrhoxxxAe
\else
\ifnum#1=36 %
\hatcurPPrhoxxxAer
\else
\ifnum#1=33 %
\hatcurPPrhoxxxBbs
\else
\ifnum#1=35 %
\hatcurPPrhoxxxBebs
\else
??????\fi
\fi
\fi
\fi
\fi
}
\newcommand{\hatcurPPrlong}[1]{\ifnum#1=32 %
\hatcurPPrlongxxxA
\else
\ifnum#1=34 %
\hatcurPPrlongxxxAe
\else
\ifnum#1=36 %
\hatcurPPrlongxxxAer
\else
\ifnum#1=33 %
\hatcurPPrlongxxxBbs
\else
\ifnum#1=35 %
\hatcurPPrlongxxxBebs
\else
??????\fi
\fi
\fi
\fi
\fi
}
\newcommand{\hatcurPPrshort}[1]{\ifnum#1=32 %
\hatcurPPrshortxxxA
\else
\ifnum#1=34 %
\hatcurPPrshortxxxAe
\else
\ifnum#1=36 %
\hatcurPPrshortxxxAer
\else
\ifnum#1=33 %
\hatcurPPrshortxxxBbs
\else
\ifnum#1=35 %
\hatcurPPrshortxxxBebs
\else
??????\fi
\fi
\fi
\fi
\fi
}
\newcommand{\hatcurPPtcirc}[1]{\ifnum#1=32 %
\hatcurPPtcircxxxA
\else
\ifnum#1=34 %
\hatcurPPtcircxxxAe
\else
\ifnum#1=36 %
\hatcurPPtcircxxxAer
\else
\ifnum#1=33 %
\hatcurPPtcircxxxBbs
\else
\ifnum#1=35 %
\hatcurPPtcircxxxBebs
\else
??????\fi
\fi
\fi
\fi
\fi
}
\newcommand{\hatcurPPteff}[1]{\ifnum#1=32 %
\hatcurPPteffxxxA
\else
\ifnum#1=34 %
\hatcurPPteffxxxAe
\else
\ifnum#1=36 %
\hatcurPPteffxxxAer
\else
\ifnum#1=33 %
\hatcurPPteffxxxBbs
\else
\ifnum#1=35 %
\hatcurPPteffxxxBebs
\else
??????\fi
\fi
\fi
\fi
\fi
}
\newcommand{\hatcurPPtheta}[1]{\ifnum#1=32 %
\hatcurPPthetaxxxA
\else
\ifnum#1=34 %
\hatcurPPthetaxxxAe
\else
\ifnum#1=36 %
\hatcurPPthetaxxxAer
\else
\ifnum#1=33 %
\hatcurPPthetaxxxBbs
\else
\ifnum#1=35 %
\hatcurPPthetaxxxBebs
\else
??????\fi
\fi
\fi
\fi
\fi
}
\newcommand{\hatcurPPtinfall}[1]{\ifnum#1=32 %
\hatcurPPtinfallxxxA
\else
\ifnum#1=34 %
\hatcurPPtinfallxxxAe
\else
\ifnum#1=36 %
\hatcurPPtinfallxxxAer
\else
\ifnum#1=33 %
\hatcurPPtinfallxxxBbs
\else
\ifnum#1=35 %
\hatcurPPtinfallxxxBebs
\else
??????\fi
\fi
\fi
\fi
\fi
}
\newcommand{\hatcurRVBScorr}[1]{\ifnum#1=33 %
\hatcurRVBScorrxxxBbs
\else
\ifnum#1=35 %
\hatcurRVBScorrxxxBebs
\else
??????\fi
\fi
}
\newcommand{\hatcurRVeccen}[1]{\ifnum#1=32 %
\hatcurRVeccenxxxA
\else
\ifnum#1=34 %
\hatcurRVeccenxxxAe
\else
\ifnum#1=36 %
\hatcurRVeccenxxxAer
\else
\ifnum#1=33 %
\hatcurRVeccenxxxBbs
\else
\ifnum#1=35 %
\hatcurRVeccenxxxBebs
\else
??????\fi
\fi
\fi
\fi
\fi
}
\newcommand{\hatcurRVfitrms}[1]{\ifnum#1=32 %
\hatcurRVfitrmsxxxA
\else
\ifnum#1=34 %
\hatcurRVfitrmsxxxAe
\else
\ifnum#1=36 %
\hatcurRVfitrmsxxxAer
\else
\ifnum#1=33 %
\hatcurRVfitrmsxxxBbs
\else
\ifnum#1=35 %
\hatcurRVfitrmsxxxBebs
\else
??????\fi
\fi
\fi
\fi
\fi
}
\newcommand{\hatcurRVgamma}[1]{\ifnum#1=32 %
\hatcurRVgammaxxxA
\else
\ifnum#1=34 %
\hatcurRVgammaxxxAe
\else
\ifnum#1=36 %
\hatcurRVgammaxxxAer
\else
\ifnum#1=33 %
\hatcurRVgammaxxxBbs
\else
\ifnum#1=35 %
\hatcurRVgammaxxxBebs
\else
??????\fi
\fi
\fi
\fi
\fi
}
\newcommand{\hatcurRVh}[1]{\ifnum#1=32 %
\hatcurRVhxxxA
\else
\ifnum#1=34 %
\hatcurRVhxxxAe
\else
\ifnum#1=36 %
\hatcurRVhxxxAer
\else
\ifnum#1=33 %
\hatcurRVhxxxBbs
\else
\ifnum#1=35 %
\hatcurRVhxxxBebs
\else
??????\fi
\fi
\fi
\fi
\fi
}
\newcommand{\hatcurRVjitter}[1]{\ifnum#1=32 %
\hatcurRVjitterxxxA
\else
\ifnum#1=34 %
\hatcurRVjitterxxxAe
\else
\ifnum#1=36 %
\hatcurRVjitterxxxAer
\else
\ifnum#1=33 %
\hatcurRVjitterxxxBbs
\else
\ifnum#1=35 %
\hatcurRVjitterxxxBebs
\else
??????\fi
\fi
\fi
\fi
\fi
}
\newcommand{\hatcurRVk}[1]{\ifnum#1=32 %
\hatcurRVkxxxA
\else
\ifnum#1=34 %
\hatcurRVkxxxAe
\else
\ifnum#1=36 %
\hatcurRVkxxxAer
\else
\ifnum#1=33 %
\hatcurRVkxxxBbs
\else
\ifnum#1=35 %
\hatcurRVkxxxBebs
\else
??????\fi
\fi
\fi
\fi
\fi
}
\newcommand{\hatcurRVK}[1]{\ifnum#1=32 %
\hatcurRVKxxxA
\else
\ifnum#1=34 %
\hatcurRVKxxxAe
\else
\ifnum#1=36 %
\hatcurRVKxxxAer
\else
\ifnum#1=33 %
\hatcurRVKxxxBbs
\else
\ifnum#1=35 %
\hatcurRVKxxxBebs
\else
??????\fi
\fi
\fi
\fi
\fi
}
\newcommand{\hatcurRVomega}[1]{\ifnum#1=32 %
\hatcurRVomegaxxxA
\else
\ifnum#1=34 %
\hatcurRVomegaxxxAe
\else
\ifnum#1=36 %
\hatcurRVomegaxxxAer
\else
\ifnum#1=33 %
\hatcurRVomegaxxxBbs
\else
\ifnum#1=35 %
\hatcurRVomegaxxxBebs
\else
??????\fi
\fi
\fi
\fi
\fi
}
\newcommand{\hatcurSMEiilogg}[1]{\ifnum#1=32 %
\hatcurSMEiiloggxxxA
\else
\ifnum#1=34 %
\hatcurSMEiiloggxxxAe
\else
\ifnum#1=36 %
\hatcurSMEiiloggxxxAer
\else
\ifnum#1=33 %
\hatcurSMEiiloggxxxBbs
\else
\ifnum#1=35 %
\hatcurSMEiiloggxxxBebs
\else
??????\fi
\fi
\fi
\fi
\fi
}
\newcommand{\hatcurSMEiiteff}[1]{\ifnum#1=32 %
\hatcurSMEiiteffxxxA
\else
\ifnum#1=34 %
\hatcurSMEiiteffxxxAe
\else
\ifnum#1=36 %
\hatcurSMEiiteffxxxAer
\else
\ifnum#1=33 %
\hatcurSMEiiteffxxxBbs
\else
\ifnum#1=35 %
\hatcurSMEiiteffxxxBebs
\else
??????\fi
\fi
\fi
\fi
\fi
}
\newcommand{\hatcurSMEiivmac}[1]{\ifnum#1=32 %
\hatcurSMEiivmacxxxA
\else
\ifnum#1=34 %
\hatcurSMEiivmacxxxAe
\else
\ifnum#1=36 %
\hatcurSMEiivmacxxxAer
\else
\ifnum#1=33 %
\hatcurSMEiivmacxxxBbs
\else
\ifnum#1=35 %
\hatcurSMEiivmacxxxBebs
\else
??????\fi
\fi
\fi
\fi
\fi
}
\newcommand{\hatcurSMEiivmic}[1]{\ifnum#1=32 %
\hatcurSMEiivmicxxxA
\else
\ifnum#1=34 %
\hatcurSMEiivmicxxxAe
\else
\ifnum#1=36 %
\hatcurSMEiivmicxxxAer
\else
\ifnum#1=33 %
\hatcurSMEiivmicxxxBbs
\else
\ifnum#1=35 %
\hatcurSMEiivmicxxxBebs
\else
??????\fi
\fi
\fi
\fi
\fi
}
\newcommand{\hatcurSMEiivsin}[1]{\ifnum#1=32 %
\hatcurSMEiivsinxxxA
\else
\ifnum#1=34 %
\hatcurSMEiivsinxxxAe
\else
\ifnum#1=36 %
\hatcurSMEiivsinxxxAer
\else
\ifnum#1=33 %
\hatcurSMEiivsinxxxBbs
\else
\ifnum#1=35 %
\hatcurSMEiivsinxxxBebs
\else
??????\fi
\fi
\fi
\fi
\fi
}
\newcommand{\hatcurSMEiizfeh}[1]{\ifnum#1=32 %
\hatcurSMEiizfehxxxA
\else
\ifnum#1=34 %
\hatcurSMEiizfehxxxAe
\else
\ifnum#1=36 %
\hatcurSMEiizfehxxxAer
\else
\ifnum#1=33 %
\hatcurSMEiizfehxxxBbs
\else
\ifnum#1=35 %
\hatcurSMEiizfehxxxBebs
\else
??????\fi
\fi
\fi
\fi
\fi
}
\newcommand{\hatcurSMEiizfehshort}[1]{\ifnum#1=32 %
\hatcurSMEiizfehshortxxxA
\else
\ifnum#1=34 %
\hatcurSMEiizfehshortxxxAe
\else
\ifnum#1=36 %
\hatcurSMEiizfehshortxxxAer
\else
\ifnum#1=33 %
\hatcurSMEiizfehshortxxxBbs
\else
\ifnum#1=35 %
\hatcurSMEiizfehshortxxxBebs
\else
??????\fi
\fi
\fi
\fi
\fi
}
\newcommand{\hatcurSMEilogg}[1]{\ifnum#1=32 %
\hatcurSMEiloggxxxA
\else
\ifnum#1=34 %
\hatcurSMEiloggxxxAe
\else
\ifnum#1=36 %
\hatcurSMEiloggxxxAer
\else
\ifnum#1=33 %
\hatcurSMEiloggxxxBbs
\else
\ifnum#1=35 %
\hatcurSMEiloggxxxBebs
\else
??????\fi
\fi
\fi
\fi
\fi
}
\newcommand{\hatcurSMEiteff}[1]{\ifnum#1=32 %
\hatcurSMEiteffxxxA
\else
\ifnum#1=34 %
\hatcurSMEiteffxxxAe
\else
\ifnum#1=36 %
\hatcurSMEiteffxxxAer
\else
\ifnum#1=33 %
\hatcurSMEiteffxxxBbs
\else
\ifnum#1=35 %
\hatcurSMEiteffxxxBebs
\else
??????\fi
\fi
\fi
\fi
\fi
}
\newcommand{\hatcurSMEivmac}[1]{\ifnum#1=32 %
\hatcurSMEivmacxxxA
\else
\ifnum#1=34 %
\hatcurSMEivmacxxxAe
\else
\ifnum#1=36 %
\hatcurSMEivmacxxxAer
\else
\ifnum#1=33 %
\hatcurSMEivmacxxxBbs
\else
\ifnum#1=35 %
\hatcurSMEivmacxxxBebs
\else
??????\fi
\fi
\fi
\fi
\fi
}
\newcommand{\hatcurSMEivmic}[1]{\ifnum#1=32 %
\hatcurSMEivmicxxxA
\else
\ifnum#1=34 %
\hatcurSMEivmicxxxAe
\else
\ifnum#1=36 %
\hatcurSMEivmicxxxAer
\else
\ifnum#1=33 %
\hatcurSMEivmicxxxBbs
\else
\ifnum#1=35 %
\hatcurSMEivmicxxxBebs
\else
??????\fi
\fi
\fi
\fi
\fi
}
\newcommand{\hatcurSMEivsin}[1]{\ifnum#1=32 %
\hatcurSMEivsinxxxA
\else
\ifnum#1=34 %
\hatcurSMEivsinxxxAe
\else
\ifnum#1=36 %
\hatcurSMEivsinxxxAer
\else
\ifnum#1=33 %
\hatcurSMEivsinxxxBbs
\else
\ifnum#1=35 %
\hatcurSMEivsinxxxBebs
\else
??????\fi
\fi
\fi
\fi
\fi
}
\newcommand{\hatcurSMEizfeh}[1]{\ifnum#1=32 %
\hatcurSMEizfehxxxA
\else
\ifnum#1=34 %
\hatcurSMEizfehxxxAe
\else
\ifnum#1=36 %
\hatcurSMEizfehxxxAer
\else
\ifnum#1=33 %
\hatcurSMEizfehxxxBbs
\else
\ifnum#1=35 %
\hatcurSMEizfehxxxBebs
\else
??????\fi
\fi
\fi
\fi
\fi
}
\newcommand{\hatcurSMEizfehshort}[1]{\ifnum#1=32 %
\hatcurSMEizfehshortxxxA
\else
\ifnum#1=34 %
\hatcurSMEizfehshortxxxAe
\else
\ifnum#1=36 %
\hatcurSMEizfehshortxxxAer
\else
\ifnum#1=33 %
\hatcurSMEizfehshortxxxBbs
\else
\ifnum#1=35 %
\hatcurSMEizfehshortxxxBebs
\else
??????\fi
\fi
\fi
\fi
\fi
}
\newcommand{\hatcurTRESgamma}[1]{\ifnum#1=32 %
\hatcurTRESgammaxxxA
\else
\ifnum#1=34 %
\hatcurTRESgammaxxxAe
\else
\ifnum#1=36 %
\hatcurTRESgammaxxxAer
\else
\ifnum#1=33 %
\hatcurTRESgammaxxxBbs
\else
\ifnum#1=35 %
\hatcurTRESgammaxxxBebs
\else
??????\fi
\fi
\fi
\fi
\fi
}
\newcommand{\hatcurTRESlogg}[1]{\ifnum#1=32 %
\hatcurTRESloggxxxA
\else
\ifnum#1=34 %
\hatcurTRESloggxxxAe
\else
\ifnum#1=36 %
\hatcurTRESloggxxxAer
\else
\ifnum#1=33 %
\hatcurTRESloggxxxBbs
\else
\ifnum#1=35 %
\hatcurTRESloggxxxBebs
\else
??????\fi
\fi
\fi
\fi
\fi
}
\newcommand{\hatcurTRESnumspec}[1]{\ifnum#1=32 %
\hatcurTRESnumspecxxxA
\else
\ifnum#1=34 %
\hatcurTRESnumspecxxxAe
\else
\ifnum#1=36 %
\hatcurTRESnumspecxxxAer
\else
\ifnum#1=33 %
\hatcurTRESnumspecxxxBbs
\else
\ifnum#1=35 %
\hatcurTRESnumspecxxxBebs
\else
??????\fi
\fi
\fi
\fi
\fi
}
\newcommand{\hatcurTRESrvrms}[1]{\ifnum#1=32 %
\hatcurTRESrvrmsxxxA
\else
\ifnum#1=34 %
\hatcurTRESrvrmsxxxAe
\else
\ifnum#1=36 %
\hatcurTRESrvrmsxxxAer
\else
\ifnum#1=33 %
\hatcurTRESrvrmsxxxBbs
\else
\ifnum#1=35 %
\hatcurTRESrvrmsxxxBebs
\else
??????\fi
\fi
\fi
\fi
\fi
}
\newcommand{\hatcurTRESspan}[1]{\ifnum#1=32 %
\hatcurTRESspanxxxA
\else
\ifnum#1=34 %
\hatcurTRESspanxxxAe
\else
\ifnum#1=36 %
\hatcurTRESspanxxxAer
\else
\ifnum#1=33 %
\hatcurTRESspanxxxBbs
\else
\ifnum#1=35 %
\hatcurTRESspanxxxBebs
\else
??????\fi
\fi
\fi
\fi
\fi
}
\newcommand{\hatcurTRESteff}[1]{\ifnum#1=32 %
\hatcurTRESteffxxxA
\else
\ifnum#1=34 %
\hatcurTRESteffxxxAe
\else
\ifnum#1=36 %
\hatcurTRESteffxxxAer
\else
\ifnum#1=33 %
\hatcurTRESteffxxxBbs
\else
\ifnum#1=35 %
\hatcurTRESteffxxxBebs
\else
??????\fi
\fi
\fi
\fi
\fi
}
\newcommand{\hatcurTRESvsini}[1]{\ifnum#1=32 %
\hatcurTRESvsinixxxA
\else
\ifnum#1=34 %
\hatcurTRESvsinixxxAe
\else
\ifnum#1=36 %
\hatcurTRESvsinixxxAer
\else
\ifnum#1=33 %
\hatcurTRESvsinixxxBbs
\else
\ifnum#1=35 %
\hatcurTRESvsinixxxBebs
\else
??????\fi
\fi
\fi
\fi
\fi
}
\newcommand{\hatcurTRESzfeh}[1]{\ifnum#1=32 %
\hatcurTRESzfehxxxA
\else
\ifnum#1=34 %
\hatcurTRESzfehxxxAe
\else
\ifnum#1=36 %
\hatcurTRESzfehxxxAer
\else
\ifnum#1=33 %
\hatcurTRESzfehxxxBbs
\else
\ifnum#1=35 %
\hatcurTRESzfehxxxBebs
\else
??????\fi
\fi
\fi
\fi
\fi
}
\newcommand{\hatcurXdist}[1]{\ifnum#1=32 %
\hatcurXdistxxxA
\else
\ifnum#1=34 %
\hatcurXdistxxxAe
\else
\ifnum#1=36 %
\hatcurXdistxxxAer
\else
\ifnum#1=33 %
\hatcurXdistxxxBbs
\else
\ifnum#1=35 %
\hatcurXdistxxxBebs
\else
??????\fi
\fi
\fi
\fi
\fi
}
\newcommand{\hatcurXsecdur}[1]{\ifnum#1=32 %
\hatcurXsecdurxxxA
\else
\ifnum#1=34 %
\hatcurXsecdurxxxAe
\else
\ifnum#1=36 %
\hatcurXsecdurxxxAer
\else
\ifnum#1=33 %
\hatcurXsecdurxxxBbs
\else
\ifnum#1=35 %
\hatcurXsecdurxxxBebs
\else
??????\fi
\fi
\fi
\fi
\fi
}
\newcommand{\hatcurXsecingdur}[1]{\ifnum#1=32 %
\hatcurXsecingdurxxxA
\else
\ifnum#1=34 %
\hatcurXsecingdurxxxAe
\else
\ifnum#1=36 %
\hatcurXsecingdurxxxAer
\else
\ifnum#1=33 %
\hatcurXsecingdurxxxBbs
\else
\ifnum#1=35 %
\hatcurXsecingdurxxxBebs
\else
??????\fi
\fi
\fi
\fi
\fi
}
\newcommand{\hatcurXsecondary}[1]{\ifnum#1=32 %
\hatcurXsecondaryxxxA
\else
\ifnum#1=34 %
\hatcurXsecondaryxxxAe
\else
\ifnum#1=36 %
\hatcurXsecondaryxxxAer
\else
\ifnum#1=33 %
\hatcurXsecondaryxxxBbs
\else
\ifnum#1=35 %
\hatcurXsecondaryxxxBebs
\else
??????\fi
\fi
\fi
\fi
\fi
}
\newcommand{\hatcurXsecphase}[1]{\ifnum#1=32 %
\hatcurXsecphasexxxA
\else
\ifnum#1=34 %
\hatcurXsecphasexxxAe
\else
\ifnum#1=36 %
\hatcurXsecphasexxxAer
\else
\ifnum#1=33 %
\hatcurXsecphasexxxBbs
\else
\ifnum#1=35 %
\hatcurXsecphasexxxBebs
\else
??????\fi
\fi
\fi
\fi
\fi
}
\newcommand{\hatcurxxxxA}{HAT-P-32}
\newcommand{\hatcurbxxxxA}{HAT-P-32b}
\newcommand{\hatcurcxxxxA}{HAT-P-32c}

\newcommand{\hatcurplanetnumxxxxA}{32}

\newcommand{\hatcurRVgammaabsxxxxA}{\hatcurDSgamma{\hatcurplanetnumxxxxA}}                           

\newcommand{\hatcurRVgammarelxxxxA}{\hatcurRVgamma{\hatcurplanetnumxxxxA}}                           

\newcommand{\hatcurCCtassvixxxxA}{\ensuremath{0.62\pm0.10}}                  

\newcommand{\hatcurSMEversionxxxxA}{ii}                                       

\newcommand{\hatcurisoshortxxxxA}{YY}
\newcommand{\hatcurisofullxxxxA}{Yonsei-Yale}
\newcommand{\hatcurisocitexxxxA}{yi:2001}

\newcommand{\hatcurlumindxxxxA}{\arstar}

\newcommand{\hatcurjhkfilsetxxxxA}{ESO}

\newcommand{\hatcurSMEteffxxxxA}{\ifthenelse{\equal{\hatcurSMEversionxxxxA}{i}}{\hatcurSMEiteff{\hatcurplanetnumxxxxA}}{\hatcurSMEiiteff{\hatcurplanetnumxxxxA}}}
\newcommand{\hatcurSMEzfehxxxxA}{\ifthenelse{\equal{\hatcurSMEversionxxxxA}{i}}{\hatcurSMEizfeh{\hatcurplanetnumxxxxA}}{\hatcurSMEiizfeh{\hatcurplanetnumxxxxA}}}
\newcommand{\hatcurSMEzfehshortxxxxA}{\ifthenelse{\equal{\hatcurSMEversionxxxxA}{i}}{\hatcurSMEizfehshort{\hatcurplanetnumxxxxA}}{\hatcurSMEiizfehshort{\hatcurplanetnumxxxxA}}}
\newcommand{\hatcurSMEloggxxxxA}{\ifthenelse{\equal{\hatcurSMEversionxxxxA}{i}}{\hatcurSMEilogg{\hatcurplanetnumxxxxA}}{\hatcurSMEiilogg{\hatcurplanetnumxxxxA}}}
\newcommand{\hatcurSMEvsinxxxxA}{\ifthenelse{\equal{\hatcurSMEversionxxxxA}{i}}{\hatcurSMEivsin{\hatcurplanetnumxxxxA}}{\hatcurSMEiivsin{\hatcurplanetnumxxxxA}}}
\newcommand{\hatcurSMEvmacxxxxA}{\ifthenelse{\equal{\hatcurSMEversionxxxxA}{i}}{\hatcurSMEivmac{\hatcurplanetnumxxxxA}}{\hatcurSMEiivmac{\hatcurplanetnumxxxxA}}}
\newcommand{\hatcurSMEvmicxxxxA}{\ifthenelse{\equal{\hatcurSMEversionxxxxA}{i}}{\hatcurSMEivmic{\hatcurplanetnumxxxxA}}{\hatcurSMEiivmic{\hatcurplanetnumxxxxA}}}

\newcommand{\hatcurxxxxB}{HAT-P-33}
\newcommand{\hatcurbxxxxB}{HAT-P-33b}
\newcommand{\hatcurcxxxxB}{HAT-P-33c}

\newcommand{\hatcurplanetnumxxxxB}{33}

\newcommand{\hatcurRVgammaabsxxxxB}{\hatcurDSgamma{\hatcurplanetnumxxxxB}}                           

\newcommand{\hatcurRVgammarelxxxxB}{\hatcurRVgamma{\hatcurplanetnumxxxxB}}                           

\newcommand{\hatcurCCtassvixxxxB}{\ensuremath{0.577\pm0.087}}                  

\newcommand{\hatcurSMEversionxxxxB}{ii}                                       

\newcommand{\hatcurisoshortxxxxB}{YY}
\newcommand{\hatcurisofullxxxxB}{Yonsei-Yale (YY)}
\newcommand{\hatcurisocitexxxxB}{yi:2001}

\newcommand{\hatcurlumindxxxxB}{\arstar}

\newcommand{\hatcurjhkfilsetxxxxB}{ESO}

\newcommand{\hatcurSMEteffxxxxB}{\ifthenelse{\equal{\hatcurSMEversionxxxxB}{i}}{\hatcurSMEiteff{\hatcurplanetnumxxxxB}}{\hatcurSMEiiteff{\hatcurplanetnumxxxxB}}}
\newcommand{\hatcurSMEzfehxxxxB}{\ifthenelse{\equal{\hatcurSMEversionxxxxB}{i}}{\hatcurSMEizfeh{\hatcurplanetnumxxxxB}}{\hatcurSMEiizfeh{\hatcurplanetnumxxxxB}}}
\newcommand{\hatcurSMEzfehshortxxxxB}{\ifthenelse{\equal{\hatcurSMEversionxxxxB}{i}}{\hatcurSMEizfehshort{\hatcurplanetnumxxxxB}}{\hatcurSMEiizfehshort{\hatcurplanetnumxxxxB}}}
\newcommand{\hatcurSMEloggxxxxB}{\ifthenelse{\equal{\hatcurSMEversionxxxxB}{i}}{\hatcurSMEilogg{\hatcurplanetnumxxxxB}}{\hatcurSMEiilogg{\hatcurplanetnumxxxxB}}}
\newcommand{\hatcurSMEvsinxxxxB}{\ifthenelse{\equal{\hatcurSMEversionxxxxB}{i}}{\hatcurSMEivsin{\hatcurplanetnumxxxxB}}{\hatcurSMEiivsin{\hatcurplanetnumxxxxB}}}
\newcommand{\hatcurSMEvmacxxxxB}{\ifthenelse{\equal{\hatcurSMEversionxxxxB}{i}}{\hatcurSMEivmac{\hatcurplanetnumxxxxB}}{\hatcurSMEiivmac{\hatcurplanetnumxxxxB}}}
\newcommand{\hatcurSMEvmicxxxxB}{\ifthenelse{\equal{\hatcurSMEversionxxxxB}{i}}{\hatcurSMEivmic{\hatcurplanetnumxxxxB}}{\hatcurSMEiivmic{\hatcurplanetnumxxxxB}}}

\newcommand{\hatcur}[1]{\ifnum#1=32 %
\hatcurxxxxA
\else
\ifnum#1=33 %
\hatcurxxxxB
\else
??????\fi
\fi
}
\newcommand{\hatcurb}[1]{\ifnum#1=32 %
\hatcurbxxxxA
\else
\ifnum#1=33 %
\hatcurbxxxxB
\else
??????\fi
\fi
}
\newcommand{\hatcurc}[1]{\ifnum#1=32 %
\hatcurcxxxxA
\else
\ifnum#1=33 %
\hatcurcxxxxB
\else
??????\fi
\fi
}
\newcommand{\hatcurCCtassvi}[1]{\ifnum#1=32 %
\hatcurCCtassvixxxxA
\else
\ifnum#1=33 %
\hatcurCCtassvixxxxB
\else
??????\fi
\fi
}
\newcommand{\hatcurisocite}[1]{\ifnum#1=32 %
\hatcurisocitexxxxA
\else
\ifnum#1=33 %
\hatcurisocitexxxxB
\else
??????\fi
\fi
}
\newcommand{\hatcurisofull}[1]{\ifnum#1=32 %
\hatcurisofullxxxxA
\else
\ifnum#1=33 %
\hatcurisofullxxxxB
\else
??????\fi
\fi
}
\newcommand{\hatcurisoshort}[1]{\ifnum#1=32 %
\hatcurisoshortxxxxA
\else
\ifnum#1=33 %
\hatcurisoshortxxxxB
\else
??????\fi
\fi
}
\newcommand{\hatcurjhkfilset}[1]{\ifnum#1=32 %
\hatcurjhkfilsetxxxxA
\else
\ifnum#1=33 %
\hatcurjhkfilsetxxxxB
\else
??????\fi
\fi
}
\newcommand{\hatcurlumind}[1]{\ifnum#1=32 %
\hatcurlumindxxxxA
\else
\ifnum#1=33 %
\hatcurlumindxxxxB
\else
??????\fi
\fi
}
\newcommand{\hatcurplanetnum}[1]{\ifnum#1=32 %
\hatcurplanetnumxxxxA
\else
\ifnum#1=33 %
\hatcurplanetnumxxxxB
\else
??????\fi
\fi
}
\newcommand{\hatcurRVgammaabs}[1]{\ifnum#1=32 %
\hatcurRVgammaabsxxxxA
\else
\ifnum#1=33 %
\hatcurRVgammaabsxxxxB
\else
??????\fi
\fi
}
\newcommand{\hatcurRVgammarel}[1]{\ifnum#1=32 %
\hatcurRVgammarelxxxxA
\else
\ifnum#1=33 %
\hatcurRVgammarelxxxxB
\else
??????\fi
\fi
}
\newcommand{\hatcurSMElogg}[1]{\ifnum#1=32 %
\hatcurSMEloggxxxxA
\else
\ifnum#1=33 %
\hatcurSMEloggxxxxB
\else
??????\fi
\fi
}
\newcommand{\hatcurSMEteff}[1]{\ifnum#1=32 %
\hatcurSMEteffxxxxA
\else
\ifnum#1=33 %
\hatcurSMEteffxxxxB
\else
??????\fi
\fi
}
\newcommand{\hatcurSMEversion}[1]{\ifnum#1=32 %
\hatcurSMEversionxxxxA
\else
\ifnum#1=33 %
\hatcurSMEversionxxxxB
\else
??????\fi
\fi
}
\newcommand{\hatcurSMEvmac}[1]{\ifnum#1=32 %
\hatcurSMEvmacxxxxA
\else
\ifnum#1=33 %
\hatcurSMEvmacxxxxB
\else
??????\fi
\fi
}
\newcommand{\hatcurSMEvmic}[1]{\ifnum#1=32 %
\hatcurSMEvmicxxxxA
\else
\ifnum#1=33 %
\hatcurSMEvmicxxxxB
\else
??????\fi
\fi
}
\newcommand{\hatcurSMEvsin}[1]{\ifnum#1=32 %
\hatcurSMEvsinxxxxA
\else
\ifnum#1=33 %
\hatcurSMEvsinxxxxB
\else
??????\fi
\fi
}
\newcommand{\hatcurSMEzfeh}[1]{\ifnum#1=32 %
\hatcurSMEzfehxxxxA
\else
\ifnum#1=33 %
\hatcurSMEzfehxxxxB
\else
??????\fi
\fi
}
\newcommand{\hatcurSMEzfehshort}[1]{\ifnum#1=32 %
\hatcurSMEzfehshortxxxxA
\else
\ifnum#1=33 %
\hatcurSMEzfehshortxxxxB
\else
??????\fi
\fi
}

\newcounter{planetcounter}

\newboolean{emulateapj}
\setboolean{emulateapj}{true}

\newboolean{rvtablelong}
\setboolean{rvtablelong}{true}

\newboolean{astroph}
\setboolean{astroph}{true}

\shortauthors{Hartman et al.}
\shorttitle{\hatcur{32}\lowercase{b} and \hatcur{33}\lowercase{b}}
\ifthenelse{\boolean{emulateapj}}{
    \newcommand{\titledag}{$\dagger$}
}{
    \newcommand{\titledag}{\dagger}
}

\begin{document}

\title{
\hatcur{32}\lowercase{b} and \hatcur{33}\lowercase{b}: 
Two Highly Inflated Hot Jupiters\\
Transiting High-Jitter Stars\altaffilmark{\titledag}
}

\author{
    J.~D.~Hartman\altaffilmark{1},
    G.~\'A.~Bakos\altaffilmark{1},
    G.~Torres\altaffilmark{1},
    D.~W.~Latham\altaffilmark{1},
    G\'eza.~Kov\'acs\altaffilmark{2},
    B.~B\'eky\altaffilmark{1},
    S.~N.~Quinn\altaffilmark{1},
    T.~Mazeh\altaffilmark{3},
    A.~Shporer\altaffilmark{4},
    G.~W.~Marcy\altaffilmark{5},
    A.~W.~Howard\altaffilmark{5},
    D.~A.~Fischer\altaffilmark{6},
    J.~A.~Johnson\altaffilmark{7},
    G.~A.~Esquerdo\altaffilmark{1},
    R.~W.~Noyes\altaffilmark{1},
    D.~D.~Sasselov\altaffilmark{1},
    R.~P.~Stefanik\altaffilmark{1},
    J.~M.~Fernandez\altaffilmark{8},
    T.~Szklen\'ar\altaffilmark{1},
    J.~L\'az\'ar\altaffilmark{9},
    I.~Papp\altaffilmark{9},
    P.~S\'ari\altaffilmark{9}
}
\altaffiltext{1}{Harvard-Smithsonian Center for Astrophysics,
    Cambridge, MA; email: gbakos@cfa.harvard.edu}

\altaffiltext{2}{Konkoly Observatory, Budapest, Hungary}

\altaffiltext{3}{School of Physics and Astronomy, Raymond \& Beverly
  Sackler Faculty of Exact Sciences, Tel Aviv University, Tel Aviv
  69978, Israel}

\altaffiltext{4}{LCOGT, 6740 Cortona Drive, Santa Barbara, CA, \&
  Department of Physics, Broida Hall, UC Santa Barbara, CA}

\altaffiltext{5}{Department of Astronomy, University of California,
    Berkeley, CA}

\altaffiltext{6}{Department of Astronomy, Yale University, 
  New Haven, CT}

\altaffiltext{7}{Department of Astrophysics, California Institute of Technology, 
  Pasadena, CA}

\altaffiltext{8}{Georg-August-Universit\"at G\"ottingen, Institut f\"ur Astrophysik, G\"ottingen, Germany}

\altaffiltext{9}{Hungarian Astronomical Association, Budapest, 
    Hungary}

\altaffiltext{$\dagger$}{
    Based in part on observations obtained at the W.~M.~Keck
    Observatory, which is operated by the University of California and
    the California Institute of Technology. Keck time has been granted
    by NOAO (A285Hr, A146Hr, A201Hr, A289Hr), NASA (N128Hr, N145Hr,
    N049Hr, N018Hr, N167Hr, N029Hr), and the NOAO Gemini/Keck
    time-exchange program (G329Hr).
}


\begin{abstract}

\setcounter{footnote}{10}
We report the discovery of two exoplanets transiting high-jitter stars.
\hatcurb{32} orbits the bright V=\hatcurCCtassmv{32}\ late-F--early-G
dwarf star \hatcurCCgsc{32}, with a period $P=\hatcurLCP{32}$\,d. The
stellar and planetary masses and radii depend on the eccentricity of
the system, which is poorly constrained due to the high velocity
jitter ($\sim 80$\,\ms). Assuming a circular orbit, the star has a
mass of \hatcurISOm{32}\,\msun, and radius of \hatcurISOr{32}\,\rsun,
while the planet has a mass of \hatcurPPmlong{32}\,\mjup, and a radius
of \hatcurPPrlong{32}\,\rjup. When the eccentricity is allowed to
vary, the best-fit model has $e = \hatcurRVeccen{34}$ and results in a
planet which is close to filling its Roche Lobe. We determine an
analytic approximation for the transit-inferred radius of an eccentric
planet which fills its Roche Lobe; including the constraint that the
planet cannot exceed its Roche Lobe results in the following best-fit
parameters: $e = \hatcurRVeccen{36}$, $\mpl = \hatcurPPm{36}$\,\mjup,
$\rpl = \hatcurPPr{36}$\,\rjup, $\mstar = \hatcurISOm{36}$\,\msun\ and
$\rstar = \hatcurISOr{36}$\,\rsun. The second planet, \hatcurb{33},
orbits the bright V=\hatcurCCtassmv{33}\ late-F dwarf star
\hatcurCCgsc{33}, with a period $P=\hatcurLCP{33}$\,d. As for
\hatcur{32}, the stellar and planetary masses and radii of \hatcur{33}
depend on the eccentricity, which is poorly constrained due to the
high jitter ($\sim 50$\,\ms). In this case spectral line bisector
spans are significantly anti-correlated with the radial velocity residuals,
and we are able to use this correlation to reduce the residual rms to
$\sim 35$\,\ms. We find the star has a mass of either
\hatcurISOm{33}\,\msun\ or \hatcurISOm{35}\,\msun, and a radius of
either \hatcurISOr{33}\,\rsun\ or \hatcurISOr{35}\,\rsun, while the
planet has a mass of either \hatcurPPmlong{33}\,\mjup\ or
\hatcurPPmlong{35}\,\mjup, and a radius of either
\hatcurPPrlong{33}\,\rjup\ or \hatcurPPrlong{35}\,\rjup, for an
assumed circular orbit or for the best-fit eccentric orbit
respectively. Due to the large bisector span variations exhibited by
both stars we rely on detailed modeling of the photometric light curves to rule out blend scenarios.
Both planets are among the largest radii transiting planets
discovered to date.
\setcounter{footnote}{0}
\end{abstract}

\keywords{
    planetary systems ---
    stars: individual (
\setcounter{planetcounter}{1}
\hatcur{32},
\hatcurCCgsc{32}\loopcommanoperiod
\setcounter{planetcounter}{2}
\hatcur{33},
\hatcurCCgsc{33}\loopcommanoperiod
) 
    techniques: spectroscopic, photometric
}


\section{Introduction}
\label{sec:introduction}

One of the most significant findings that has resulted from the study
of transiting exoplanets (TEPs) over the past decade is the discovery
that some close-in ``hot Jupiters'' have radii that are substantially
larger than what was thought to be theoretically
possible. The three most inflated TEPs, including WASP-17b ($R = 1.99
\pm 0.08$\,\rjup; \citealp{anderson:2011}), WASP-12b ($R = 1.79 \pm
0.09$\,\rjup; \citealp{hebb:2009}), and TrES-4b ($R = 1.78 \pm
0.09$\,\rjup; \citealp{sozzetti:2009}) have radii that are as much as
50\% larger than expected from, for example, the coreless
\citet{fortney:2007} models. Recently it has become clear that the
degree to which TEPs are inflated is correlated with the planet
equilibrium temperature
\citep{fortney:2007,enoch:2011,kovacs:2010,faedi:2011,beky:2011,laughlin:2011}
and anti-correlated with stellar metallicity
\citep{guillot:2006,fortney:2007,burrows:2007,enoch:2011,beky:2011}. Several
mechanisms that might explain the correlation with equilibrium
temperature in particular have been proposed
\citep{bodenheimer:2001,guillot:2002,batygin:2010}, though to date
this issue remains unresolved.

In this work we report the discovery of two new TEPs, \hatcurb{32} and
\hatcurb{33} (orbiting the stars \hatcurCCgsc{32} and \hatcurCCgsc{33},
respectively), with radii among the largest found to date. These
planets have high equilibrium temperatures, supporting the
aforementioned correlation. Both of these highly inflated hot Jupiters
were discovered by the Hungarian-made Automated Telescope Network
\citep[HATNet;][]{bakos:2004} survey for TEPs orbiting bright stars
($9 \lesssim r \lesssim 14.5$). HATNet operates six wide-field
instruments, including four at the Fred Lawrence Whipple Observatory
(FLWO) in Arizona, and two on the roof of the hangar servicing the
Smithsonian Astrophysical Observatory's Submillimeter Array, in
Hawaii.

Although the two planets presented here were among the first
candidates identified by HATNet, with discovery observations dating
back to 2004, they proved to be difficult to confirm due to the
significant radial velocity (RV) jitter exhibited by the stellar hosts
(\hatcurRVjitter{32}\,\ms\ and $55.1$\,\ms\ for \hatcur{32} and
\hatcur{33}, respectively), which limits the power of the traditional
spectral line bisector technique used to rule out blend scenarios. We
argue that the jitter is astrophysical in origin, and likely related
to convective inhomogeneities which vary in time, perhaps due to
time-varying photospheric magnetic fields, and we conduct detailed
blend modeling of the observations to confirm the planetary nature of
these systems. The high jitter values for both stars result in poor
constraints on the orbital eccentricities of the two systems. While we
could assume circular orbits as has been done for other TEPs, we
choose not to do so because several eccentric short-period TEPs have
been discovered (e.g. XO-3b, \citealp{johnskrull:2008}; WASP-14b,
\citealp{joshi:2009}; HAT-P-21b, \citealp{bakos:2010b}), so the
possibility that either planet is eccentric should be included in the
parameter uncertainties. Importantly the inferred planetary radii
depend strongly on the orbital eccentricities. In particular, if
\hatcurb{32} has an eccentric orbit its radius may be larger than any
other known TEP.

The structure of the paper is as follows. In \refsecl{obs} we
summarize the detection of the photometric transit signal and the
subsequent spectroscopic and photometric observations of each star to
confirm the planets. In \refsecl{analysis} we analyze the data to rule
out false positive scenarios, and to determine the stellar and
planetary parameters. Our conclusions are discussed in
\refsecl{discussion}.

\section{Observations}
\label{sec:obs}

\subsection{Photometric detection}
\label{sec:detection}

\begin{figure*}[!ht]
\plottwo{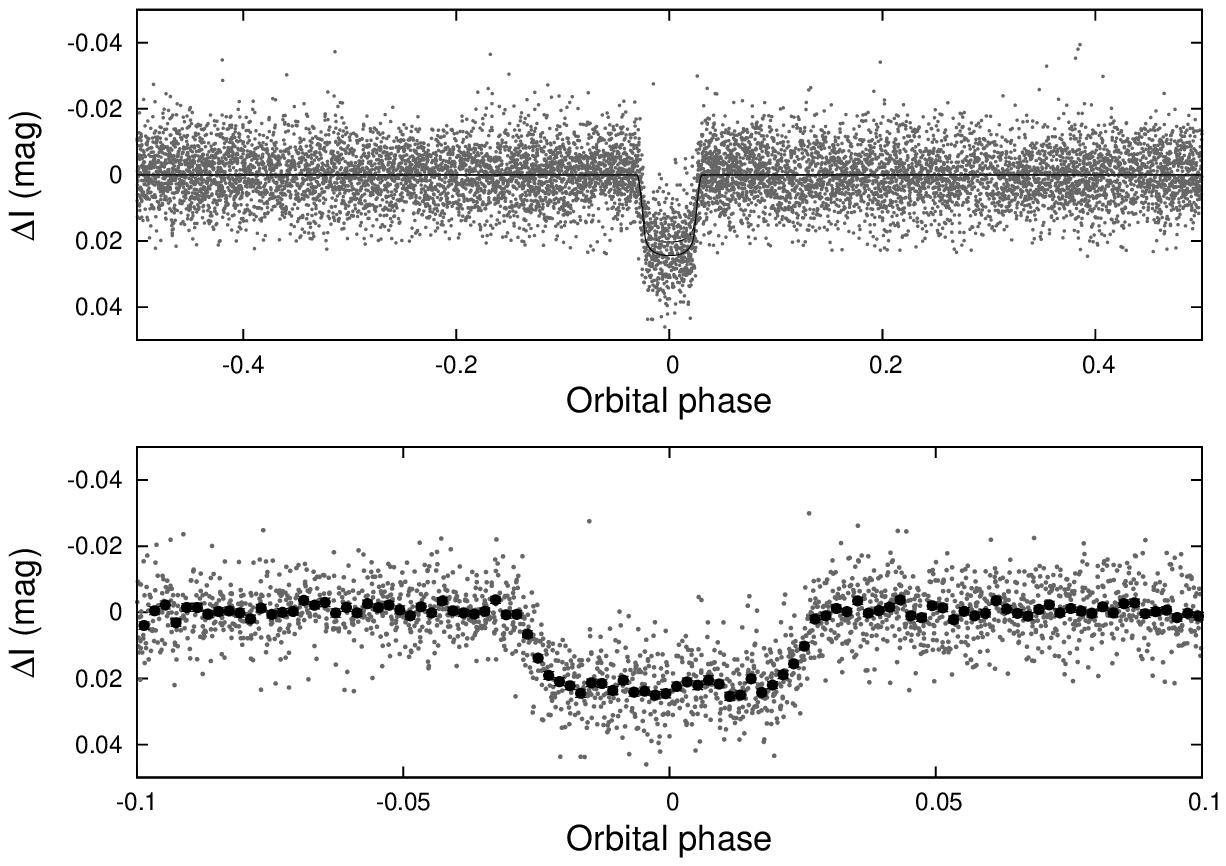}{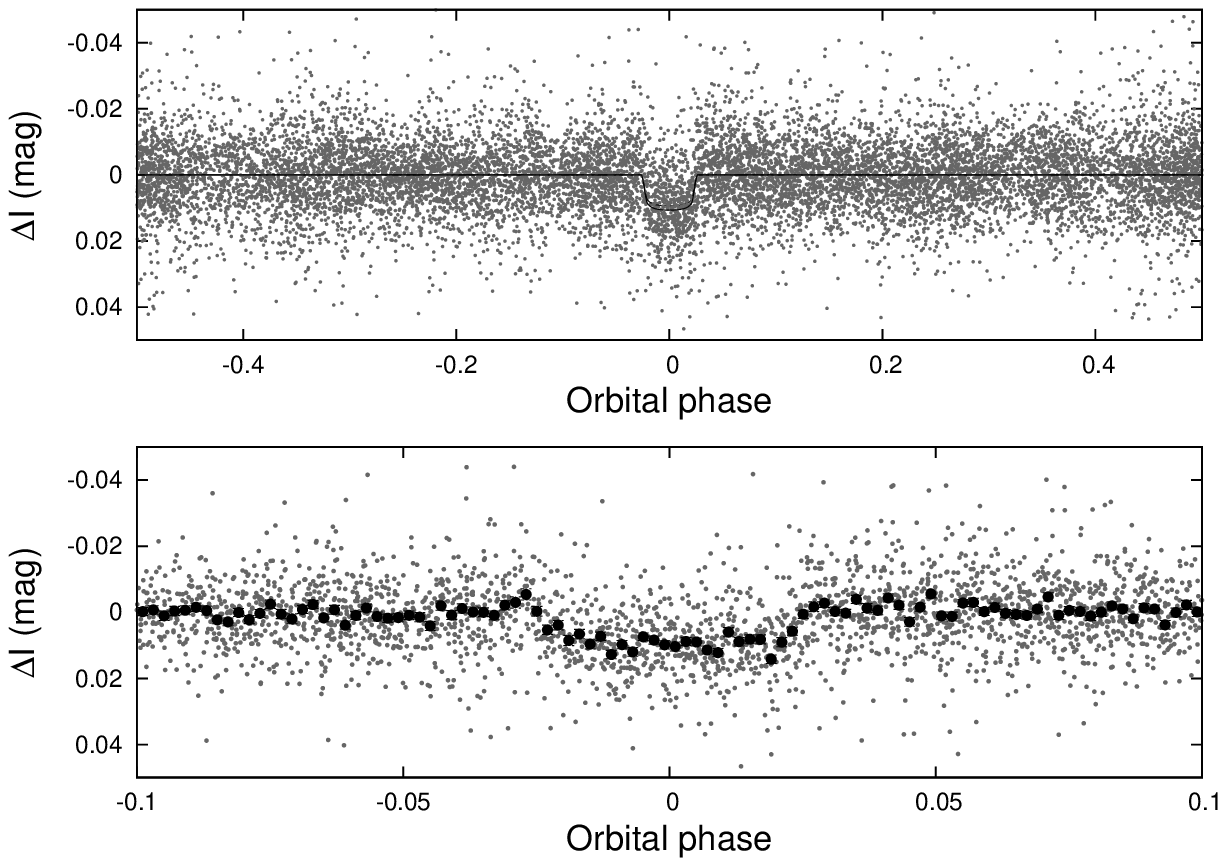}
\caption[]{
    Left: unbinned \lc{} of \hatcur{32} from HATNet (top, see the text
    for details), folded with the period $P =
    \hatcurLCPprec{32}$\,days. The solid line shows a transit-model
    fit to the light curve (\refsecl{globmod}). We also show a
    zoomed-in view of the transit (bottom), with the dark points
    showing the light curve binned in phase with a binsize of
    0.002. Right: same as left, here we show \hatcur{33}.
\label{fig:hatnet32}}
\end{figure*}

\reftabl{photobs} summarizes the HATNet discovery observations of each
new planetary system. The HATNet images were processed and reduced to
trend-filtered light curves following the procedure described by
\cite{bakos:2010}.  The \lcs{} were searched for periodic box-shaped
signals using the Box Least-Squares \citep[BLS; see][]{kovacs:2002}
method. For \hatcur{33} supporting observations of the discovery were
obtained with the Wise-HAT \citep[WHAT;][]{shporer:2009} telescope at Wise
Observatory in Israel; these were analyzed in the same manner as the
HATNet images. We detected significant signals in the \lcs\ of the
stars summarized below:

\begin{itemize}
\item {\em \hatcur{32}} -- \hatcurCCgsc{32} (also known as
  \hatcurCCtwomass{32}; $\alpha = \hatcurCCra{32}$, $\delta =
  \hatcurCCdec{32}$; J2000; V=\hatcurCCtassmv{32}
  \citealp{droege:2006}). A signal was detected for this star with an
  apparent depth of $\sim$\hatcurLCdip{32}\,mmag, and a period of
  $P=$\hatcurLCPshort{32}\,days (see \reffigl{hatnet32}, left).
\item {\em \hatcur{33}} -- \hatcurCCgsc{33} (also known as
  \hatcurCCtwomass{33}; $\alpha = \hatcurCCra{33}$, $\delta =
  \hatcurCCdec{33}$; J2000; V=\hatcurCCtassmv{33}
  \citealp{droege:2006}). A signal was detected for this star with an
  apparent depth of $\sim$\hatcurLCdip{33}\,mmag, and a period of
  $P=$\hatcurLCPshort{33}\,days (see \reffigl{hatnet32}, right).
\end{itemize}

%
%
%

\ifthenelse{\boolean{emulateapj}}{
    \begin{deluxetable*}{llrrr}
}{
    \begin{deluxetable}{llrrr}
}
\tablewidth{0pc}
\tabletypesize{\scriptsize}
\tablecaption{
    Summary of photometric observations
    \label{tab:photobs}
}
\tablehead{
    \colhead{~~~~~~~~Instrument/Field~~~~~~~~}  &
    \colhead{Date(s)} &
    \colhead{Number of Images\tablenotemark{a}} &
    \colhead{Mode Cadence (s)} &
    \colhead{Filter}
}
\startdata
\sidehead{\textbf{\hatcur{32}}}
~~~~HAT-7/G125 & 2005 Sep--2006 Feb & 7077 & 330 & \band{I} \\
~~~~HAT-8/G125 & 2005 Sep--2005 Oct & 816 & 330 & \band{I} \\
~~~~HAT-6/G126 & 2004 Dec--2005 Mar & 1365 & 330 & \band{I} \\
~~~~HAT-9/G126 & 2004 Dec--2005 Mar & 1505 & 330 & \band{I} \\
~~~~KeplerCam         & 2007 Sep 24        & 602 & 33 & \band{z} \\
~~~~KeplerCam         & 2007 Oct 22        & 489 & 53 & \band{z} \\
~~~~KeplerCam\tablenotemark{b}         & 2007 Oct 23        & 161 & 38 & \band{z} \\
~~~~KeplerCam         & 2007 Nov 06        & 759 & 28 & \band{z} \\
~~~~KeplerCam\tablenotemark{b}         & 2007 Nov 18        & 551 & 38 & \band{z} \\
~~~~KeplerCam         & 2007 Nov 19        & 665 & 38 & \band{z} \\
~~~~KeplerCam\tablenotemark{b}         & 2007 Dec 03        & 817 & 33 & \band{z} \\
~~~~KeplerCam         & 2007 Dec 04        & 596 & 33 & \band{g} \\
\sidehead{\textbf{\hatcur{33}}}
~~~~HAT-6/G176 & 2004 Nov--2005 Oct & 2754 & 330 & \band{I} \\
~~~~HAT-9/G176 & 2004 Nov--2005 Oct & 4383 & 330 & \band{I} \\
~~~~WHAT/G221 & 2004 Jan--2005 May & 5439 & 330 & \band{I} \\
~~~~KeplerCam         & 2006 Dec 29        & 413 & 33 & \band{i} \\
~~~~KeplerCam         & 2007 Nov 24        & 274 & 57 & \band{z} \\
~~~~KeplerCam         & 2008 Mar 25        & 260 & 29 & \band{z} \\
~~~~KeplerCam         & 2008 Nov 23        & 668 & 28 & \band{i} \\
~~~~KeplerCam         & 2008 Nov 30        & 257 & 32 & \band{i} \\
~~~~KeplerCam         & 2011 Feb 14        & 346 & 57 & \band{g} \\
~~~~KeplerCam         & 2011 Feb 21        & 336 & 58 & \band{g} \\
\enddata
\tablenotetext{a}{For HATNet and WHAT observations this includes images which were rejected by the photometric reduction pipeline. For KeplerCam observations this excludes images which were rejected by the photometric reduction pipeline, but includes images which were rejected via $\sigma$-clipping during the fitting procedure.}
\tablenotetext{b}{Out of transit observation used to constrain the presence of a secondary transit event during blend modeling.}
\ifthenelse{\boolean{emulateapj}}{
    \end{deluxetable*}
}{
    \end{deluxetable}
}

\subsection{Reconnaissance Spectroscopy}
\label{sec:recspec}

High-resolution, low-S/N ``reconnaissance'' spectra were obtained for
both \hatcur{32} and \hatcur{33} using the Harvard-Smithsonian Center
for Astrophysics (CfA) Digital Speedometer \citep[DS;][]{latham:1992}
on the \flwos\ telescope. These observations, which are summarized in
\reftabl{reconspecobs}, were reduced and analyzed following the
procedure described by \cite{torres:2002} \citep[see
  also][]{latham:2009}. We find that both stars show no velocity
variation at the 1\,\kms\ precision of the observations, and all
spectra are consistent with single, moderately-rotating dwarf
stars. For \hatcur{32} the stellar atmospheric parameters that we
find, assuming solar composition, are
$\teffstar=\hatcurDSteff{32}$\,K, $\loggstar=\hatcurDSlogg{32}$,
$\vsini=\hatcurDSvsini{32}$\,\kms, and $\gamma_{\rm
  RV}=\hatcurDSgamma{32}$\,\kms. For \hatcur{33} we find
$\teffstar=\hatcurDSteff{33}$\,K, $\loggstar=\hatcurDSlogg{33}$,
$\vsini=\hatcurDSvsini{33}$\,\kms, and $\gamma_{\rm
  RV}=\hatcurDSgamma{33}$\,\kms.

In addition to the DS observations for \hatcur{33}, we also obtained
several initial reconnaissance observations of this target with the
SOPHIE spectrograph on the Observatoire de Haute-Provence 1.93\,m
telescope. These observations showed $\sim 100$\,\ms\ scatter, with
only a very faint hint of phasing with the photometric ephemeris,
hinting on the possibility that the system is a blend. Based on these
observations we postponed further follow-up of the target for several
years.

\ifthenelse{\boolean{emulateapj}}{
    \begin{deluxetable}{lrr}
}{
    \begin{deluxetable}{lrr}
}
\tablewidth{0pc}
\tabletypesize{\scriptsize}
\tablecaption{
    DS reconnaissance spectroscopy observations
    \label{tab:reconspecobs}
}
\tablehead{
    \multicolumn{1}{c}{JD - 2400000}             &
    \multicolumn{1}{c}{RV\tablenotemark{a}} &
    \multicolumn{1}{c}{$\sigma_{\rm RV}$\tablenotemark{b}}         \\
    &
    \multicolumn{1}{c}{(\kms)}              &
    \multicolumn{1}{c}{(\kms)}
}
\startdata
\sidehead{\textbf{\hatcur{32}}}
~~~~53988.0008 &  -24.58 &  1.37 \\
~~~~53992.9400 &  -24.02 &  1.05 \\
~~~~54016.7473 &  -23.51 &  0.75 \\
~~~~54070.7298 &  -23.80 &  1.08 \\
~~~~54072.7003 &  -24.07 &  0.71 \\
~~~~54075.8081 &  -22.86 &  0.62 \\
~~~~54077.7263 &  -23.25 &  0.75 \\
~~~~54421.8025 &  -21.91 &  0.81 \\
~~~~54422.7370 &  -23.67 &  0.92 \\
~~~~54423.7596 &  -23.03 &  0.86 \\
~~~~54424.7471 &  -23.80 &  0.78 \\
~~~~54425.8210 &  -22.60 &  0.77 \\
~~~~54427.7552 &  -22.34 &  1.00 \\
~~~~54430.6748 &  -20.83 &  1.70 \\
~~~~54726.8681 &  -23.50 &  0.90 \\
\sidehead{\textbf{\hatcur{33}}}
~~~~53864.6428 &   22.10 &  0.62 \\
~~~~53865.6274 &   21.91 &  0.48 \\
~~~~53866.6480 &   22.41 &  0.48 \\
~~~~53873.6285 &   24.07 &  0.63 \\
~~~~54041.9844 &   23.21 &  0.76 \\
~~~~54043.9879 &   24.41 &  0.79 \\
~~~~54047.0178 &   22.75 &  0.64 \\
~~~~54047.9772 &   23.54 &  0.50 \\
~~~~54048.9252 &   22.84 &  0.66 \\
\enddata 
\tablenotetext{a}{
    The measured heliocentric RV of the target quoted on the native
    CfA system. Our best guess is that $0.14$\,\kms\ should be added
    to put the CfA velocities onto an absolute system defined by
    observations of minor planets.
}
\tablenotetext{b}{
    The RV measurement uncertainty.
}
\ifthenelse{\boolean{emulateapj}}{
    \end{deluxetable}
}{
    \end{deluxetable}
}

\subsection{High resolution, high S/N spectroscopy}
\label{sec:hispec}

\begin{figure} [ht]
\plotone{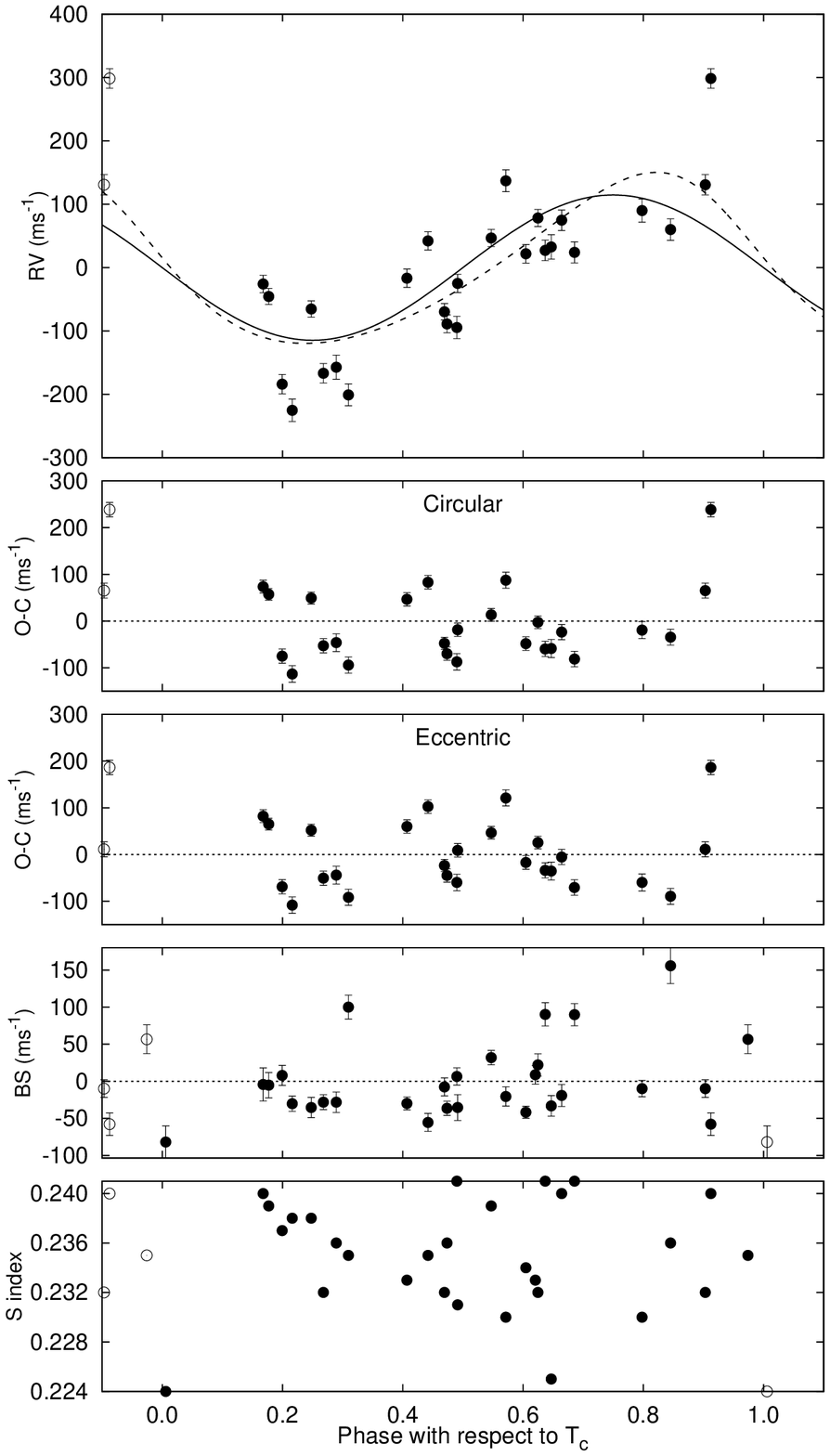}
\ifthenelse{\value{planetcounter}=1}{
\caption{
    {\em Top panel:} Keck/HIRES RV measurements for
    \hbox{\hatcur{32}{}} shown as a function of orbital phase, along
    with our best-fit circular (solid line) and eccentric (dashed
    line) models as determined from our global modeling procedure
    (\refsecl{globmod}); see \reftabl{planetparam32}.  Zero phase
    corresponds to the time of mid-transit.  The center-of-mass
    velocity has been subtracted.
    {\em Second panel:} Velocity $O\!-\!C$ residuals from the best fit
    circular orbit. Jitter is not
    included in the error bars.
    {\em Third panel:} Velocity $O\!-\!C$ residuals from the best
        fit eccentric orbit.
    {\em Fourth panel:} Bisector spans (BS), with the mean value
        subtracted. The measurement from the template spectrum is
        included (see \refsecl{blend}).
    {\em Bottom panel:} Chromospheric activity index $S$
        measured from the Keck spectra.
    Note the different vertical scales of the panels. Observations
    shown twice are represented with open symbols.
}}{
\caption{
    Keck/HIRES observations of \hatcur{32}. The panels are as in
    \reffigl{rvbis32}.  The parameters used in the
    best-fit model are given in \reftabl{planetparam32}.
}}
\label{fig:rvbis32}
\end{figure}

\begin{figure} [ht]
\plotone{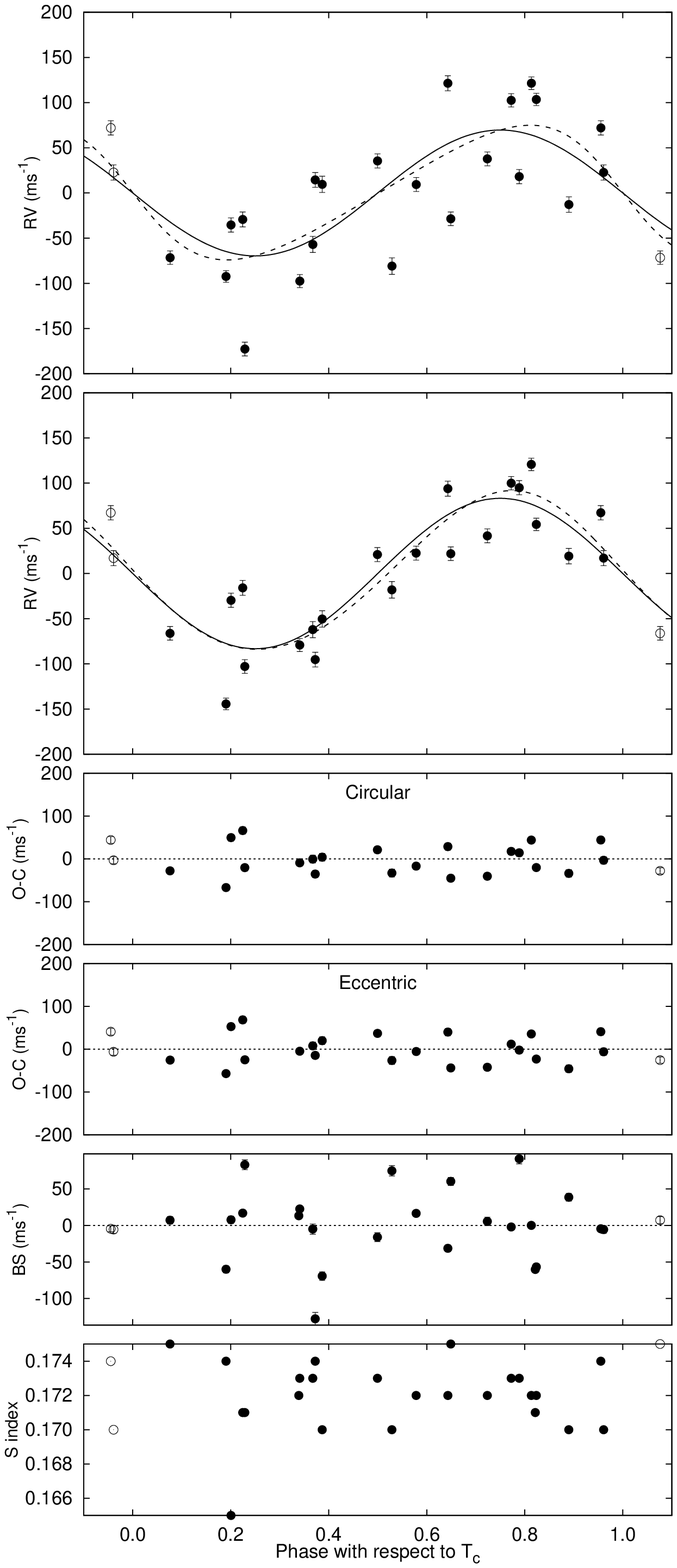}
\caption{
    {\em Top panel:} Keck/HIRES RV measurements for
    \hbox{\hatcur{33}{}} shown as a function of orbital phase, along
    with our best-fit circular (solid line) and eccentric (dashed
    line) models as determined from our global modeling procedure
    (\refsecl{globmod}); see \reftabl{planetparam33}.  Zero phase
    corresponds to the time of mid-transit.  The center-of-mass
    velocity has been subtracted.
    {\em Second panel:} Same as top panel, here we have subtracted a
    linear correlation with the spectral line bisector spans (BS) from
    the RV measurements. This correlation was determined
    simultaneously with the fit, and significantly reduces the
    residual RMS.  For the displayed points we subtract the
    correlation determined in the circular orbit fit.
    {\em Third panel:} Velocity $O\!-\!C$ residuals from the best fit
    circular orbit including the BS correlation. Jitter is not
    included in the error bars.
    {\em Fourth panel:} Velocity $O\!-\!C$ residuals from the best
        fit eccentric orbit including the BS correlation.
    {\em Fifth panel:} BS, with the mean value
        subtracted. The measurement from the template spectrum is
        included (see \refsecl{blend}).
    {\em Bottom panel:} Chromospheric activity index $S$
        measured from the Keck spectra.
    Note the different vertical scales of the panels. Observations
    shown twice are represented with open symbols.
}
\label{fig:rvbis33}
\end{figure}

We obtained high-resolution, high-S/N spectra for both stars using
HIRES \citep{vogt:1994} on the Keck-I telescope in Hawaii. For
\hatcur{32} we gathered 28 spectra with the $\mathrm{I}_2$ absorption
cell \citep{marcy:1992} between 2007 August and 2010 December,
together with a single $\mathrm{I}_{2}$-free template spectrum. We
rejected one low-S/N spectrum for which we were unable to obtain a
high-precision RV measurement, and exclude from the analysis two
spectra which were obtained during transit and thus may be affected by
the Rossiter-McLaughlin effect \citep[e.g.][]{queloz:2000}. For
\hatcur{33} we gathered 22 spectra with the $\mathrm{I}_{2}$ cell
between 2008 September and 2010 December, and two template
spectra. The HIRES spectra were reduced to barycentric RVs following
\cite{butler:1996}. The resulting measurements are given in
Tables~\ref{tab:rvs32} and~\ref{tab:rvs33} for \hatcur{32} and
\hatcur{33} respectively. The phased data, along with our best fit
models for both circular and eccentric orbits, are displayed in
Figures~\ref{fig:rvbis32} and~\ref{fig:rvbis33}. 

For both candidates the RV residuals from the best-fit model show
significant scatter greatly in excess of what is expected based on the
measurement uncertainties. For \hatcur{32} the residual RMS is
\hatcurRVfitrms{32}\,\ms\ or \hatcurRVfitrms{36}\,\ms\ for circular
and eccentric models respectively, while the RMS expected from
instrumental variations plus photon noise is only 15.7\,\ms. For
\hatcur{33} the residual RMS is $55.7$\,\ms\ or $54.1$\,\ms, again for
circular and eccentric models respectively, while the expected RMS is
7.9\,\ms. Because of this high ``jitter'' we gathered substantially
more high-S/N spectra for these targets than we do for typical HATNet
candidates. The reason for this is two-fold: the high jitter could
have been due to additional Keplerian motion due to the presence of
additional planets in the systems which would be revealed by further
observations, and in the presence of high jitter more observations are
needed to precisely determine the orbital parameters of the system. We
consider further the origin of the jitter for each object in
\refsecl{jitter} concluding that it is most likely not due to
additional planets, but rather due to stellar activity.

For each spectrum we also calculated the spectral line bisector span
(BS) following the method described in \S 5 of \cite{bakos:2007} and
the $S$ activity index calibrated to the scale of \cite{vaughan:1978}
following the procedure of \cite{isaacson:2010}. Note that for the BS
we follow the convention ${\rm BS} = v_{1} - v_{2}$ where $v_{1}$ is
the velocity of a point on the bisector near the continuum level and
$v_{2}$ is the velocity of a point on the bisector near the line
core. The BS and S values are plotted in Figures~\ref{fig:rvbis32}
and~\ref{fig:rvbis33}. For both stars the BS and S values do not phase
with the photometric ephemeris; however the scatters in the BS values
are comparable to the RV semiamplitudes, so we are not able to rule
out blend scenarios for either object based on the bisectors. Instead
we rely on detailed blend-modeling of the observations to rule out
these scenarios as described in \refsecl{blend}. For \hatcur{33} we
found that the RV residuals are correlated with the BS (see also
\refsecl{jitter}), and were thus able to use the BS to correct the
RVs, reducing the effective jitter to $\sim 35$\,\ms.

\setcounter{planetcounter}{1}
%
\setcounter{planetcounter}{2}
%

\ifthenelse{\boolean{emulateapj}}{
    \begin{deluxetable*}{lrrrrrr}
}{
    \begin{deluxetable}{lrrrrrr}
}
\tablewidth{0pc}
\tablecaption{
    Relative radial velocities, bisector spans, and activity index
    measurements of \hatcur{32}.
    \label{tab:rvs32}
}
\tablehead{
    \colhead{BJD\tablenotemark{a}} &
    \colhead{RV\tablenotemark{b}} &
    \colhead{\ensuremath{\sigma_{\rm RV}}\tablenotemark{c}} &
    \colhead{BS} &
    \colhead{\ensuremath{\sigma_{\rm BS}}} &
    \colhead{S\tablenotemark{d}} &
    \colhead{Phase}\\
    \colhead{\hbox{(2,454,000$+$)}} &
    \colhead{(\ms)} &
    \colhead{(\ms)} &
    \colhead{(\ms)} &
    \colhead{(\ms)} &
    \colhead{} &
    \colhead{}
}
\startdata
$ 336.95660 $ & $   -12.33 $ & $    13.77 $ & $    -4.23 $ & $    22.35 $ & $    0.240 $ & $   0.168 $\\
$ 337.12852 $ & $   -51.77 $ & $    12.83 $ & $   -35.38 $ & $    13.61 $ & $    0.238 $ & $   0.248 $\\
$ 337.93027 $ & \nodata      & \nodata      & $     8.77 $ & $    12.61 $ & $    0.233 $ & $   0.621 $\\
$ 337.93908 $ & $    91.89 $ & $    13.58 $ & $    22.03 $ & $    15.06 $ & $    0.232 $ & $   0.625 $\\
$ 339.12681 $ & $   -32.01 $ & $    12.59 $ & $    -5.23 $ & $    17.10 $ & $    0.239 $ & $   0.177 $\\
$ 339.92225 $ & $    60.33 $ & $    13.50 $ & $    31.99 $ & $     9.65 $ & $    0.239 $ & $   0.547 $\\
$ 344.05490 $ & $   -56.05 $ & $    12.94 $ & $    -7.56 $ & $    12.17 $ & $    0.232 $ & $   0.469 $\\
$ 345.14032\tablenotemark{e} $ & $ -7.96 $      & $12.44$  & $    56.66 $ & $    19.42 $ & $    0.235 $ & $   0.974 $\\
$ 397.81437 $ & $   -75.23 $ & $    14.22 $ & $   -36.38 $ & $     9.63 $ & $    0.236 $ & $   0.474 $\\
$ 398.09647 $ & $    35.38 $ & $    14.70 $ & $   -41.76 $ & $     8.10 $ & $    0.234 $ & $   0.605 $\\
$ 427.84636 $ & $    55.80 $ & $    14.45 $ & $   -55.48 $ & $    12.13 $ & $    0.235 $ & $   0.442 $\\
$ 428.85791 $ & $   312.16 $ & $    15.38 $ & $   -57.89 $ & $    15.15 $ & $    0.240 $ & $   0.912 $\\
$ 429.92102 $ & $    -3.17 $ & $    14.43 $ & $   -29.84 $ & $     8.73 $ & $    0.233 $ & $   0.407 $\\
$ 455.90267 $ & $   -11.30 $ & $    14.42 $ & $   -35.44 $ & $    17.49 $ & $    0.231 $ & $   0.491 $\\
$ 458.93801 $ & $   144.46 $ & $    16.08 $ & $    -9.85 $ & $    11.93 $ & $    0.232 $ & $   0.903 $\\
$ 460.86208 $ & $   103.67 $ & $    18.28 $ & $    -9.84 $ & $    11.05 $ & $    0.230 $ & $   0.798 $\\
$ 548.72493 $ & $    88.38 $ & $    16.21 $ & $   -19.13 $ & $    14.89 $ & $    0.240 $ & $   0.664 $\\
$ 635.11482 $ & $    73.67 $ & $    16.94 $ & $   155.85 $ & $    24.14 $ & $    0.236 $ & $   0.845 $\\
$ 636.11275 $ & $  -187.30 $ & $    17.27 $ & $   100.10 $ & $    16.18 $ & $    0.235 $ & $   0.310 $\\
$ 724.02617 $ & $  -170.37 $ & $    15.27 $ & $     7.97 $ & $    13.53 $ & $    0.237 $ & $   0.199 $\\
$ 725.07143 $ & $    37.66 $ & $    16.62 $ & $    89.82 $ & $    14.96 $ & $    0.241 $ & $   0.686 $\\
$ 727.11674 $ & $    40.79 $ & $    16.16 $ & $    90.24 $ & $    15.73 $ & $    0.241 $ & $   0.637 $\\
$ 777.92375 $ & $  -153.16 $ & $    15.28 $ & $   -28.12 $ & $    10.06 $ & $    0.232 $ & $   0.268 $\\
$ 810.82619 $ & $   150.64 $ & $    17.27 $ & $   -20.45 $ & $    12.98 $ & $    0.230 $ & $   0.571 $\\
$ 838.93902 $ & $    46.30 $ & $    19.26 $ & $   -33.20 $ & $    13.87 $ & $    0.225 $ & $   0.647 $\\
$ 1192.92153 $ & $  -143.71 $ & $    19.01 $ & $   -28.23 $ & $    13.83 $ & $    0.236 $ & $   0.289 $\\
$ 1250.81416 $ & $  -211.65 $ & $    17.65 $ & $   -30.15 $ & $    10.29 $ & $    0.238 $ & $   0.216 $\\
$ 1376.10348 $ & $   -80.92 $ & $    17.65 $ & $     6.54 $ & $    11.60 $ & $    0.241 $ & $   0.490 $\\
$ 1544.91306\tablenotemark{e} $ & $-186.72$      & $18.93$  & $   -81.83 $ & $    21.62 $ & $    0.224 $ & $   0.006 $\\
\enddata
\tablenotetext{a}{
    Barycentric Julian dates throughout the paper are calculated from 
    Coordinated Universal Time (UTC).
}
\tablenotetext{b}{
    The zero-point of these velocities is arbitrary. An overall offset
    $\gamma_{\rm rel}$ fitted to these velocities in \refsecl{globmod}
    has {\em not} been subtracted.
}
\tablenotetext{c}{
    Internal errors excluding the component of astrophysical jitter
    considered in \refsecl{globmod}.
}
\tablenotetext{d}{
    Chromospheric activity index calibrated to the scale
    of \citet{vaughan:1978} following \citet{isaacson:2010}.
}
\tablenotetext{e}{
    Observation during transit which was excluded from the analysis.
}
\ifthenelse{\boolean{rvtablelong}}{
    \tablecomments{
        For the iodine-free template exposures we do not
        measure the RV but do measure the BS and S index.  Such
        template exposures can be distinguished by the missing RV
        value.
    }
}{
    \tablecomments{
        For the iodine-free template exposures we do not
        measure the RV but do measure the BS and S index.  Such
        template exposures can be distinguished by the missing RV
        value.  This table is presented in its entirety in the
        electronic edition of the Astrophysical Journal.  A portion is
        shown here for guidance regarding its form and content.
    }
} 
\ifthenelse{\boolean{emulateapj}}{
    \end{deluxetable*}
}{
    \end{deluxetable}
}
%
\ifthenelse{\boolean{emulateapj}}{
    \begin{deluxetable*}{lrrrrrr}
}{
    \begin{deluxetable}{lrrrrrr}
}
\tablewidth{0pc}
\tablecaption{
    Relative radial velocities, bisector spans, and activity index
    measurements of \hatcur{33}.
    \label{tab:rvs33}
}
\tablehead{
    \colhead{BJD\tablenotemark{a}} &
    \colhead{RV\tablenotemark{b}} &
    \colhead{\ensuremath{\sigma_{\rm RV}}\tablenotemark{c}} &
    \colhead{BS} &
    \colhead{\ensuremath{\sigma_{\rm BS}}} &
    \colhead{S\tablenotemark{d}} &
    \colhead{Phase}\\
    \colhead{\hbox{(2,454,000$+$)}} &
    \colhead{(\ms)} &
    \colhead{(\ms)} &
    \colhead{(\ms)} &
    \colhead{(\ms)} &
    \colhead{} &
    \colhead{}
}
\startdata
$ 728.11380 $ & \nodata      & \nodata      & $   -60.23 $ & $     2.93 $ & $    0.171 $ & $   0.822 $\\
$ 728.12071 $ & $   107.70 $ & $     6.84 $ & $   -56.67 $ & $     4.32 $ & $    0.172 $ & $   0.824 $\\
$ 778.03749 $ & $   -88.19 $ & $     6.47 $ & $   -59.99 $ & $     4.67 $ & $    0.174 $ & $   0.190 $\\
$ 779.11168 $ & $    39.67 $ & $     7.79 $ & $   -16.02 $ & $     6.12 $ & $    0.173 $ & $   0.499 $\\
$ 780.05973 $ & $   106.79 $ & $     7.40 $ & $    -2.11 $ & $     4.05 $ & $    0.173 $ & $   0.772 $\\
$ 791.11863 $ & $    76.24 $ & $     7.97 $ & $    -4.56 $ & $     4.00 $ & $    0.174 $ & $   0.955 $\\
$ 805.96712 $ & $  -168.48 $ & $     7.63 $ & $    83.04 $ & $     6.79 $ & $    0.171 $ & $   0.229 $\\
$ 807.00978 $ & $   -76.74 $ & $     9.16 $ & $    74.87 $ & $     7.06 $ & $    0.170 $ & $   0.529 $\\
$ 807.99835 $ & $   125.68 $ & $     6.92 $ & $     0.18 $ & $     3.79 $ & $    0.172 $ & $   0.813 $\\
$ 809.98993 $ & $    13.78 $ & $     9.01 $ & $   -69.27 $ & $     5.86 $ & $    0.170 $ & $   0.387 $\\
$ 810.88079 $ & $   125.60 $ & $     8.34 $ & $   -31.31 $ & $     4.57 $ & $    0.172 $ & $   0.643 $\\
$ 838.95648 $ & $    41.95 $ & $     7.70 $ & $     5.62 $ & $     5.71 $ & $    0.172 $ & $   0.724 $\\
$ 865.01789 $ & $   -25.12 $ & $     8.25 $ & $    16.88 $ & $     4.31 $ & $    0.171 $ & $   0.224 $\\
$ 954.83936 $ & $   -67.35 $ & $     7.49 $ & $     7.29 $ & $     4.92 $ & $    0.175 $ & $   0.076 $\\
$ 955.85129 $ & $   -52.75 $ & $     8.95 $ & $    -4.96 $ & $     7.02 $ & $    0.173 $ & $   0.367 $\\
$ 1192.01643 $ & \nodata      & \nodata      & $    13.37 $ & $     2.83 $ & $    0.172 $ & $   0.339 $\\
$ 1192.02324 $ & $   -93.30 $ & $     7.19 $ & $    22.62 $ & $     3.46 $ & $    0.173 $ & $   0.341 $\\
$ 1193.09374 $ & $   -24.35 $ & $     7.52 $ & $    60.32 $ & $     5.56 $ & $    0.175 $ & $   0.649 $\\
$ 1193.93089 $ & $    -8.61 $ & $     8.57 $ & $    38.64 $ & $     5.33 $ & $    0.170 $ & $   0.890 $\\
$ 1251.91465 $ & $    13.52 $ & $     7.72 $ & $    16.56 $ & $     4.23 $ & $    0.172 $ & $   0.578 $\\
$ 1468.06245 $ & $    22.33 $ & $     7.86 $ & $    91.25 $ & $     6.82 $ & $    0.173 $ & $   0.789 $\\
$ 1470.09079 $ & $    18.65 $ & $     8.16 $ & $  -127.71 $ & $     8.71 $ & $    0.174 $ & $   0.372 $\\
$ 1545.09930 $ & $    26.89 $ & $     8.37 $ & $    -5.60 $ & $     5.38 $ & $    0.170 $ & $   0.961 $\\
$ 1545.93211 $ & $   -31.18 $ & $     7.91 $ & $     7.80 $ & $     4.89 $ & $    0.165 $ & $   0.201 $\\
\enddata
\tablenotetext{a}{
    Barycentric Julian dates throughout the paper are calculated from 
    Coordinated Universal Time (UTC).
}
\tablenotetext{b}{
    The zero-point of these velocities is arbitrary. An overall offset
    $\gamma_{\rm rel}$ fitted to these velocities in \refsecl{globmod}
    has {\em not} been subtracted.
}
\tablenotetext{c}{
    Internal errors excluding the component of astrophysical jitter
    considered in \refsecl{globmod}.
}
\tablenotetext{d}{
    Chromospheric activity index calibrated to the scale
    of \citet{vaughan:1978} following \citet{isaacson:2010}.
}
\ifthenelse{\boolean{rvtablelong}}{
    \tablecomments{
        For the iodine-free template exposures we do not
        measure the RV but do measure the BS and S index.  Such
        template exposures can be distinguished by the missing RV
        value.
    }
}{
    \tablecomments{
        For the iodine-free template exposures we do not
        measure the RV but do measure the BS and S index.  Such
        template exposures can be distinguished by the missing RV
        value.  This table is presented in its entirety in the
        electronic edition of the Astrophysical Journal.  A portion is
        shown here for guidance regarding its form and content.
    }
} 
\ifthenelse{\boolean{emulateapj}}{
    \end{deluxetable*}
}{
    \end{deluxetable}
}

\subsection{Photometric follow-up observations}
\label{sec:phot}

%
\setcounter{planetcounter}{1}
%
\begin{figure*}[!ht]
\plottwo{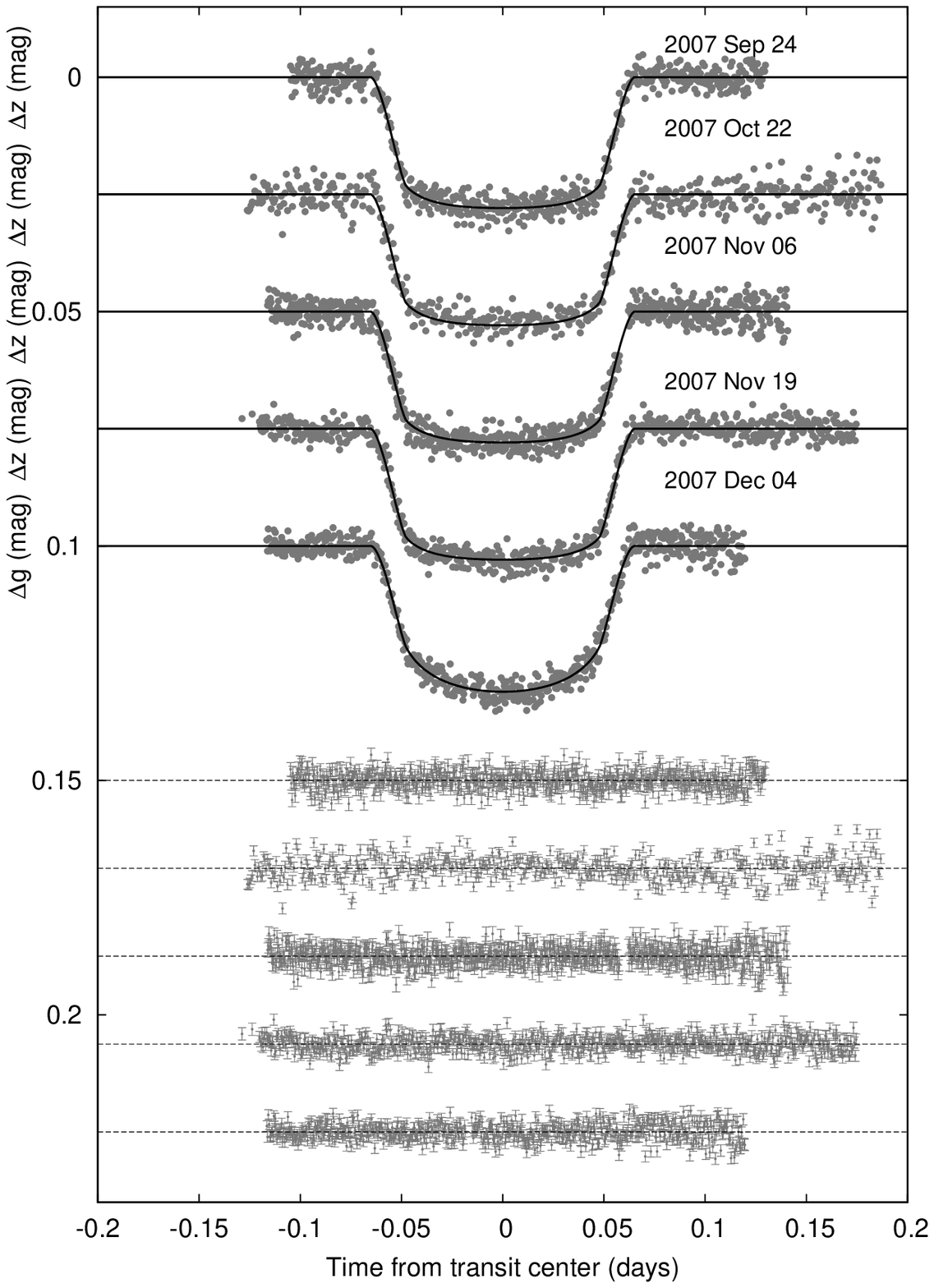}{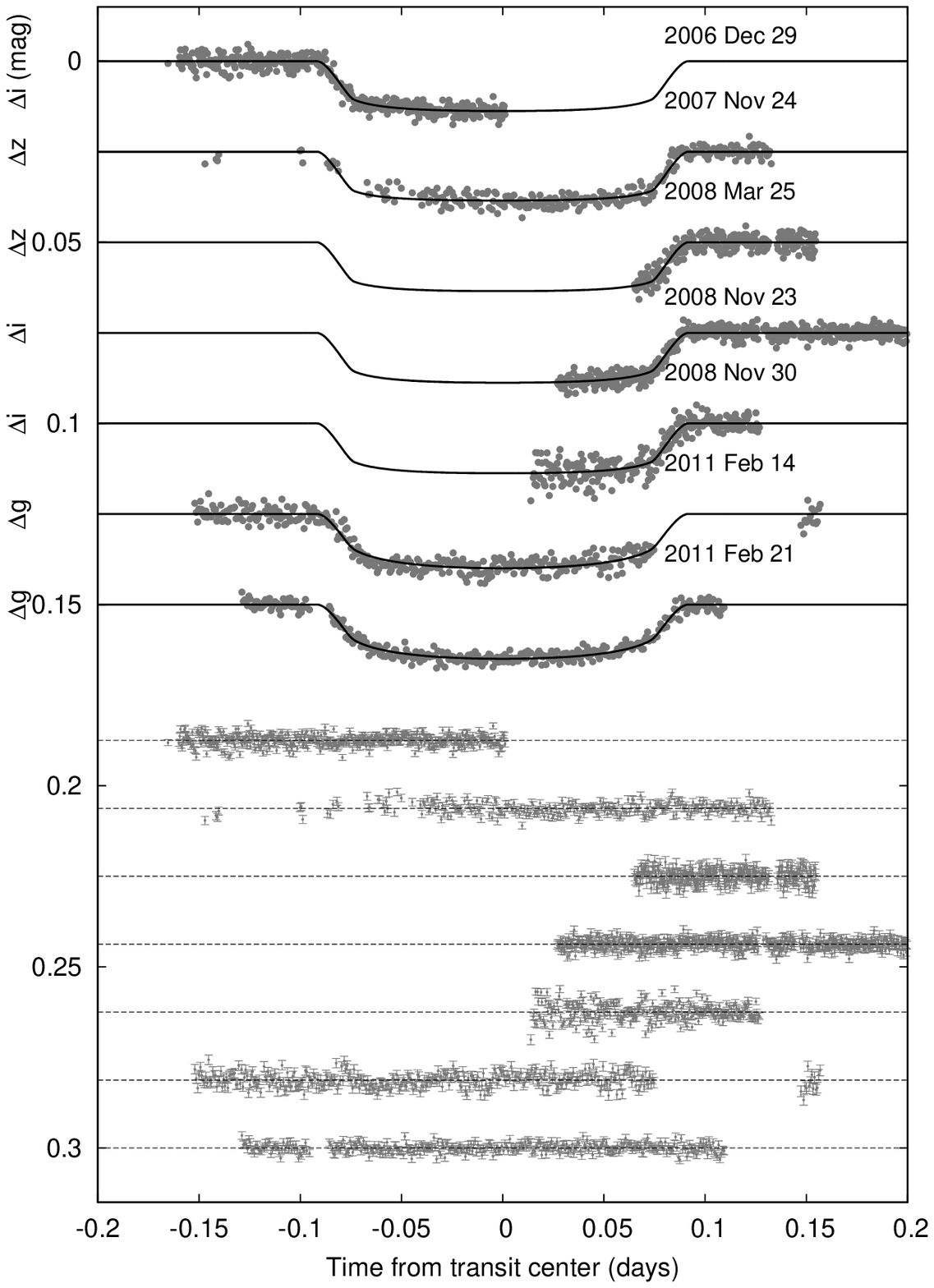}
\caption{
    Unbinned transit \lcs{} for \hatcur{32} (left) and \hatcur{33}
    (right), acquired with KeplerCam at the \flwof{} telescope.  The
    light curves have been EPD and TFA processed, as described in
    \cite{bakos:2010}.
    The dates of the events are indicated.  Curves after the first are
    displaced vertically for clarity.  Our best fits from the global
    modeling described in \refsecl{globmod} are shown by the solid
    lines.  Residuals from the fits are displayed at the bottom, in the
    same order as the top curves.  The error bars represent the photon
    and background shot noise, plus the readout noise.
}
\label{fig:lc32}
\end{figure*}

We conducted high-precision photometric observations of \hatcur{32}
and \hatcur{33} using the KeplerCam CCD camera on the \flwof{}
telescope.  The observations for each target are summarized in
\reftabl{photobs}. For \hatcur{32}, in addition to observations taken
during transit, we also obtained three sets of observations taken out
of transit at the predicted times of secondary eclipse (assuming a
circular orbit). These data are excluded from the global analysis
described in \refsecl{globmod}, but are used in \refsecl{blend} in
ruling out blend-scenarios.

The reduction of the KeplerCam images was performed as described by
\cite{bakos:2010}.  We performed External Parameter Decorrelation
(EPD) and used the Trend Filtering Algorithm \citep[TFA;][]{kovacs:2005} to
remove trends simultaneously with the light curve modeling (for more
details, see \citealp{bakos:2010}).  The final time series, together
with our best-fit transit \lc{} models, are shown in \reffigl{lc32};
the individual measurements are reported in \reftabl{phfu32} and
\reftabl{phfu33} for \hatcur{32} and \hatcur{33} respectively.

\begin{deluxetable}{lrrrr}
\tablewidth{0pc}
\tablecaption{
    High-precision differential photometry of
    \hatcur{32}\label{tab:phfu32}.
}
\tablehead{
    \colhead{BJD} & 
    \colhead{Mag\tablenotemark{a}} & 
    \colhead{\ensuremath{\sigma_{\rm Mag}}} &
    \colhead{Mag(orig)\tablenotemark{b}} & 
    \colhead{Filter} \\
    \colhead{\hbox{~~~~(2,400,000$+$)~~~~}} & 
    \colhead{} & 
    \colhead{} &
    \colhead{} & 
    \colhead{}
}
\startdata
$ 54368.74140 $ & $  -0.00204 $ & $   0.00143 $ & $  10.54500 $ & $ z$\\
$ 54368.74180 $ & $   0.00090 $ & $   0.00144 $ & $  10.54840 $ & $ z$\\
$ 54368.74217 $ & $   0.00417 $ & $   0.00143 $ & $  10.55120 $ & $ z$\\
$ 54368.74255 $ & $  -0.00158 $ & $   0.00143 $ & $  10.54590 $ & $ z$\\
$ 54368.74296 $ & $   0.00016 $ & $   0.00142 $ & $  10.54740 $ & $ z$\\
$ 54368.74335 $ & $   0.00079 $ & $   0.00143 $ & $  10.54810 $ & $ z$\\
$ 54368.74373 $ & $   0.00331 $ & $   0.00144 $ & $  10.55080 $ & $ z$\\
$ 54368.74411 $ & $   0.00071 $ & $   0.00142 $ & $  10.54810 $ & $ z$\\
$ 54368.74449 $ & $  -0.00132 $ & $   0.00142 $ & $  10.54610 $ & $ z$\\
$ 54368.74489 $ & $  -0.00129 $ & $   0.00141 $ & $  10.54610 $ & $ z$\\
\enddata
\tablenotetext{a}{
    The out-of-transit level has been subtracted. These magnitudes have
    been subjected to the EPD and TFA procedures, carried out
    simultaneously with the transit fit.
}
\tablenotetext{b}{
    Raw magnitude values without application of the EPD and TFA
    procedures.
}
\tablecomments{
    This table is available in a machine-readable form in the online
    journal.  A portion is shown here for guidance regarding its form
    and content.
}
\end{deluxetable}
%
\begin{deluxetable}{lrrrr}
\tablewidth{0pc}
\tablecaption{
    High-precision differential photometry of
    \hatcur{33}\label{tab:phfu33}.
}
\tablehead{
    \colhead{BJD} & 
    \colhead{Mag\tablenotemark{a}} & 
    \colhead{\ensuremath{\sigma_{\rm Mag}}} &
    \colhead{Mag(orig)\tablenotemark{b}} & 
    \colhead{Filter} \\
    \colhead{\hbox{~~~~(2,400,000$+$)~~~~}} & 
    \colhead{} & 
    \colhead{} &
    \colhead{} & 
    \colhead{}
}
\startdata
$ 54099.68885 $ & $   0.00058 $ & $   0.00085 $ & $   9.51222 $ & $ i$\\
$ 54099.69425 $ & $   0.00071 $ & $   0.00083 $ & $   9.51224 $ & $ i$\\
$ 54099.69465 $ & $  -0.00123 $ & $   0.00083 $ & $   9.51016 $ & $ i$\\
$ 54099.69502 $ & $  -0.00319 $ & $   0.00083 $ & $   9.50882 $ & $ i$\\
$ 54099.69541 $ & $   0.00122 $ & $   0.00083 $ & $   9.51279 $ & $ i$\\
$ 54099.69579 $ & $   0.00204 $ & $   0.00083 $ & $   9.51399 $ & $ i$\\
$ 54099.69619 $ & $   0.00045 $ & $   0.00083 $ & $   9.51219 $ & $ i$\\
$ 54099.69659 $ & $  -0.00038 $ & $   0.00083 $ & $   9.51120 $ & $ i$\\
$ 54099.69697 $ & $  -0.00305 $ & $   0.00083 $ & $   9.50858 $ & $ i$\\
$ 54099.69735 $ & $  -0.00342 $ & $   0.00083 $ & $   9.50826 $ & $ i$\\
\enddata
\tablenotetext{a}{
    The out-of-transit level has been subtracted. These magnitudes have
    been subjected to the EPD and TFA procedures, carried out
    simultaneously with the transit fit.
}
\tablenotetext{b}{
    Raw magnitude values without application of the EPD and TFA
    procedures.
}
\tablecomments{
    This table is available in a machine-readable form in the online
    journal.  A portion is shown here for guidance regarding its form
    and content.
}
\end{deluxetable}

\section{Analysis}
\label{sec:analysis}

\subsection{Properties of the parent stars}
\label{sec:stelparam}

Planetary parameters, such as the mass and radius, depend strongly on
the stellar mass and radius, which in turn are constrained by the
observed stellar spectra as well as the light curves and RV curves. We
followed an iterative procedure, described by \cite{bakos:2010}, to
determine the relevant stellar parameters. The procedure involves
iterating between infering stellar atmospheric parameters (including
the effective temperature $\teffstar$, surface gravity $\loggstar$,
metallicity $\feh$, and projected rotation velocity $\vsini$) from the
Keck/HIRES template spectrum using the Spectroscopy Made Easy package
\citep[SME;][]{valenti:1996} and the \cite{valenti:2005} atomic line
database, and modeling the light curves and RV curves (see
\refsecl{globmod}) to determine the stellar density $\rhostar$. At a
given cycle in the iteration we combine our estimates of $\teffstar$,
$\feh$ and $\rhostar$ with the \hatcurisofull{32}
\citep[\hatcurisoshort{32};][]{\hatcurisocite{32}} series of stellar
evolution models to determine the stellar mass, radius, and surface
gravity among other parameters. If the resulting surface gravity
differs significantly from the value determined from the spectrum with
SME, we repeat the analysis fixing the surface gravity in SME to the
new value.

For each star, the {\em initial} SME analysis, in which the surface
gravity was allowed to vary, yielded the following values and
uncertainties:
\begin{itemize}
\item {\em \hatcur{32}}:
$\teffstar=\hatcurSMEiteff{32}$\,K, 
$\feh=\hatcurSMEizfeh{32}$\,dex,
$\loggstar=\hatcurSMEilogg{32}$\,(cgs), and
$\vsini=\hatcurSMEivsin{32}$\,\kms.
\item {\em \hatcur{33}}:
$\teffstar=\hatcurSMEiteff{33}$\,K, 
$\feh=\hatcurSMEizfeh{33}$\,dex,
$\loggstar=\hatcurSMEilogg{33}$\,(cgs), and
$\vsini=\hatcurSMEivsin{33}$\,\kms.
\end{itemize}

As described in \refsecl{globmod} the inferred stellar densities, and
hence radii, for \hatcur{32} and \hatcur{33} depend strongly on the
orbital eccentricities, which due to the high RV jitters are poorly
constrained. For each system we conducted separate analyses, first
assuming a circular orbit and then allowing the eccentricity to
vary. In each case we obtained a new value of \loggstar\ which we held
fixed during a second SME iteration.
The final stellar parameters for each star, assuming both
circular and eccentric models, are listed in \reftabl{stellar32}.

\begin{figure*}[!ht]
\plottwo{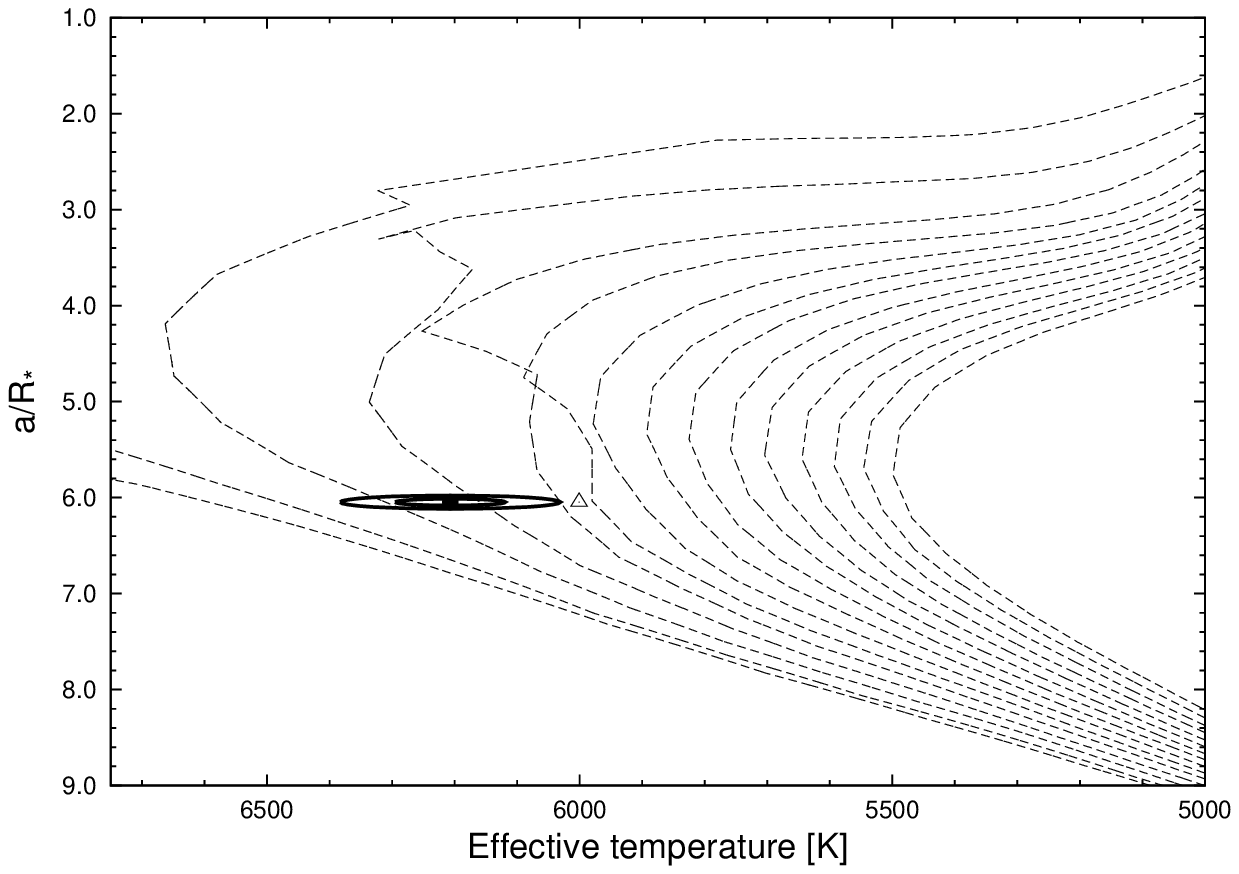}{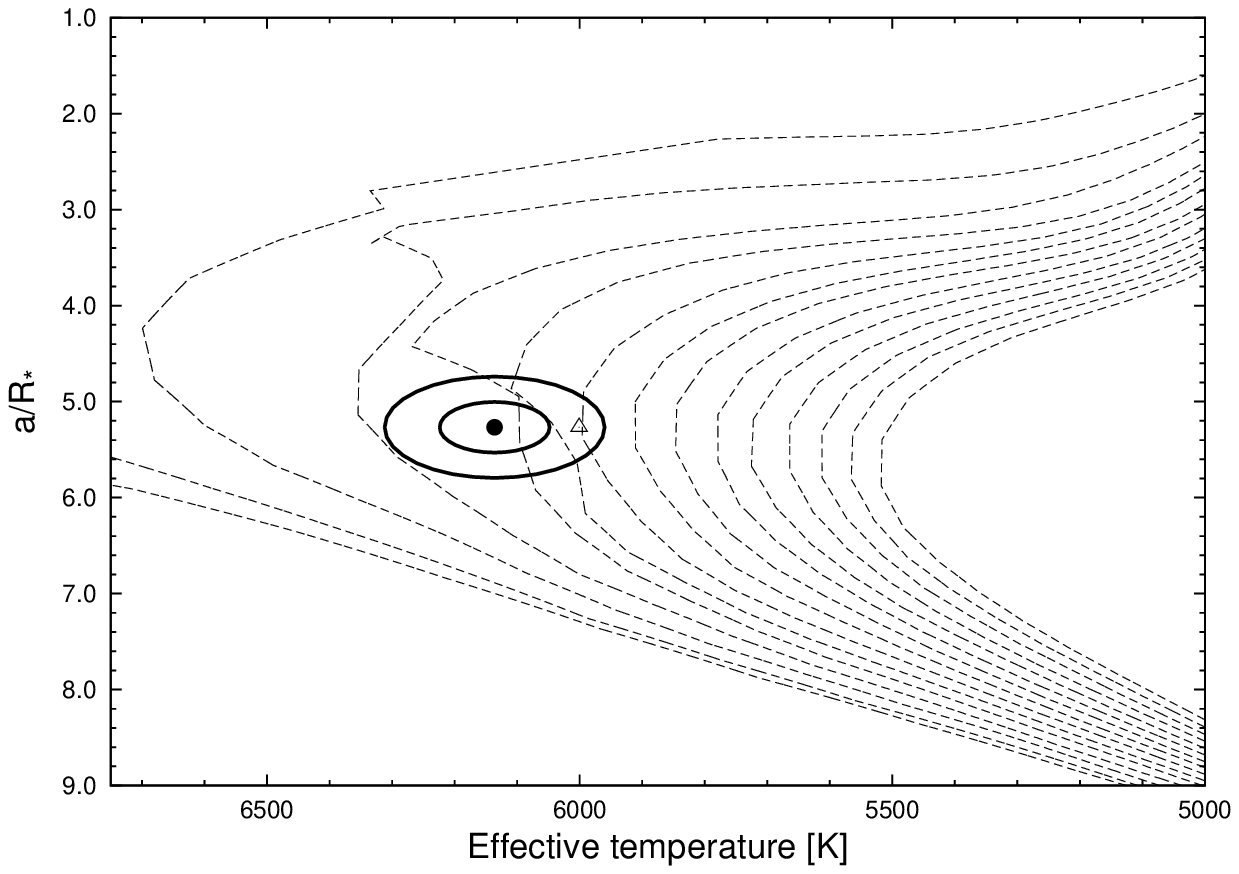}
\caption{
    Model isochrones from \cite{\hatcurisocite{32}} assuming a
    circular orbit (left) and the best-fit eccentric orbit (right) for
    the measured metallicity of \hatcur{32}, and ages of 0.5\,Gyr and
    1--14\,Gyr in 1\,Gyr steps (left to right in each plot).  The
    adopted values of $\teffstar$ and \arstar\ are shown together with
    their 1$\sigma$ and 2$\sigma$ confidence ellipsoids.  The initial
    values of \teffstar\ and \arstar\ from the first SME and
    \lc\ analyses are represented with a triangle. 
}
\label{fig:iso32}
\end{figure*}

The inferred location of each star for both the circular and eccentric
models, in diagrams of \arstar\ (which is related to \rhostar) versus
\teffstar, analogous to the classical H-R diagram, is shown in
\reffigls{iso32}{iso33}.
In each case the stellar properties and their 1$\sigma$ and 2$\sigma$
confidence ellipsoids are displayed against the backdrop of model
isochrones for a range of ages, and the appropriate stellar
metallicity. 

\setcounter{planetcounter}{1}
%
\setcounter{planetcounter}{2}
%
\begin{figure*}[!ht]
\plottwo{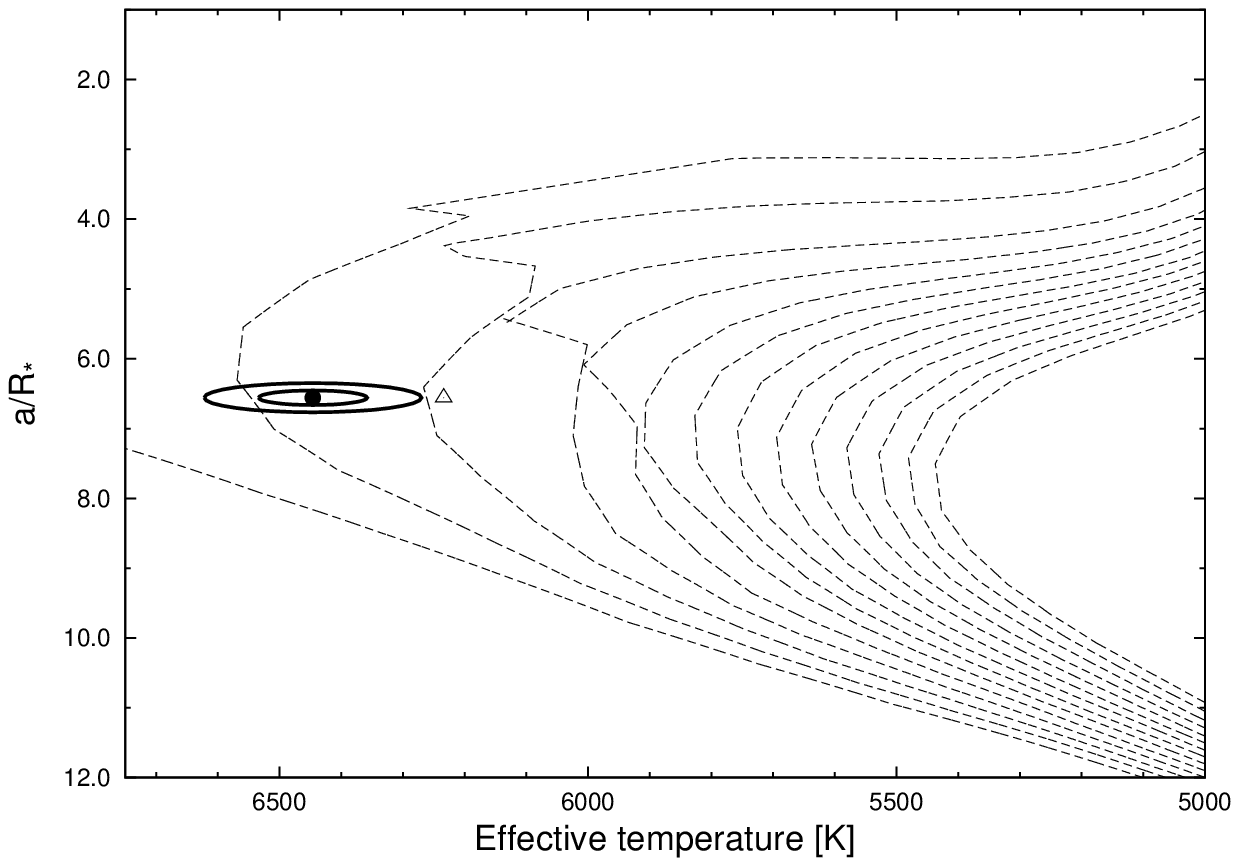}{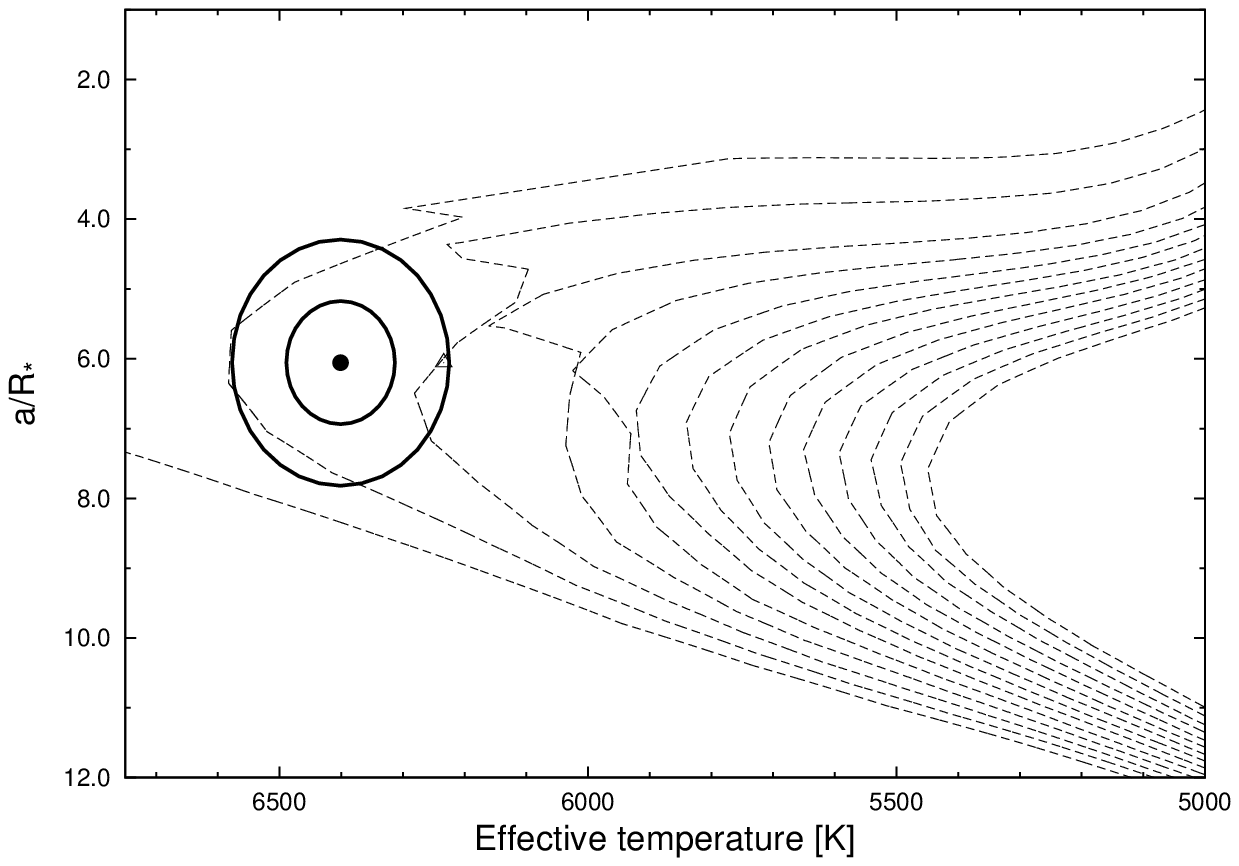}
\caption{
    Same as \reffigl{iso32}, here we show the results for
    \hatcur{33}. In this case the 0.5\,Gyr isochrone is not included.
}
\label{fig:iso33}
\end{figure*}

As a check on our stellar parameter determinations we compare the
measured photometric colors of the two stars to the predicted values
based on the models. \hatcur{32} has $J-K =
\hatcurCCesoJKmag{32}$\footnote{Taken from 2MASS
  \citep{skrutskie:2006} and converted to the ESO photometric system
  using the transformations by \citet{carpenter:2001}.} and $V-I_{C} =
\hatcurCCtassvi{32}$ \citep[from TASS;][]{droege:2006}, whereas the
predicted values are $J-K = \hatcurISOJK{32}$ and $V-I_{C} =
\hatcurISOvi{32}$ for the circular model, and $J-K = \hatcurISOJK{36}$
and $V-I_{C} = \hatcurISOvi{36}$ for the eccentric model. In both
cases the model $V-I_{C}$ colors are consistent with the measurements,
but the model $J-K$ colors are slightly ($\sim 2\sigma$) redder than
the measurements. \hatcur{33} has $J-K = \hatcurCCesoJKmag{33}$ and
$V-I_{C} = \hatcurCCtassvi{33}$ (also from 2MASS and TASS), and has
predicted values of $J-K = \hatcurISOJK{33}$ and $V-I_{C} =
\hatcurISOvi{33}$ for the circular model, and $J-K = \hatcurISOJK{35}$
and $V-I_{C} = \hatcurISOvi{35}$ for the eccentric model. The measured
photometric colors of \hatcur{33} are within 1$\sigma$ of the
predicted values for both models.

Neither \hatcur{32} nor \hatcur{33} shows significant chromospheric
emission in their Ca~II~H and K line cores. Following the procedure of
\cite{isaacson:2010}, we find median $\log R^{\prime}_{\rm HK}$
\citep{noyes:1984} values of $-4.62$ and $-4.88$ for \hatcur{32} and
\hatcur{33}, respectively.

%
%
\ifthenelse{\boolean{emulateapj}}{
    \begin{deluxetable*}{lccccl}
}{
  \begin{deluxetable}{lccccl}
}
\tablewidth{0pc}
\tabletypesize{\scriptsize}
\tablecaption{
    Stellar parameters for \hatcur{32} and \hatcur{33}
    \label{tab:stellar32}
}
\tablehead{
    &
    \multicolumn{2}{c}{\bf HAT-P-32} &
    \multicolumn{2}{c}{\bf HAT-P-33} &
    \\
    \multicolumn{1}{c}{~~~~~~~~Parameter~~~~~~~~} &
    \multicolumn{1}{c}{Value\tablenotemark{a}}                     &
    \multicolumn{1}{c}{Value}                     &
    \multicolumn{1}{c}{Value\tablenotemark{a}}                     &
    \multicolumn{1}{c}{Value}                     &
    \multicolumn{1}{c}{Source} \\
    &
    \multicolumn{1}{c}{Circular} &
    \multicolumn{1}{c}{Eccentric} &
    \multicolumn{1}{c}{Circular} &
    \multicolumn{1}{c}{Eccentric} &
}
\startdata
\noalign{\vskip -3pt}
\sidehead{Spectroscopic properties}
~~~~$\teffstar$ (K)\dotfill         &  \hatcurSMEteff{32} & \hatcurSMEiteff{36}   &  \hatcurSMEteff{33} & \hatcurSMEiiteff{35}   & SME\tablenotemark{b}\\
~~~~$\feh$\dotfill                  &  \hatcurSMEzfeh{32} & \hatcurSMEizfeh{36} &  \hatcurSMEzfeh{33} & \hatcurSMEiizfeh{35}  & SME                 \\
~~~~$\vsini$ (\kms)\dotfill         &  \hatcurSMEvsin{32} & \hatcurSMEivsin{36}  &  \hatcurSMEvsin{33} & \hatcurSMEiivsin{35}  & SME                 \\
~~~~$\vmac$ (\kms)\dotfill          &  \hatcurSMEvmac{32} & \hatcurSMEivmac{36}  &  \hatcurSMEvmac{33} & \hatcurSMEiivmac{35}  & SME                 \\
~~~~$\vmic$ (\kms)\dotfill          &  \hatcurSMEvmic{32} & \hatcurSMEivmic{36}  &  \hatcurSMEvmic{33} & \hatcurSMEiivmic{35}  & SME                 \\
~~~~$\gamma_{\rm RV}$ (\kms)\dotfill&  \hatcurDSgamma{32} & $\cdots$ & \hatcurDSgamma{33}  & $\cdots$ & DS\tablenotemark{c}                  \\
\sidehead{Photometric properties}
~~~~$V$ (mag)\dotfill               &  \hatcurCCtassmv{32}  &  $\cdots$ & \hatcurCCtassmv{33}  & $\cdots$ & TASS                \\
~~~~$\vic$ (mag)\dotfill            &  \hatcurCCtassvi{32}  &  $\cdots$ & \hatcurCCtassvi{33}  & $\cdots$ & TASS                \\
~~~~$J$ (mag)\dotfill               &  \hatcurCCtwomassJmag{32} &  $\cdots$ & \hatcurCCtwomassJmag{33} & $\cdots$ & 2MASS           \\
~~~~$H$ (mag)\dotfill               &  \hatcurCCtwomassHmag{32} &  $\cdots$ & \hatcurCCtwomassHmag{33} & $\cdots$ & 2MASS           \\
~~~~$K_s$ (mag)\dotfill             &  \hatcurCCtwomassKmag{32} &  $\cdots$ & \hatcurCCtwomassKmag{33} & $\cdots$ & 2MASS           \\
\sidehead{Derived properties}
~~~~$\mstar$ ($\msun$)\dotfill      &  \hatcurISOmlong{32} &  \hatcurISOmlong{36}   &  \hatcurISOmlong{33} & \hatcurISOmlong{35}   & \hatcurisoshort{33}+\hatcurlumind{33}+SME \tablenotemark{d}\\
~~~~$\rstar$ ($\rsun$)\dotfill      &  \hatcurISOrlong{32} & \hatcurISOrlong{36}  &  \hatcurISOrlong{33} & \hatcurISOrlong{35}  & \hatcurisoshort{33}+\hatcurlumind{33}+SME         \\
~~~~$\loggstar$ (cgs)\dotfill       &  \hatcurISOlogg{32} & \hatcurISOlogg{36}   &  \hatcurISOlogg{33} & \hatcurISOlogg{35}   & \hatcurisoshort{33}+\hatcurlumind{33}+SME         \\
~~~~$\lstar$ ($\lsun$)\dotfill      &  \hatcurISOlum{32} & \hatcurISOlum{36}    &  \hatcurISOlum{33} & \hatcurISOlum{35}    & \hatcurisoshort{33}+\hatcurlumind{33}+SME         \\
~~~~$M_V$ (mag)\dotfill             &  \hatcurISOmv{32} & \hatcurISOmv{36}     &  \hatcurISOmv{33} & \hatcurISOmv{35}     & \hatcurisoshort{33}+\hatcurlumind{33}+SME         \\
~~~~$M_K$ (mag,\hatcurjhkfilset{33})\dotfill &  \hatcurISOMK{32} & \hatcurISOMK{36} &  \hatcurISOMK{33} & \hatcurISOMK{35} & \hatcurisoshort{33}+\hatcurlumind{33}+SME         \\
~~~~Age (Gyr)\dotfill               &  \hatcurISOage{32} & \hatcurISOage{36}    &  \hatcurISOage{33} & \hatcurISOage{35}    & \hatcurisoshort{33}+\hatcurlumind{33}+SME         \\
~~~~Distance (pc)\dotfill           &  \hatcurXdist{32}\phn & \hatcurXdist{36}\phn &  \hatcurXdist{33}\phn & \hatcurXdist{35}\phn & \hatcurisoshort{33}+\hatcurlumind{33}+SME\\ [-1.5ex]
\enddata
\tablenotetext{a}{
    The eccentricities of both
    \hatcur{32} and \hatcur{33} are poorly constrained--both planets
    are consistent with circular orbits but may also have significant
    eccentricities. We list separately the parameters obtained when a
    circular orbit is fixed, and when the eccentricity is allowed to
    vary.
}
\tablenotetext{b}{
    SME = ``Spectroscopy Made Easy'' package for the analysis of
    high-resolution spectra \citep{valenti:1996}.  These parameters
    rely primarily on SME, but have a small dependence also on the
    iterative analysis incorporating the isochrone search and global
    modeling of the data, as described in the text.
}
\tablenotetext{c}{
    The mean heliocentric velocity as measured by the CfA Digital
    Speedometer. The uncertainty is the rms of the measured velocities
    of the star rather than the uncertainty on the mean.
}
\tablenotetext{d}{
    \hatcurisoshort{33}+\hatcurlumind{33}+SME = Based on the \hatcurisoshort{33}
    isochrones \citep{\hatcurisocite{33}}, \hatcurlumind{33} as a luminosity
    indicator, and the SME results.
}
\ifthenelse{\boolean{emulateapj}}{
    \end{deluxetable*}
}{
    \end{deluxetable}
}

\subsection{Stellar Jitter}
\label{sec:jitter}

Both \hatcur{32} and \hatcur{33} exhibit notably high scatter in their
velocity residuals. Stellar jitter values of
\hatcurRVjitter{32}\,\ms\ and $55.1$\,\ms\ for \hatcur{32} and
\hatcur{33} respectively must be added in quadrature to their velocity
errors to achieve reduced $\chi^{2}$ values of unity for the best-fit
circular orbits (see \refsecl{globmod}).

\begin{figure*}[!ht]
\plotone{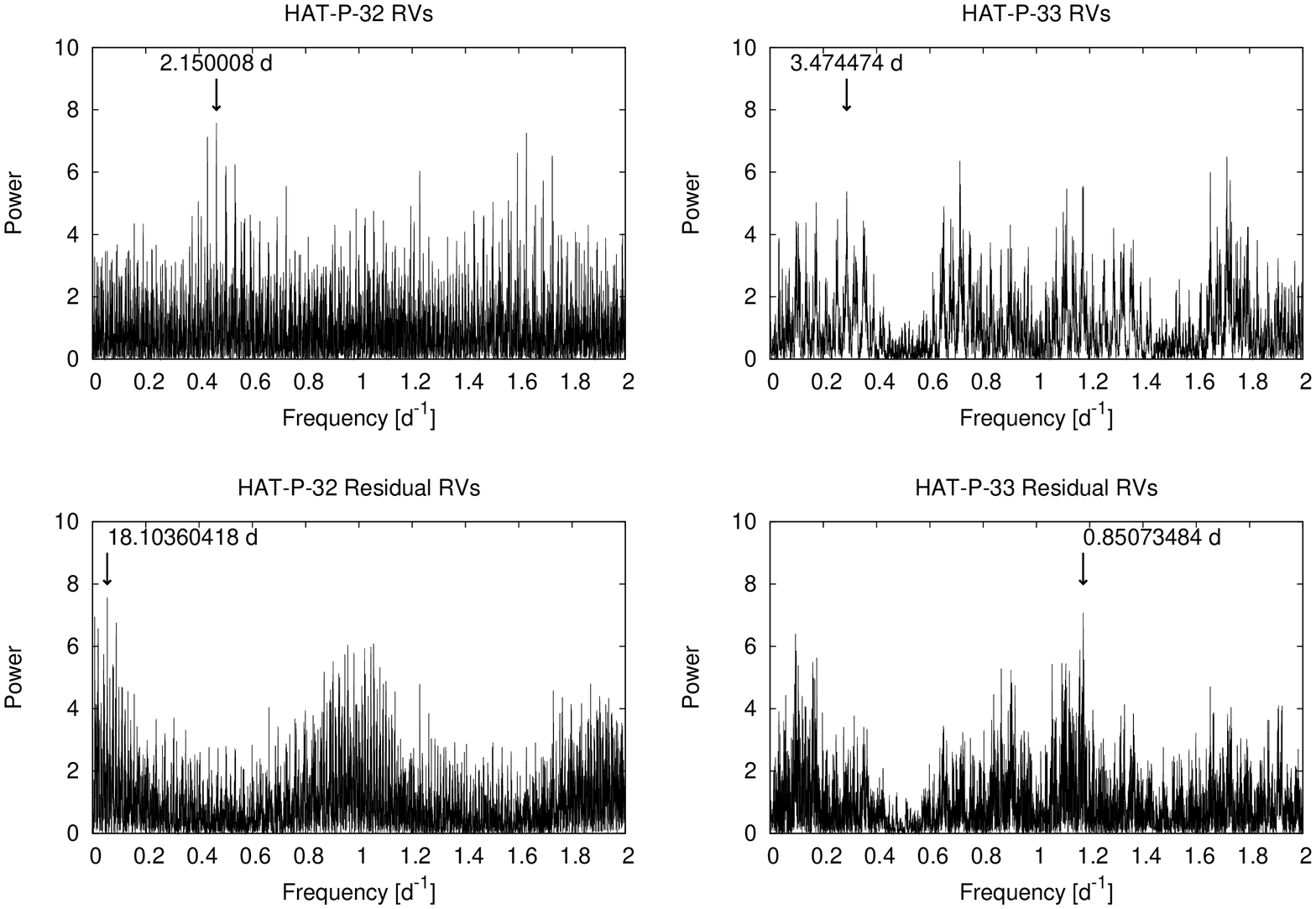}
\caption[]{
    Lomb-Scargle frequency spectra for the RVs (top) and residual RVs
    from the best-fit circular orbit models (bottom) for \hatcur{32}
    (left) and \hatcur{33} (right). In the top panels we mark the
    transit frequencies, while in the bottom panels we mark the
    highest peaks in the spectra. In all cases the false alarm
    probability of finding a peak as high the marked peak in a
    Gaussian white-noise RV curve is greater than 8\%.
\label{fig:rvls}}
\end{figure*}

A possible cause of the jitter is the presence of one or more
additional planets in either system, though this would not explain the
high scatter seen in the BS measurements. \reffigl{rvls} shows the
Lomb-Scargle frequency spectra
\citep[L-S;][]{lomb:1976,scargle:1982,press:1989} of the RV
observations for both stars. In both cases there is a peak in the
spectrum at the transit frequency; in the case of \hatcur{32} this is
the highest peak in the spectrum, while in the case of \hatcur{33} the
highest peak is at a high frequency alias of the transit frequency (if
we restrict the search to frequencies shorter than $0.49$\,d$^{-1}$
the transit frequency is the highest peak). We note that in neither
case would the planet be detectable from the RV data alone. For
\hatcur{32}, the false alarm probability of detecting a peak in the
L-S periodogram with a height greater than or equal to the measured
peak height is $\sim 8\%$ (this is determined by applying L-S to
simulated Gaussian white noise RV curves with the same time sampling
as the observations). For \hatcur{33} the false alarm probability of
the transit peak is $79$\% (the false alarm probability for the
highest peak in the spectrum is 30\%). We also show the L-S spectra of
the RV residuals from the best-fit circular orbit models, in order to
see if there is evidence for additional short period planets in the
systems. The highest peaks in these spectra are at periods of
$18.104$\,d and $0.8507$\,d for \hatcur{32} and \hatcur{33}
respectively (for each system there is some ambiguity in the peak that
depends on the frequency range searched and whether or not a
multiharmonic period search is used). If we assume that each system
has an additional planet at the above stated periods, the RMS of the
RV residuals decreases to $\sim 40$\,\ms\ and $\sim 30$\,\ms\ for
\hatcur{32} and \hatcur{33} respecitvely. However, these peaks have
corresponding false alarm probabilities of $8$\% and $8.9$\%, so they
are not statistically significant detections. We conclude that while
either system may host additional short period planets, there are not
at present enough RV observations to make a statistically significant
detection.

%

\begin{figure*}[!ht]
\plotone{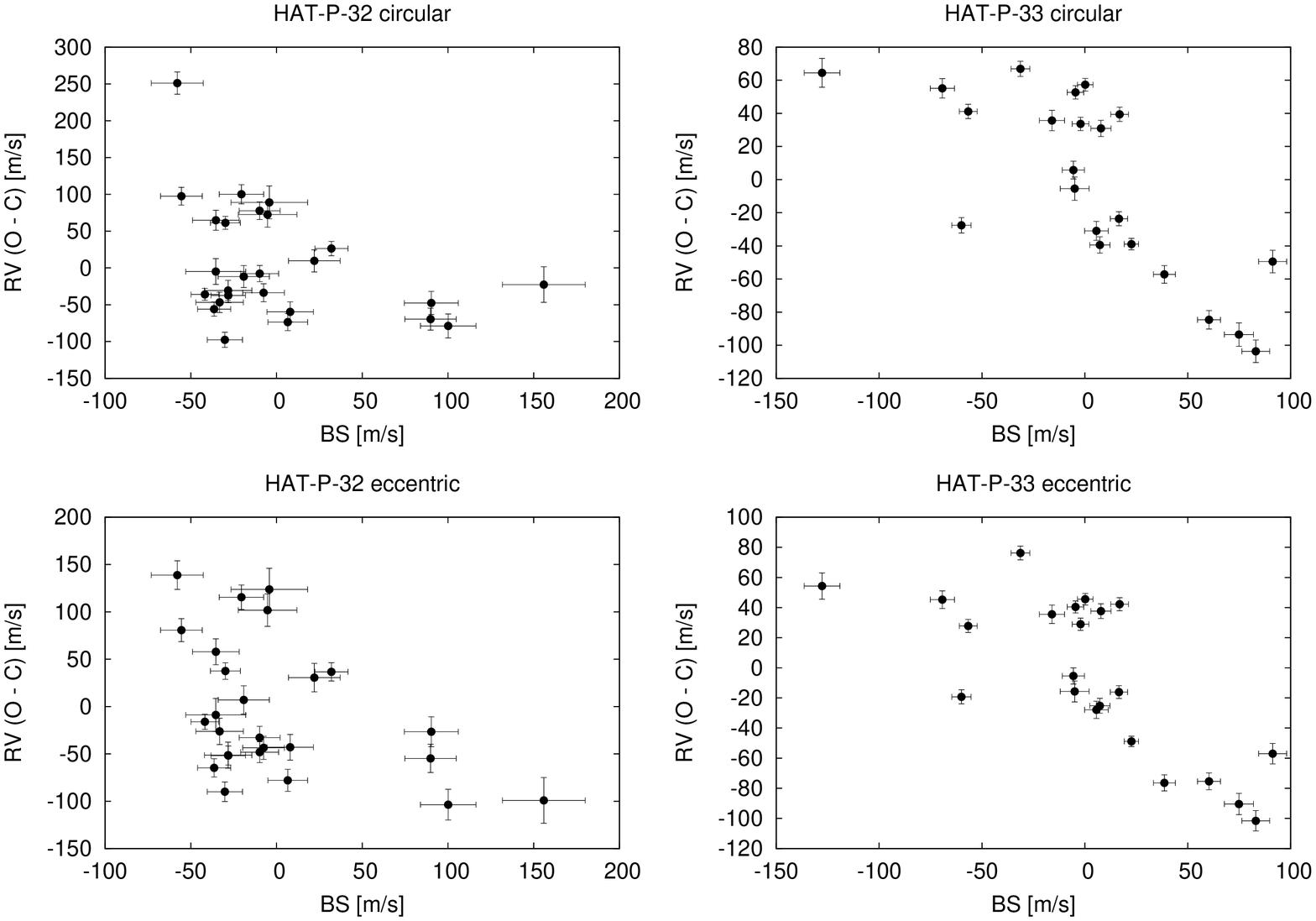}
\caption[]{
    RV residuals from the best-fit circular (top) and eccentric
    (bottom) model orbits vs. BS for \hatcur{32} (left) and
    \hatcur{33} (right). There is a hint of an anti-correlation
    between these quantities for \hatcur{32}, though it is not
    statistically significant. For \hatcur{33} the quantities are
    clearly anti-correlated. See the discussion in \refsecl{jitter}.
\label{fig:rvresidvsbs}}
\end{figure*}
%

Another potential source of RV and BS jitter is varying scattered
moonlight contaminating the spectra. Following \cite{kovacs:2010} we
investigated the possibility that the BS and RV may be affected by
varying sky contamination, and found no evidence that this is the
case. Alternatively, RV and BS jitter might be caused by variable
contamination from a nearby star. Based on our KeplerCam observations
of \hatcur{32} and \hatcur{33}, we can rule out contaminating
neighbors with $\Delta i \la 5$\,mag at a separation greater than
$3\arcsec$, or neighbors with $\Delta i \la 2$\,mag at a separation
greater than $1\arcsec$ of either star. We cannot rule out fainter
close neighbors which could be responsible for at least some of the
jitter. Constraints on potential neighbors based on modeling the
photometric light curves are further considered in \refsecl{blend}.

\reffigl{rvresidvsbs} compares the RV residuals to the BSs for
\hatcur{32} and \hatcur{33}. For \hatcur{33} there is a clear
anti-correlation between these quantities (the Spearman rank-order
correlation test \citep[e.g.][]{press:1992} yields a correlation
coefficient of $r_{s} = -0.77$ with a false-alarm probability of
$0.04\%$ assuming a circular orbit, while for the best-fit eccentric
orbit $r_{s} = -0.72$ with a $0.1\%$ false alarm probability). For
\hatcur{32} there is a hint of an anti-correlation, but it is not
statistically significant ($r_{s} = -0.30$ with a false-alarm
probability of $14\%$ assuming a circular orbit, and $r_{s} = -0.39$
with $4.9\%$ false alarm probability for the best-fit eccentric
orbit). The anti-correlation for \hatcur{33} is an indication that the
measured jitter for this star is dominated by intrinsic stellar
variations, rather than being due to another planet in the system, or
to the instrumental errors being underestimated. We make use of this
correlation in \refsecl{globmod} to reduce the scatter in the RV
residuals for \hatcur{33} and improve the precision of the fitted
parameters.

\cite{saar:1998} argue that the RV jitter of F stars is due primarily
to convective inhomogeneities (due for example to regions of strong
magnetic fields which locally suppress convection) which vary in time,
rather than temperature inhomogeneities (due for example to cool
starspots) which vary in time. The latter may be more important for G
and K stars. Both sources of jitter would result in significant BS
variations that may be correlated with the RV variations, but the lack
of detectable out-of-transit photometric variations in the HATNet
light curves of \hatcur{32} and \hatcur{33}, and the fact the neither
star shows particularly high chromospheric emission in their Ca II H
and K line cores, is consistent with the expectation that temporally
changing convective inhomogeneities, rather than temporally changing
temperature inhomogeneities, are responsible for the jitter of these
two stars.

Jitter values as high as those found for \hatcur{32} and \hatcur{33}
are typical of stars with similar spectral types and rotation
velocities. From the $\vsini$--jitter correlation measured by
\cite{saar:2003}, the expected jitter for an F dwarf with $\vsini =
20$\,\kms\ is $\sim 50$\,\ms, while for an F dwarf with $\vsini =
14$\,\kms\ the expected jitter is $\sim 30$\,\ms. The sample used by
\cite{saar:2003} to determine this correlation includes only a handful
of stars with $\vsini > 10$\,\kms, and the scatter about the relation
is fairly significant---one star with $\vsini \sim 15$\,\kms\ was
found to have a jitter in excess of $100$\,\ms. The planet hosting
star HAT-P-2, which has a similar temperature and rotation velocity to
\hatcur{32} and \hatcur{33} ($\teffstar = 6290 \pm 60$\,K, $\vsini =
20.8 \pm 0.3$\,\kms; \citealp{pal:2010}) has been reported to have a
high jitter of $\sim 60$\,\ms\ based on data from Keck and Lick
\citep{bakos:2007}. A subsequent analysis by \cite{winn:2007}, using
only Keck/HIRES data, found a somewhat lower jitter of $\sim 30$\,\ms,
though this is still quite a bit higher than most planet-hosting stars
discovered to date (e.g. the median jitter of the previously published
HATNet planets is $\sim 7$\,\ms). The primary difference between
HAT-P-2 and the two systems presented here is that the planet HAT-P-2b
is significantly more massive than either \hatcurb{32} or \hatcurb{33}
(HAT-P-2b has $\mpl = 9.09 \pm 0.24$\,\mjup; \citealp{pal:2010}); as a
result, the RV semiamplitude of HAT-P-2b ($K = 984 \pm 17$\,\ms)
greatly exceeds the jitter, making this planet more straightforward to
confirm than either \hatcurb{32} or \hatcurb{33}.

\subsection{Excluding blend scenarios}
\label{sec:blend}

Both \hatcur{32} and \hatcur{33} exhibit significant spectral line
bisector span (BS) variations (\reffigls{rvbis32}{rvbis33}; the RMS of
the BS is $53\,\ms$ and $51\,\ms$ for \hatcur{32} and \hatcur{33}
respectively). In neither case are the variations in phase with the
transit ephemeris, as one might expect if the observed RV variation
were due to a blend between an eclipsing binary system and another
star. As discussed in \refsecl{jitter}, these variations are likely
due to temporally changing convective inhomogeneities, perhaps created
by variable photospheric magnetic faculae as in the Sun, which we
suspect are responsible for the significant RV jitter seen in both of
these stars. Nonetheless the large BS variations, which are comparable
to the semiamplitudes of the RV signals, prevent us from using the BSs
to rule out the possibility that either of these systems is a blend.

To rule out blend scenarios we made use of the {\sc blendanal} program
(\citealp{hartman:2011b}; see also, \citealp{hartman:2011a}) which
models the observed light curves, stellar atmospheric
parameters, and calibrated photometric magnitudes using various blended
eclipsing binary scenarios as well as scenarios involving a transiting
planet system potentially blended with light from another star. The
program relies on a combination of the Padova \citep{girardi:2000} and
\cite{baraffe:1998} stellar evolution models, the Eclipsing Binary
Orbit Program \citep[EBOP]{popper:1981,etzel:1981,nelson:1972} as
modified by \citet{southworth:2004a,southworth:2004b}, and stellar
limb darkening parameters from \citet{claret:2004}. It is similar to
the {\sc blender} program \citep{torres:2005} which has been used to
confirm {\em Kepler} planets \citep[e.g.][]{torres:2011}, but with a
number of technical differences which are described by
\cite{hartman:2011b}.

\begin{figure}[!ht]
\plotone{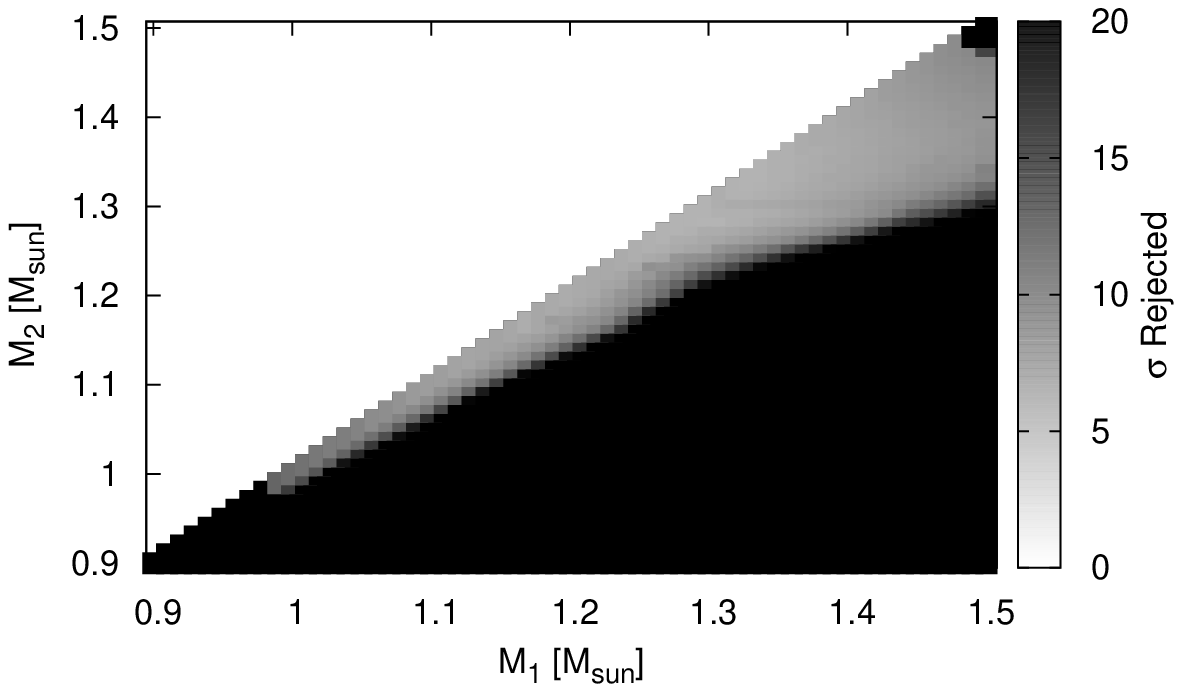}
\caption[]{
    The $\sigma$-level at which the class 3 blend model (hierarchical
    triple stellar system) can be rejected for \hatcur{33} as a
    function of the masses of the two largest stars in the system:
    $M_{1}$ is the mass of the uneclipsed star, and $M_{2}$ is the
    mass of the primary star in the eclipsing system. Note that the
    lack of several \kms\ RV variations leads to the constraint $M_{1}
    > M_{2}$. The best-fit model is rejected with $\sim 6\sigma$
    confidence.
\label{fig:hsssigma}}
\end{figure}

For each object we fit 4 classes of models:
\begin{enumerate}
\item A single star with a transiting planet.
\item A planet transiting one component of a binary star system.
\item A hierarchical triple stellar system.
\item A blend between a bright stationary star, and a fainter,
  physically unrelated eclipsing binary.
\end{enumerate}
Initially we assume that the eclipsing components have a circular
orbit. For both \hatcur{32} and \hatcur{33} we find that the class 1
model (a single star with a transiting planet), or the class 2 model
with a planet-host star that is much brighter than its binary star
companion, provide better fits (lower $\chi^2$) than the class 3 and
class 4 models. To evaluate the statistical significance with which
the class 3 and class 4 models may be rejected, we follow the Monte
Carlo procedure described in \cite{hartman:2011a}. We find that for
\hatcur{32} we may reject both the class 3 and 4 models with $\sim
13\sigma$ confidence (the best-fit class 4 model consists of a group
of stars with similar parameters to the best-fit class 3 model), while
for \hatcur{33} we may reject the class 3 and 4 models with $\sim
6\sigma$ confidence (e.g. \reffigl{hsssigma}).

\begin{figure*}[!ht]
\plotone{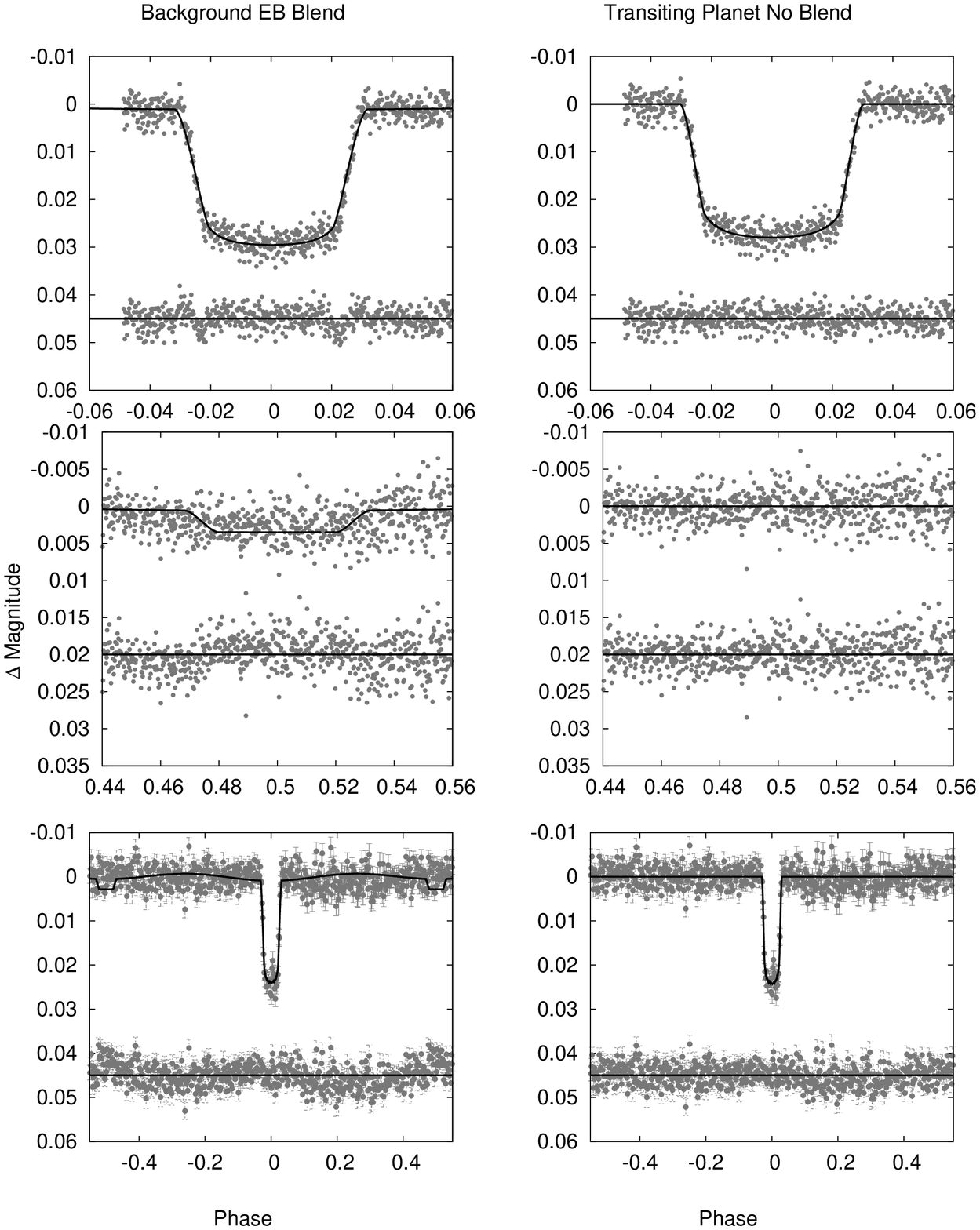}
\caption[]{
    Comparison of the best-fit class 4 (background eclipsing binary;
    left) and class 1 (single star with a transiting planet; right)
    blend-models for \hatcur{32}. The results are shown for three
    illustrative light curves: a KeplerCam \band{z} primary transit
    from 2007 September 24 (top), a KeplerCam \band{z} out-of-transit
    light curve from 2007 December 3 (middle), and the \band{I_{C}}
    HATNet field G125 light curve (bottom). The HATNet light curve is
    folded and binned in phase, using a binsize of 0.002 (note that
    this is only for display purposes, we do not bin the data in the
    modeling). In each panel the light curve is shown at top together
    with the model, and the residual is shown below. Note that for the
    KeplerCam observations we plot the EPD/TFA-corrected light curve,
    this correction is determined simultaneously with the fit, as a
    result there are slight differences in the plotted KeplerCam light
    curves for the two classes of blend models. The class 4 blend
    model provides a notably poorer fit to the light curves than the
    class 1 model--this includes slight differences in the
    ingress/egress of the primary transit, and a secondary transit and
    out of transit variation which are not seen in the KeplerCam or
    HATNet data.
\label{fig:exampleblendlc}}
\end{figure*}

For both \hatcur{32} and \hatcur{33} the eclipsing binary star blend
scenarios (classes 3 and 4) are excluded in part due to the lack of an
apparent secondary eclipse or out of transit variation that are
predicted by models capable of fitting the observed primary
transits. For \hatcur{33} the HATNet light curve provides these
constraints, while for \hatcur{32} three sets of KeplerCam
observations collected during predicted secondary eclipses (assuming a
circular orbit) augment the constraints provided by the HATNet light
curve. \reffigl{exampleblendlc} shows a few example model light curves
for the best-fit class 4 model for \hatcur{32} which illustrate this.

Because the time of secondary eclipse depends on the eccentricity,
which is poorly constrained by the RV observations,
we repeat the blend analysis for both systems fixing the
eccentricities to the best-fit values as determined in
\refsecl{globmod}. While for \hatcur{32} this scenario causes the
secondary eclipses to not occur during the out of transit KeplerCam
observations, the eccentric eclipsing binary results in stronger
out of transit variations which are ruled out by the
HATNet data. In this case for \hatcur{32} we may reject the class 3
and 4 models with $> 11\sigma$ confidence. For \hatcur{33} we may
reject the class 3 and 4 models with $> 7\sigma$ confidence.

For the class 2 models we consider two cases, one in which the
transiting planet orbits the brighter binary star component, and the
other in which the planet orbits the fainter binary star component. In
both cases the spectroscopic temperature and the photometric colors
constrain the brighter star to have a mass of $\ga 1\,\msun$ for both
systems. We find that the case of the planet orbiting the fainter
component can be rejected outright with $> 4\sigma$ confidence for
both \hatcur{32} and \hatcur{33}. The case of a planet orbiting the
brighter binary star component amounts to including third light in the
fit. When the secondary star contributes negligible light to the
system the model becomes indistinguishable from the best-fit case 1
model, such a model cannot be ruled out with the available
data. Instead we may place an upper limit on the mass of any binary
star companion. For \hatcur{32} we find that a putative companion must
have $M < 0.5\,\msun$ with $5\sigma$ confidence, while for \hatcur{33}
a companion must have $M < 0.55\,\msun$ with $5\sigma$
confidence. These translate into upper limits on the secondary to
primary $V$-band luminosity ratio of $\sim 1\%$ for both \hatcur{32} and
\hatcur{33}.

Based on the above discussion we conclude that the signals detected in
the \hatcur{32} and \hatcur{33} \lcs\ and RV curves are planetary in
nature.

\subsection{Global modeling of the data}
\label{sec:globmod}

We modeled the HATNet photometry, the follow-up photometry, and the
high-precision RV measurements using the procedure described in detail
by \citet{pal:2008,bakos:2010}. One significant difference from our
previous planet discoveries is that we use a \cite{mandel:2002}
transit model to describe the HATNet photometry, rather than a
simplified no-limb-darkening model. This was necessary due to the high
S/N HATNet detections, especially for \hatcur{32}. To describe the
follow-up light curves we use a \cite{mandel:2002} transit model
together with the simultaneous External-Parameter-Decorellation (EPD)
and Trend-Filtering Algorithm model of instrumental variations
\citep{bakos:2010}, and we use a Keplerian orbit to describe the RV
curves. For \hatcur{33} we include a term $\alpha_{\rm BS} \times {\rm
  BS}$ in the RV model, where $\alpha_{\rm BS}$ is a free parameter
describing the residual RV--BS correlation, as we found that this
significantly reduces the RV residuals. For \hatcur{32} we do not
include this correction because the residual RV--BS correlation is not
statistically significant. For each planetary system we fit two
models, one in which the eccentricity is fixed to zero, and another in
which the eccentricity varies. The parameters for each system are
listed in \reftabls{planetparam32}{planetparam33}. Neither system has
a clearly non-zero eccentricity. Using the \citet{lucy:1971} test we
find that there is a non-negligible $\sim 3\%$ probability that the
eccentricity found for \hatcur{32} arises by chance from a circular
orbit, while for \hatcur{33} the probability is $\sim 20\%$.

\begin{figure*}[!ht]
\plotone{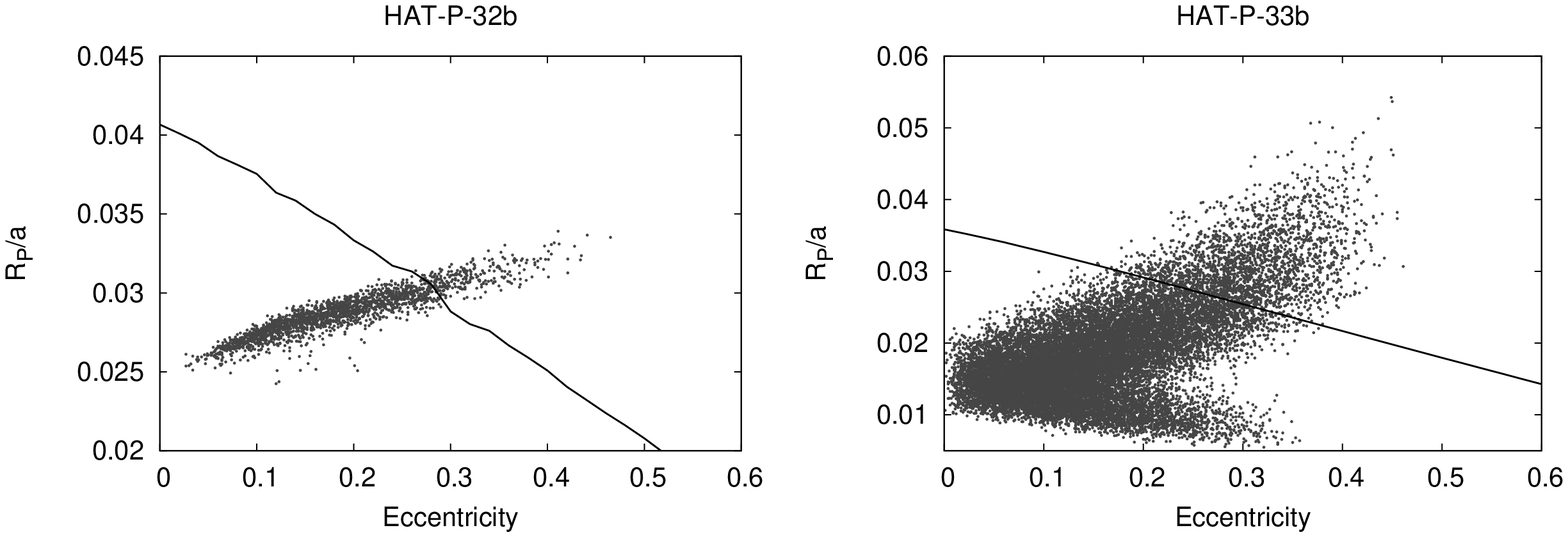}
\caption[]{
    Planet radius normalized to the semi-major axis vs.~eccentricity
    for \hatcurb{32} (left) and \hatcurb{33} (right). Each point
    corresponds to a single MCMC parameter realization. The solid
    lines show the maximum allowed transit-inferred planet radius as a
    function of eccentricity corresponding to a planet which fills its
    Roche-Lobe at periastron (\refsecl{rochelobecalc}). These are
    calculated using the median mass-ratio and argument of periastron
    for each system. The wiggles seen for \hatcurb{32} are due to
    numerical noise in the calculation.
\label{fig:rocheeccenconstraint}}
\end{figure*}

Because the resulting planets have large radii (particularly when the
eccentricity is allowed to vary), the assumption that each planet has
a spherical surface which lies well within its Roche Lobe may no
longer be valid. In particular the fact that a planet cannot exceed
its Roche Lobe places an upper limit on the transit-inferred radius of
the planet for a given semi-major axis, eccentricity, and
mass-ratio. Following the procedure described in the appendix, we
impose the constraint that the planet cannot exceed its Roche Lobe in
determining the parameters and errors for each
system. \reffigl{rocheeccenconstraint} shows an approximation of the
constraints on $\rpl/a$ and $e$ for each system. In both cases the
best-fit solutions are below this limit, however the uncertainties on
the parameters (particularly the upper uncertainties on the
eccentricities and radii) are reduced by imposing the Roche Lobe
constraint.

%
%
\ifthenelse{\boolean{emulateapj}}{
    \begin{deluxetable*}{lcc}
}{
    \begin{deluxetable}{lcc}
}
\tabletypesize{\scriptsize}
\tablecaption{Orbital and planetary parameters for \hatcurb{32}\label{tab:planetparam32}}
\tablehead{
    \multicolumn{1}{c}{~~~~~~~~~~~~~~~Parameter~~~~~~~~~~~~~~~} &
    \multicolumn{1}{c}{Value\tablenotemark{a}} &
    \multicolumn{1}{c}{Value} \\
    &
    \multicolumn{1}{c}{Circular} &
    \multicolumn{1}{c}{Eccentric}
}
\startdata
\noalign{\vskip -3pt}
\sidehead{\Lc{} parameters}
~~~$P$ (days)             \dotfill    & $\hatcurLCP{32}$  & $\hatcurLCP{36}$\\
~~~$T_c$ (${\rm BJD}$)    
      \tablenotemark{b}   \dotfill    & $\hatcurLCT{32}$  & $\hatcurLCT{36}$\\
~~~$T_{14}$ (days)
      \tablenotemark{b}   \dotfill    & $\hatcurLCdur{32}$  & $\hatcurLCdur{36}$\\
~~~$T_{12} = T_{34}$ (days)
      \tablenotemark{b}   \dotfill    & $\hatcurLCingdur{32}$  & $\hatcurLCingdur{36}$\\
~~~$\arstar$              \dotfill    & $\hatcurPPar{32}$  & $\hatcurPPar{36}$\\
~~~$\zrstar$              \dotfill    & $\hatcurLCzeta{32}$\phn & $\hatcurLCzeta{36}$\\
~~~$\rpl/\rstar$          \dotfill    & $\hatcurLCrprstar{32}$  & $\hatcurLCrprstar{36}$\\
~~~$b^2$                  \dotfill    & $\hatcurLCbsq{32}$  & $\hatcurLCbsq{36}$\\
~~~$b \equiv a \cos i/\rstar$
                          \dotfill    & $\hatcurLCimp{32}$  & $\hatcurLCimp{36}$\\
~~~$i$ (deg)              \dotfill    & $\hatcurPPi{32}$\phn & $\hatcurPPi{36}$\\

\sidehead{Limb-darkening coefficients \tablenotemark{c}}
~~~$c_1,i$ (linear term)  \dotfill    & $\hatcurLBii{32}$  & $\hatcurLBii{36}$\\
~~~$c_2,i$ (quadratic term) \dotfill  & $\hatcurLBiii{32}$  & $\hatcurLBiii{36}$\\
~~~$c_1,z$               \dotfill    & $\hatcurLBiz{32}$  & $\hatcurLBiz{36}$\\
~~~$c_2,z$               \dotfill    & $\hatcurLBiiz{32}$  & $\hatcurLBiiz{36}$\\
~~~$c_1,g$               \dotfill    & $\hatcurLBig{32}$  & $\hatcurLBig{36}$\\
~~~$c_2,g$               \dotfill    & $\hatcurLBiig{32}$  & $\hatcurLBiig{36}$\\

\sidehead{RV parameters}
~~~$K$ (\ms)              \dotfill    & $\hatcurRVK{32}$\phn\phn & $\hatcurRVK{36}$\\
~~~$e \cos (\omega)$\tablenotemark{d} 
                          \dotfill    & $\hatcurRVk{32}$\phs & $\hatcurRVk{36}$\\
~~~$e \sin (\omega)$\tablenotemark{d}
                          \dotfill    & $\hatcurRVh{32}$  & $\hatcurRVh{36}$\\
~~~$e$                    \dotfill    & $\hatcurRVeccen{32}$  & $\hatcurRVeccen{36}$\\
~~~$\omega$ (deg)         \dotfill    & $\hatcurRVomega{32}$\phn & $\hatcurRVomega{36}$\\
~~~RV jitter (\ms)        \dotfill    & \hatcurRVjitter{32} & \hatcurRVjitter{36}\\

\sidehead{Secondary eclipse parameters}
~~~$T_s$ (BJD)            \dotfill    & $\hatcurXsecondary{32}$  & $\hatcurXsecondary{36}$\\
~~~$T_{s,14}$              \dotfill   & $\hatcurXsecdur{32}$  & $\hatcurXsecdur{36}$\\
~~~$T_{s,12}$              \dotfill   & $\hatcurXsecingdur{32}$  & $\hatcurXsecingdur{36}$\\

\sidehead{Planetary parameters}
~~~$\mpl$ ($\mjup$)       \dotfill    & $\hatcurPPmlong{32}$  & $\hatcurPPmlong{36}$\\
~~~$\rpl$ ($\rjup$)       \dotfill    & $\hatcurPPrlong{32}$  & $\hatcurPPrlong{36}$\\
~~~$C(\mpl,\rpl)$
    \tablenotemark{e}     \dotfill    & $\hatcurPPmrcorr{32}$  & $\hatcurPPmrcorr{36}$\\
~~~$\rhopl$ (\gcmc)       \dotfill    & $\hatcurPPrho{32}$  & $\hatcurPPrho{36}$\\
~~~$\log g_p$ (cgs)       \dotfill    & $\hatcurPPlogg{32}$  & $\hatcurPPlogg{36}$\\
~~~$a$ (AU)               \dotfill    & $\hatcurPParel{32}$  & $\hatcurPParel{36}$\\
~~~$T_{\rm eq}$ (K)        \dotfill   & $\hatcurPPteff{32}$  & $\hatcurPPteff{36}$\\
~~~$\Theta$\tablenotemark{f} \dotfill & $\hatcurPPtheta{32}$  & $\hatcurPPtheta{36}$\\
~~~$\langle F \rangle$ ($10^{9}$\ergscmsq) \tablenotemark{g}
                          \dotfill    & $\hatcurPPfluxavg{32}$  & $\hatcurPPfluxavg{36}$\\
\enddata
\tablenotetext{a}{
    The eccentricity of \hatcur{32} is
    poorly constrained. We list separately the parameters obtained
    when a circular orbit is fixed, and when the eccentricity is
    allowed to vary.
}
\tablenotetext{b}{
    \ensuremath{T_c}: Reference epoch of mid transit that
    minimizes the correlation with the orbital period.
    \ensuremath{T_{14}}: total transit duration, time
    between first to last contact;
    \ensuremath{T_{12}=T_{34}}: ingress/egress time, time between first
    and second, or third and fourth contact.
}
\tablenotetext{c}{
    Values for a quadratic law, adopted from the tabulations by
    \cite{claret:2004} according to the spectroscopic (SME) parameters
    listed in \reftabl{stellar32}.
}
\tablenotetext{d}{
    Lagrangian orbital parameters derived from the global modeling, 
    and primarily determined by the RV data. 
}
\tablenotetext{e}{
    Correlation coefficient between the planetary mass \mpl\ and radius
    \rpl.
}
\tablenotetext{f}{
    The Safronov number is given by $\Theta = \frac{1}{2}(V_{\rm
    esc}/V_{\rm orb})^2 = (a/\rpl)(\mpl / \mstar )$
    \citep[see][]{hansen:2007}.
}
\tablenotetext{g}{
    Incoming flux per unit surface area, averaged over the orbit.
}
\ifthenelse{\boolean{emulateapj}}{
    \end{deluxetable*}
}{
    \end{deluxetable}
}
%

%
%
\ifthenelse{\boolean{emulateapj}}{
    \begin{deluxetable*}{lcc}
}{
    \begin{deluxetable}{lcc}
}
\tabletypesize{\scriptsize}
\tablecaption{Orbital and planetary parameters for \hatcurb{33}\label{tab:planetparam33}}
\tablehead{
    \multicolumn{1}{c}{~~~~~~~~~~~~~~~Parameter~~~~~~~~~~~~~~~} &
    \multicolumn{1}{c}{Value} &
    \multicolumn{1}{c}{Value} \\
    &
    \multicolumn{1}{c}{Circular} &
    \multicolumn{1}{c}{Eccentric}
}
\startdata
\noalign{\vskip -3pt}
\sidehead{\Lc{} parameters}
~~~$P$ (days)             \dotfill    & $\hatcurLCP{33}$  & $\hatcurLCP{35}$              \\
~~~$T_c$ (${\rm BJD}$)    
      \tablenotemark{b}   \dotfill    & $\hatcurLCT{33}$  & $\hatcurLCT{35}$              \\
~~~$T_{14}$ (days)
      \tablenotemark{b}   \dotfill    & $\hatcurLCdur{33}$  & $\hatcurLCdur{35}$            \\
~~~$T_{12} = T_{34}$ (days)
      \tablenotemark{b}   \dotfill    & $\hatcurLCingdur{33}$  & $\hatcurLCingdur{35}$         \\
~~~$\arstar$              \dotfill    & $\hatcurPPar{33}$  & $\hatcurPPar{35}$             \\
~~~$\zrstar$              \dotfill    & $\hatcurLCzeta{33}$\phn  & $\hatcurLCzeta{35}$\phn     \\
~~~$\rpl/\rstar$          \dotfill    & $\hatcurLCrprstar{33}$  & $\hatcurLCrprstar{35}$        \\
~~~$b^2$                  \dotfill    & $\hatcurLCbsq{33}$  & $\hatcurLCbsq{35}$            \\
~~~$b \equiv a \cos i/\rstar$
                          \dotfill    & $\hatcurLCimp{33}$  & $\hatcurLCimp{35}$            \\
~~~$i$ (deg)              \dotfill    & $\hatcurPPi{33}$\phn  & $\hatcurPPi{35}$\phn        \\

\sidehead{Limb-darkening coefficients \tablenotemark{c}}
~~~$c_1,i$ (linear term)  \dotfill    & $\hatcurLBii{33}$  & $\hatcurLBii{35}$             \\
~~~$c_2,i$ (quadratic term) \dotfill  & $\hatcurLBiii{33}$  & $\hatcurLBiii{35}$            \\
~~~$c_1,z$               \dotfill    & $\hatcurLBiz{33}$  & $\hatcurLBiz{35}$             \\
~~~$c_2,z$               \dotfill    & $\hatcurLBiiz{33}$  & $\hatcurLBiiz{35}$            \\
~~~$c_1,g$               \dotfill    & $\hatcurLBig{33}$  & $\hatcurLBig{35}$             \\
~~~$c_2,g$               \dotfill    & $\hatcurLBiig{33}$  & $\hatcurLBiig{35}$            \\

\sidehead{RV parameters}
~~~$K$ (\ms)              \dotfill    & $\hatcurRVK{33}$\phn\phn  & $\hatcurRVK{35}$\phn\phn    \\
~~~$e \cos (\omega)$\tablenotemark{d} 
                          \dotfill    & $\hatcurRVk{33}$\phs  & $\hatcurRVk{35}$\phs        \\
~~~$e \sin (\omega)$\tablenotemark{d}
                          \dotfill    & $\hatcurRVh{33}$  & $\hatcurRVh{35}$              \\
~~~$e$                    \dotfill    & $\hatcurRVeccen{33}$  & $\hatcurRVeccen{35}$          \\
~~~$\omega$ (deg)         \dotfill    & $\hatcurRVomega{33}$\phn  & $\hatcurRVomega{35}$\phn    \\
~~~RV jitter (\ms)        \dotfill    & \hatcurRVjitter{33} & \hatcurRVjitter{35}           \\
~~~$\alpha_{\rm BS}$\tablenotemark{e}        \dotfill   & \hatcurRVBScorr{33} & \hatcurRVBScorr{35} \\

\sidehead{Secondary eclipse parameters}
~~~$T_s$ (BJD)            \dotfill    & $\hatcurXsecondary{33}$  & $\hatcurXsecondary{35}$       \\
~~~$T_{s,14}$              \dotfill   & $\hatcurXsecdur{33}$  & $\hatcurXsecdur{35}$          \\
~~~$T_{s,12}$              \dotfill   & $\hatcurXsecingdur{33}$  & $\hatcurXsecingdur{35}$       \\

\sidehead{Planetary parameters}
~~~$\mpl$ ($\mjup$)       \dotfill    & $\hatcurPPmlong{33}$  & $\hatcurPPmlong{35}$          \\
~~~$\rpl$ ($\rjup$)       \dotfill    & $\hatcurPPrlong{33}$  & $\hatcurPPrlong{35}$          \\
~~~$C(\mpl,\rpl)$
    \tablenotemark{f}     \dotfill    & $\hatcurPPmrcorr{33}$  & $\hatcurPPmrcorr{35}$         \\
~~~$\rhopl$ (\gcmc)       \dotfill    & $\hatcurPPrho{33}$  & $\hatcurPPrho{35}$            \\
~~~$\log g_p$ (cgs)       \dotfill    & $\hatcurPPlogg{33}$  & $\hatcurPPlogg{35}$           \\
~~~$a$ (AU)               \dotfill    & $\hatcurPParel{33}$  & $\hatcurPParel{35}$           \\
~~~$T_{\rm eq}$ (K)        \dotfill   & $\hatcurPPteff{33}$  & $\hatcurPPteff{35}$           \\
~~~$\Theta$\tablenotemark{g} \dotfill & $\hatcurPPtheta{33}$  & $\hatcurPPtheta{35}$          \\
~~~$\langle F \rangle$ ($10^{9}$\ergscmsq) \tablenotemark{h}
                          \dotfill    & $\hatcurPPfluxavg{33}$  & $\hatcurPPfluxavg{35}$        \\ [-1.5ex]
\enddata
\tablenotetext{a}{
    The eccentricity of \hatcur{33} is
    poorly constrained. We list separately the parameters obtained
    when a circular orbit is fixed, and when the eccentricity is
    allowed to vary.
}
\tablenotetext{b}{
    \ensuremath{T_c}: Reference epoch of mid transit that
    minimizes the correlation with the orbital period.
    \ensuremath{T_{14}}: total transit duration, time
    between first to last contact;
    \ensuremath{T_{12}=T_{34}}: ingress/egress time, time between first
    and second, or third and fourth contact.
}
\tablenotetext{c}{
    Values for a quadratic law, adopted from the tabulations by
    \cite{claret:2004} according to the spectroscopic (SME) parameters
    listed in \reftabl{stellar32}.
}
\tablenotetext{d}{
    Lagrangian orbital parameters derived from the global modeling, 
    and primarily determined by the RV data. 
}
\tablenotetext{e}{
    Parameter describing a linear dependence of the RVs on the BS values.
}
\tablenotetext{f}{
    Correlation coefficient between the planetary mass \mpl\ and radius
    \rpl.
}
\tablenotetext{g}{
    The Safronov number is given by $\Theta = \frac{1}{2}(V_{\rm
    esc}/V_{\rm orb})^2 = (a/\rpl)(\mpl / \mstar )$
    \citep[see][]{hansen:2007}.
}
\tablenotetext{h}{
    Incoming flux per unit surface area, averaged over the orbit.
}
\ifthenelse{\boolean{emulateapj}}{
    \end{deluxetable*}
}{
    \end{deluxetable}
}
%



\section{Discussion}
\label{sec:discussion}

\begin{figure*}[!ht]
\plotone{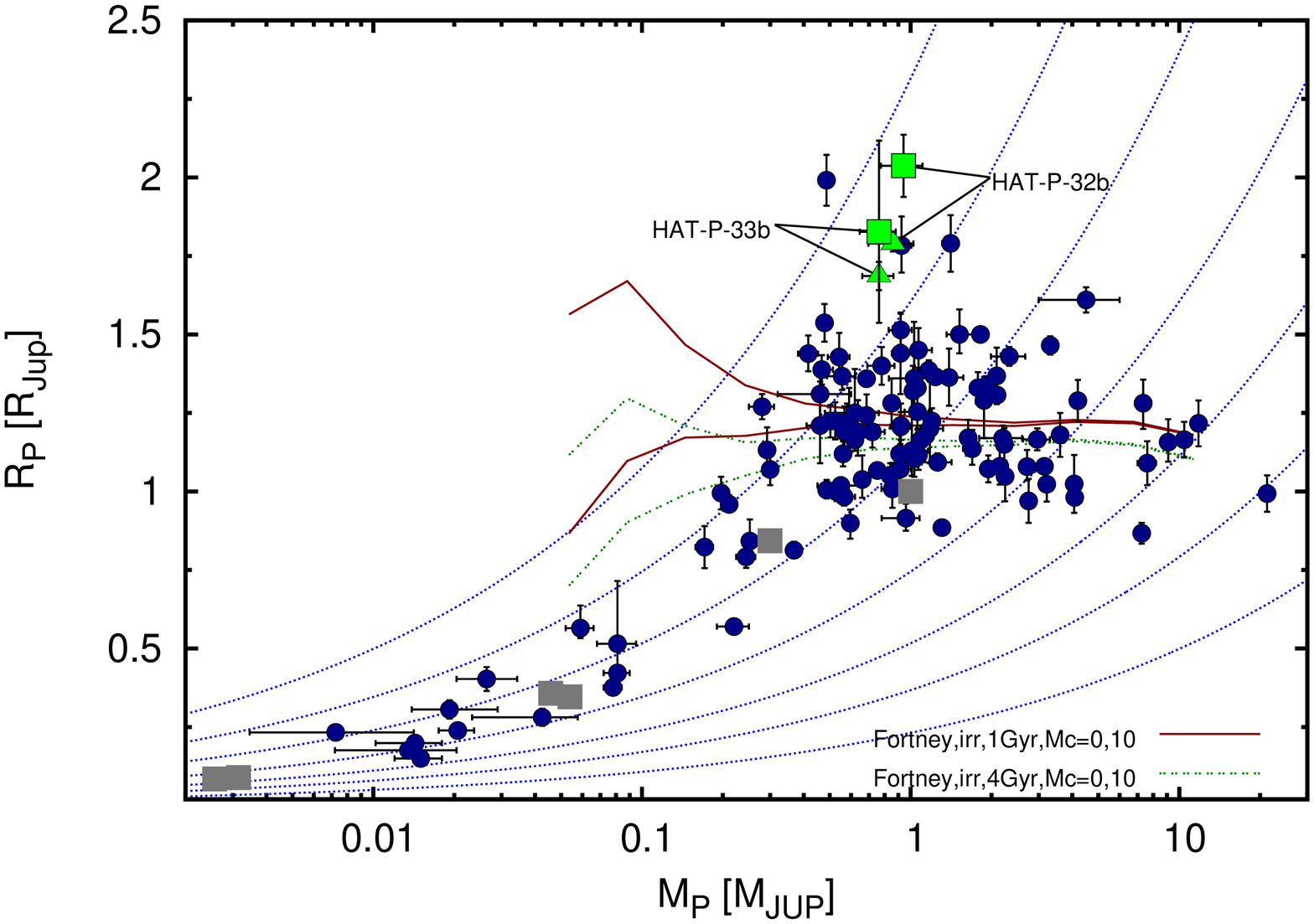}
\caption{ 
   Mass--radius diagram of TEPs. \hatcurb{32} and \hatcurb{33} are
   indicated. The triangles indicate the parameters for assumed
   circular orbits, while the squares indicate the parameters when the
   eccentricity is allowed to vary. Filled circles are all other TEPS,
   and filled squares are solar system planets. We also show lines of
   constant density (dotted lines running from lower left to upper
   right) and theoretical planet mass-radius relations from
   \cite{fortney:2007}.
\label{fig:exomr}}
\end{figure*}

\begin{figure*}[!ht]
\plotone{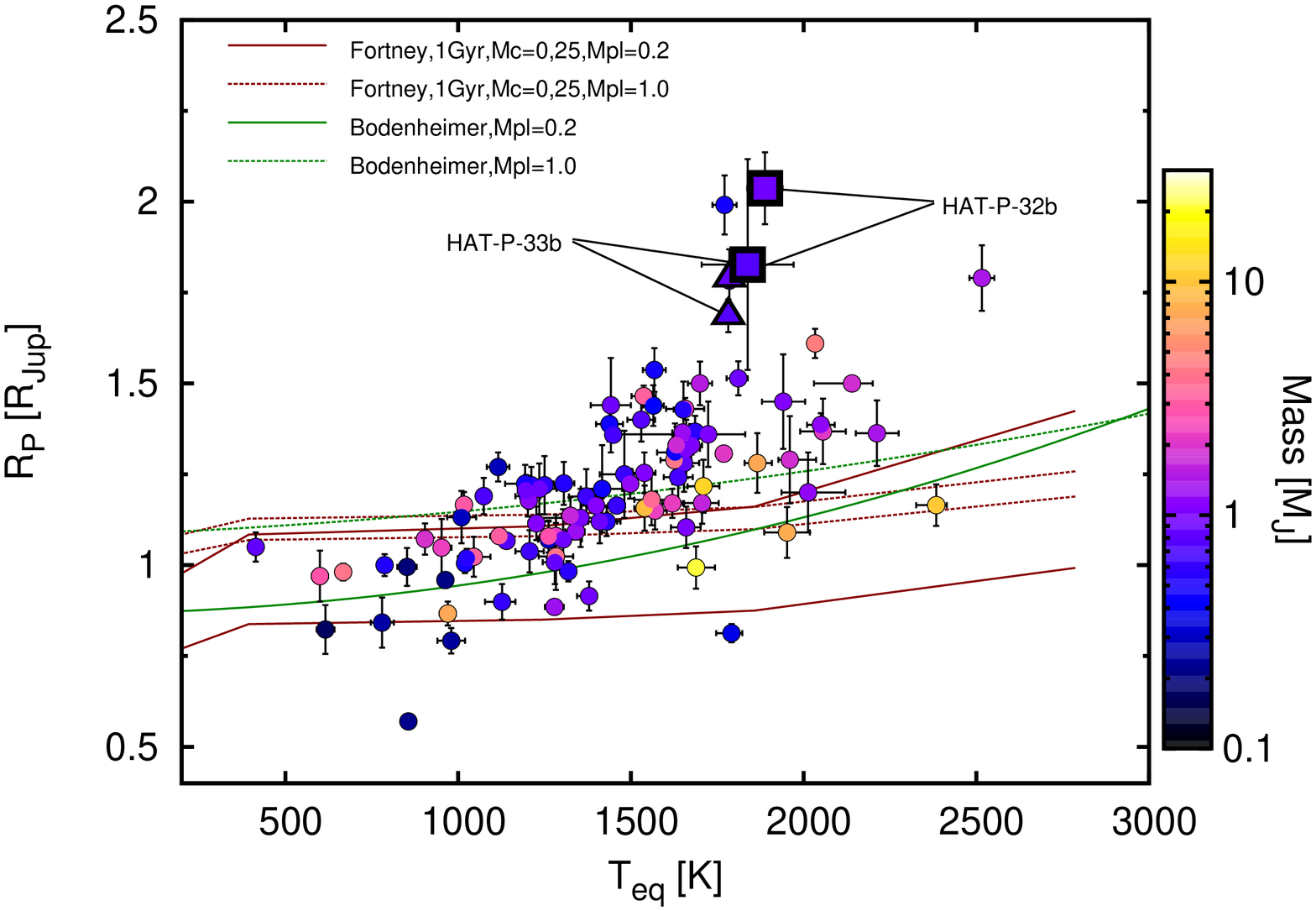}
\plotone{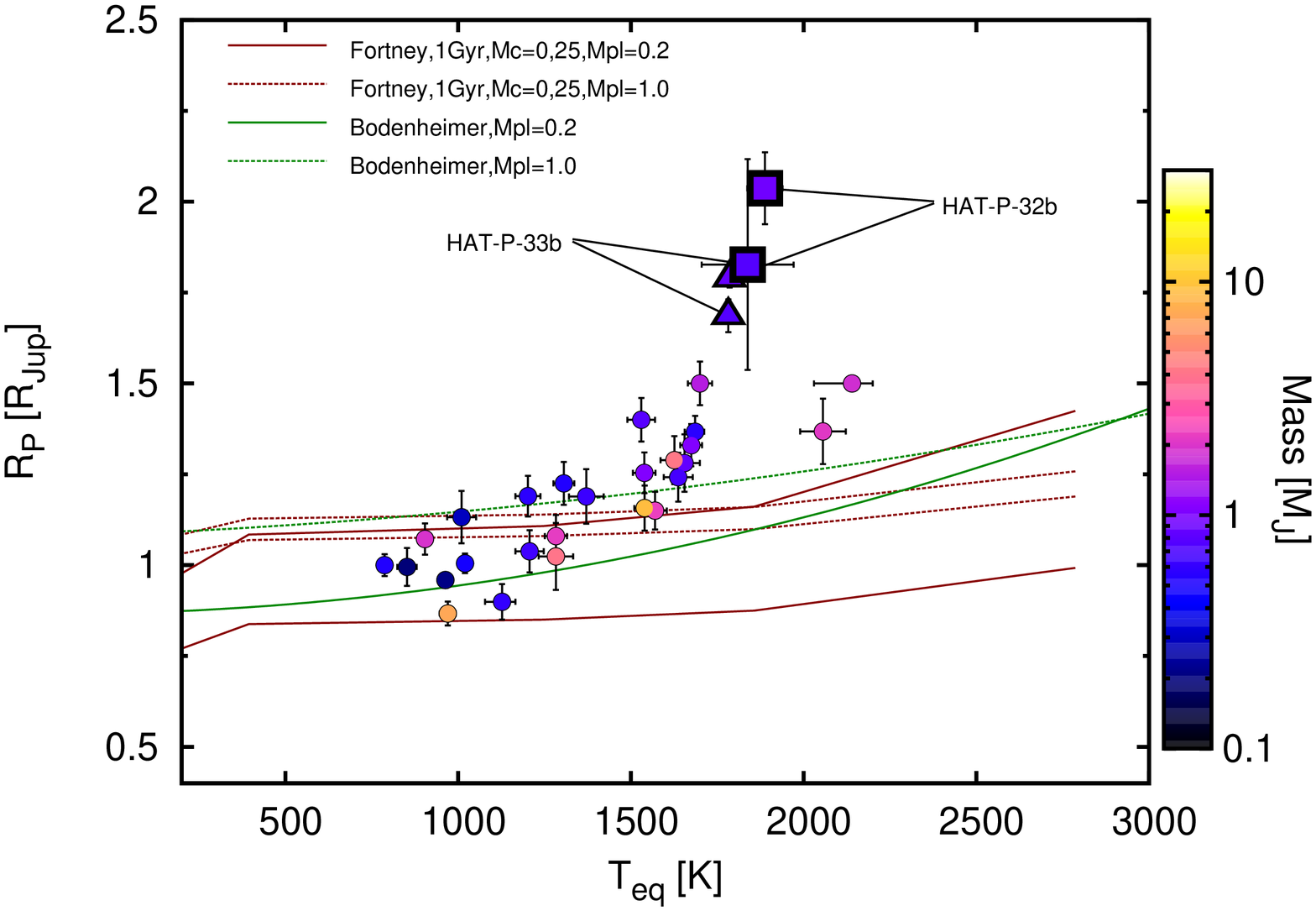}
\caption{ 
Top: T$_{\rm eq}$--radius diagram of TEPs with $0.1\,\mjup <
\mpl$. \hatcurb{32} and \hatcurb{33} are indicated with triangles
(circular models) and squares (eccentric models). The shading of the
planets indicates their masses. \hatcurb{32} and \hatcurb{33} continue
the trend of higher T$_{\rm eq}$ planets having larger radii. Bottom:
the same as above, here we only show planets discovered by the HATNet
survey. The T$_{\rm eq}$--radius correlation is clearly apparent when
restricted to this sample of planets, which we take as evidence that
the correlation is not due to combining planets found by different
surveys that have different selection biases.
\label{fig:exoteqr}}
\end{figure*}

We have presented the discovery of two planets, \hatcurb{32} and
\hatcurb{33}, which have radii that are among the largest measured to
date for all transiting exoplanets. \reffigl{exomr} shows the location
of these two planets on a mass--radius diagram, while
\reffigl{exoteqr} shows them on an equilibrium temperature--radius
diagram. For each planet the radius determination depends strongly on
the eccentricity, which is poorly constrained for both systems due to
the high stellar jitters. If \hatcurb{32} has a circular orbit, its
radius would be $\sim 1.8$\,\rjup, but it could be as large as $\sim
2.0$\,\rjup\ if the planet is eccentric. \hatcurb{33} has a slightly
smaller radius of $\sim 1.7$\,\rjup\ if it is on a circular orbit, but
its radius could also be larger ($\sim 1.8$\,\rjup) if it is
eccentric.

There are only four known transiting planets with radii that are
comparable to \hatcurb{32} or \hatcurb{33}. These are WASP-17b ($R =
1.99 \pm 0.08$\,\rjup; \citealp{anderson:2011}), WASP-12b ($R = 1.79
\pm 0.09$\,\rjup; \citealp{hebb:2009}), TrES-4b ($R = 1.78 \pm
0.09$\,\rjup; \citealp{sozzetti:2009}), and Kepler-7b ($R = 1.614 \pm
0.015$\,\rjup; \citealp{demory:2011}). These four planets orbit F
stars, as do \hatcurb{32} and \hatcurb{33}. The relatively high
luminosities of these stars result in relatively high planet
equilibrium temperatures ($T_{\rm eq} > 1600$\,K in all cases).  The
equilibrium temperature in turn is correlated with planet radius (see
references in \refsecl{introduction}). Nonetheless, there must be
additional factors which cause planets like \hatcurb{32}, \hatcurb{33}
and WASP-17b to have radii of $1.7$--$2$\,\rjup with $T_{\rm eq} \sim
1800$\,K, while the much hotter planet WASP-18b ($2380$\,K;
\citealp{hellier:2009}) has a much less inflated radius of only
$1.165$\,\rjup. One significant difference between WASP-18b and the
higher radii planets are the planet masses. While \hatcurb{32},
\hatcurb{33} and WASP-17b have sub-Jupiter masses, WASP-18b has a mass
of $\sim 10$\,\mjup. As seen in \reffigl{exoteqr} it appears that
there may be a mass dependence to the $T_{\rm eq}$--radius relation,
with lower mass planets being more strongly impacted by high
equilibrium temperature than higher mass planets. A mass-dependence of
this type has been predicted by several planet inflation mechanisms
(e.g. Fig.~10 of \citealp{guillot:2005}; see also
\citealp{batygin:2011}).

Due to the high jitters of the two stars studied in this paper,
further high-precision RV observations will not significantly
constrain the eccentricities of the planetary systems. A more
promising method would be to observe the planetary occultations with
the {\em Spitzer} space telescope. Fortunately both stars are
relatively bright, ($K_{S} \sim 10.0$\,mag in both cases), and have
expected occultations deeper than $0.1$\% in both the 3.5$\,\mu$m and
4.6$\,\mu$m bandpasses. Thus we expect that it should be possible to
obtain high S/N occultation events (S/N$> 10$) with {\em Spitzer} for
both systems.

From standard tidal theory we expect the circularization time-scales
to be much shorter than the $\ga 2$\,Gyr ages of the systems. Using
equation~1 of \cite{matsumura:2008}, and assuming planetary and
stellar tidal damping factors of $Q^{\prime}_{\rm p} = Q^{\prime}_{\rm
  \star} = 10^6$, the expected circularization timescales are $\sim
3$\,Myr and $\sim 30$\,Myr for \hatcurb{32} and \hatcurb{33}
respectively. We note, however, that \citet{penev:2011} have recently
argued that standard tidal theory, which is calibrated from
observations of binary stars, significantly overestimates the tidal
interaction between planets and stars and thereby underestimates the
true circularization time-scale. It is possible that a short period
planet may maintain an eccentric orbit for the entire life of the
system.

Finally we note that the difficulty of confirming the planets
presented here illustrates the selection bias imposed on transit
surveys by the need for RV confirmation. For both planets the HATNet
transit detection was clear and robust. This, together with the high
S/N box-shaped transits observed with KeplerCam, motivated us to
continue obtaining high-precision RVs for these objects despite the
initial RVs not phasing with the photometric ephemerides, and the
stars showing significant spectral-line bisector span variations. If
either target had a shallower, less obviously planet-like transit, it
is likely that we would not have continued the intensive RV monitoring
necessary for confirmation, and it is also likely that we would not
have been able to conclusively rule out blends based on analyzing the
light curves. Moreover, if either planet had a significantly lower
mass (Saturn-mass or smaller), such that the orbital variation could
not be detected, it is also likely that the planet would not have been
confirmed.

\acknowledgements 

HATNet operations have been funded by NASA grants NNG04GN74G,
NNX08AF23G and SAO IR\&D grants. Work of G.\'A.B.~and J.~Johnson were
supported by the Postdoctoral Fellowship of the NSF Astronomy and
Astrophysics Program (AST-0702843 and AST-0702821, respectively). GT
acknowledges partial support from NASA grant NNX09AF59G. We
acknowledge partial support also from the Kepler Mission under NASA
Cooperative Agreement NCC2-1390 (D.W.L., PI). G.K.~thanks the
Hungarian Scientific Research Foundation (OTKA) for support through
grant K-81373. This research has made use of Keck telescope time
granted through NOAO (A285Hr, A146Hr, A201Hr, A289Hr), NASA (N128Hr,
N145Hr, N049Hr, N018Hr, N167Hr, N029Hr), and the NOAO Gemini/Keck
time-exchange program (G329Hr). We gratefully acknowledge F.~Bouchy,
F.~Pont and the SOPHIE team for their efforts in gathering OHP/SOPHIE
observations of \hatcur{33}.

\begin{appendix}

\section{Calculating the Transit-Inferred Radius of an Eccentric, Roche-Lobe Filling Planet}\label{sec:rochelobecalc}

The condition that the surface of a planet cannot extend beyond its
Roche-Lobe (assuming the system is {\em not} an overcontact binary, which is
true for the systems considered in this paper since in both cases the
inferred stellar radius is well within the stellar Roche Lobe for
eccentricities that are consistent with the RV curves) sets a maximum
limit on its size for a given semi-major axis, eccentricity, and
star-planet mass ratio. This in turn places an upper limit on the
radius of a planet inferred from a transit measurement. Here we
briefly review how to calculate this constraint.

Following \cite{wilson:1979}, the binary potential for the general
case of nonsynchronously rotating components on a non-circular orbit
is given by:
\begin{equation}
\Omega = r^{-1} + q \left[ (D^2 + r^2 - 2r\lambda D)^{-1/2} -
  r\lambda/D^2 \right] + \frac{1}{2} F^{2}(1 + q)r^{2}(1 - \nu^2).
\label{eqn:roche}
\end{equation}
The polar coordinates $r$, $\theta$, and $\phi$ ($\theta$ is the polar
angle) have an origin at the center of one of the binary components
(in our case we choose the planet), distance is measured in units of
the semi-major axis of the relative orbit, $\lambda$ and $\nu$ are
direction cosines ($\lambda = \sin\theta\cos\phi$, $\nu =
\cos\theta$), $D$ is the instantaneous separation between the planet
and star ($D = 1 - e\cos E$ where $e$ is the eccentricity and $E$ is
the eccentric anomaly), $q$ is the mass ratio ($q = \mstar / \mpl$ in
our case), and $F$ is the synchronicity parameter equal to the ratio
of the angular rotation velocity of the component at the origin (the
planet in our case) to the ``average'' angular velocity of the orbit
($2\pi/ P$ where $P$ is the orbital period). For an eccentric system
the tidal interaction drives the components towards pseudo-synchronous
rotation, which is between the average angular velocity and the
angular velocity at periastron \citep[eq.~42]{hut:1981}:
\begin{equation}
F = \frac{1+\frac{15}{2}e^{2}+\frac{45}{8}e^{4}+\frac{5}{16}e^6}{(1+3e^2+\frac{3}{8}e^4)(1-e^2)^{3/2}}.
\end{equation}
We assume pseudo-synchronous rotation for the planet.

The surface of a gas giant planet is expected to follow a surface of
constant potential. For the eccentric case the value and shape of the
surface potential varies with the orbital phase, in this case
\cite{wilson:1979} argues that to a good approximation the volume of
the object is constant over the orbit, and suggests a procedure, which
we adopt, for finding the surface of a Roche-Lobe filling object at
different orbital phases. Because the volume of the Roche Lobe is
smallest at periastron, it follows that an eccentric planet which
fills its Roche Lobe will do so only at periastron. The potential $\Omega_{0}$
corresponding to the Roche Lobe of the planet at periastron
can be determined by taking $D = 1-e$ and finding $r$
between $0$ and $D$ that minimizes Eq.~\ref{eqn:roche} for $\nu = 0$,
$\lambda = 1$. The volume $V_{0}$ of this surface can be calculated
numerically. We perform a Monte Carlo integration randomly generating
points uniformly distributed within a spherical shell with an inner
radius $r_{0}$ that satisfies the condition
$\Omega(r_{0},\theta,\lambda) > \Omega_{0}$ for all $\theta$ and
$\lambda$ and an outer radius $r_{1}$ that satisfies
$\Omega(r_{1}),\theta,\lambda) < \Omega_{0}$ for all $\theta$ and
$\lambda$, and taking the volume to be equal to
\begin{equation}
V_{0} = \frac{4}{3}\pi(r_{0}^3 + f(r_{1}^3 - r_{0}^3))
\end{equation}
where $f$ is the fraction of generated points with $\Omega >
\Omega_{0}$. The surface potential during transit $\Omega_{1}$ may
then be determined by setting $D$ equal to the appropriate value at
transit and finding $\Omega$ such that $V(\Omega = \Omega_{1}) =
V_{0}$. We solve this using a simple bisection search noting that the
volume of the Roche Lobe potential at transit phase (which can be
determined as at periastron) is greater than $V_{0}$. Finally the
radius of the planet which would be inferred from a transit
observation (corresponding to the radius of a circle with area equal
to the projected area of the planet surface potential as viewed by an
Earth-bound observer) is given approximately by
\begin{equation}
R_{p,RL} = \sqrt{yz}
\end{equation}
where $y = r$ such that $\Omega(r, \theta=\pi/2, \phi = \pi/2) =
\Omega_{1}$, $z = r$ such that $\Omega(r, \theta=0, \phi = 0) =
\Omega_{1}$, and we assume for simplicity an edge-on orbit (which is
reasonable for a transiting planet). We have also conducted a Monte
Carlo integration to calculate the projected area of the planet
surface potential accounting for inclination, and found the difference
from the approximation to be negligible (less than 0.1\%).

In principle one may apply the Roche-Lobe constraint in the global
modeling of the photometry and RVs by calculating $R_{p,RL}$ for each
set of trial MCMC parameters, and rejecting the trial if it yields
$\rpl / a > R_{p,RL}$. In practice this is unwieldy because of the
slow numerical integrations required by the above procedure. We
therefore searched for an analytic approximation to $R_{p,RL}$ that
depends on the mass ratio $q$, the eccentricity $e$ and the argument
of periastron $\omega$. We numerically determined $R_{p,RL}$ over a
grid of parameters spanning $10 \leq q \leq 10^5$, $0 \leq e \leq
0.5$, and $\pi/2 \leq \omega \leq 3\pi/2$, and find that the following
empirically chosen analytic function reproduces $R_{p,RL}$ to $\sim
1\%$ accuracy over this parameter range:
\begin{equation}
R_{p,RL}(q,\omega,e) \approx f_{1}(q)g_{1}(\omega)e^2+f_{2}(q)g_{2}(\omega)e+f_{3}(q)g_{3}(\omega)
\label{eqn:rpapprox}
\end{equation}
where
\begin{eqnarray*}
f_{1}(q) & = & -\exp(a_1\ln(q)^2+a_2\ln(q)+a_3) \\
g_{1}(\omega) & = & a_4\cos(\omega - \pi/2)+1 \\
f_{2}(q) & = & -\exp(a_5\ln(q)^2+a_6\ln(q)+a_7) \\
g_{2}(\omega) & = & a_8\cos(\omega - \pi/2)+1 \\
f_{3}(q) & = & \exp(a_9\ln(q)^2+a_{10}\ln(q)+a_{11}) \\
g_{3}(\omega) & = & a_{12}\cos(\omega - \pi/2)+1 \\
\end{eqnarray*}
and $a_1$ through $a_{12}$ are fitted parameters given in
\reftabl{rpapprox}.

\ifthenelse{\boolean{emulateapj}}{
    \begin{deluxetable}{lr}
}{
    \begin{deluxetable}{lr}
}
\tablewidth{0pc}
\tabletypesize{\scriptsize}
\tablecaption{
    Best fit parameters for equation~\ref{eqn:rpapprox}
    \label{tab:rpapprox}
}
\tablehead{
    \multicolumn{1}{c}{Parameter}             &
    \multicolumn{1}{c}{Value}
}
\startdata
$a_1$ & $0.0033 \pm 0.0011$ \\
$a_2$ & $-0.385 \pm 0.016$ \\
$a_3$ & $-1.673 \pm 0.056$ \\
$a_4$ & $-0.487 \pm 0.036$ \\
$a_5$ & $-0.00222 \pm 0.00032$ \\
$a_6$ & $-0.2927 \pm 0.0046$ \\
$a_7$ & $-1.199 \pm 0.017$ \\
$a_8$ & $0.1397 \pm 0.0086$ \\
$a_9$ & $-1.434 \times 10^{-3} \pm 6.9 \times 10^{-5}$ \\
$a_{10}$ & $-0.30528 \pm 0.00097$ \\
$a_{11}$ & $-0.9238 \pm 0.0031$ \\
$a_{12}$ & $-0.00399 \pm 0.00088$ \\
\enddata 
\ifthenelse{\boolean{emulateapj}}{
    \end{deluxetable}
}{
    \end{deluxetable}
}

\end{appendix}


\end{document}